\documentclass[journal,final,letterpaper,twoside,twocolumn]{IEEEtran}

\usepackage[utf8]{inputenc} 
\usepackage[T1]{fontenc}

\usepackage{amsmath}
\usepackage{amsthm}
\usepackage{amssymb}
\usepackage{mathrsfs}
\usepackage{mathtools}
\DeclarePairedDelimiter{\floor}{\lfloor}{\rfloor}

\usepackage[ruled,vlined]{algorithm2e}

\usepackage{booktabs}
\usepackage[caption=false,font=footnotesize]{subfig} 
\usepackage{graphicx}
\DeclareGraphicsExtensions{.pdf,.jpeg,.png}
\usepackage[noadjust]{cite}
\usepackage{textcomp} 

\usepackage{url}
\usepackage{hyperref}  
\hypersetup{colorlinks,%
            citecolor=red,%
            filecolor=black,%
            linkcolor=blue,%
            urlcolor=blue}

\def\Y{\mathbf{Y}}
\def\M{\mathbf{M}}
\def\A{\mathbf{A}}
\def\dM{\mathbf{dM}}

\def\Delt{\mathbf{\Delta}}

\def\a_n{\mathbf{a}_n}
\def\R[#1]#2{\mathbb{R}^{#1 \times #2}}  
\def\Zero[#1]#2{\mathbf{0}_{#1,#2}}

\def\Norm_fro#1{\left \Vert #1 \right \Vert_{\text{F}}^2}%
\def\norm_2#1{\left \Vert #1 \right \Vert_{2}^2}
\def\norme_2#1{\left \Vert #1 \right \Vert_{2}}%
\def\argmin#1{\underset{#1}{\text{arg min }} }

\def\mat#1{\mathbf{#1}}
\def\row#1{\mathbf{\widetilde{#1}}}
\def\Tr#1{\text{Tr}\left( #1\right)}

\DeclareMathOperator{\aSAM}{aSAM}
\DeclareMathOperator{\GMSE}{GMSE}
\DeclareMathOperator{\RE}{RE}

\newcommand{\ap}{\emph{a priori}}
\newcommand{\wrt}{with respect to}

\newcommand\nbpix{N}
\newcommand\nendm{K}
\newcommand\nband{L}

\def\splitA{\mathbf{w}_n^{({\mathbf{A}})}}
\def\splitM{\mathbf{W}_{\ell}^{({\mathbf{M}})}}
\def\splitT{\mathbf{W}_{k}^{({\mathbf{T}})}}
\def\splitdM{\mathbf{W}_n^{({\mathbf{dM}})}}

\def\lagA{\boldsymbol{\lambda}_n^{(\mathbf{A})}}
\def\lagM{\mathbf{\Lambda}_{\ell}^{(\mathbf{M})}}
\def\lagT{\mathbf{\Lambda}_{k}^{(\mathbf{T})}}
\def\lagdM{\mathbf{\Lambda}_{n}^{(\mathbf{dM})}}

\def\muA{\mu_n^{({\mathbf{A})}}}
\def\muM{\mu_{\ell}^{({\mathbf{M})}}}
\def\muT{\mu_{k}^{({\mathbf{T})}}}
\def\mudM{\mu_n^{({\mathbf{dM})}}}

\def\csteA{cA_n}

\def\muAit#1{\mu_n^{(\mathbf{A})(#1)}}
\def\splitAit#1{\mathbf{w}_n^{({\mathbf{A}})(#1)}}
\def\lagAit#1{\boldsymbol{\lambda}_n^{(\mathbf{A})(#1)}}
\usepackage{xcolor}

\title{Hyperspectral Unmixing with Spectral Variability Using a Perturbed Linear Mixing Model}

\author{Pierre-Antoine Thouvenin,~\IEEEmembership{Student Member,~IEEE}, Nicolas Dobigeon,~\IEEEmembership{Senior Member,~IEEE} and Jean-Yves~Tourneret,~\IEEEmembership{Senior Member,~IEEE}
\thanks{Copyright (c) 2015 IEEE. Personal use of this material is permitted. However, permission to use this material for any other purposes must be obtained from the IEEE by sending a request to pubs-permissions@ieee.org.}
\thanks{This work is supported by the Direction G\'en\'erale de l'Armement, French Ministry of Defence.}
\thanks{The authors are with the University of Toulouse, IRIT/INP-ENSEEIHT, 2 rue Camichel, BP 7122, 31071 Toulouse cedex 7, France. (e-mail: \{pierreantoine.thouvenin, Nicolas.Dobigeon, Jean-Yves.Tourneret\}@enseeiht.fr}}

\begin{document}
\maketitle
\begin{abstract}
Given a mixed hyperspectral data set, linear unmixing aims at estimating the reference spectral signatures composing the data -- referred to as \emph{endmembers} -- their abundance fractions and their number. In practice, the identified endmembers can vary spectrally within a given image and can thus be construed as variable instances of reference endmembers. Ignoring this variability induces estimation errors that are propagated into the unmixing procedure. To address this issue, endmember variability estimation consists of estimating the reference spectral signatures from which the estimated endmembers have been derived as well as their variability \wrt{} these references. This paper introduces a new linear mixing model that explicitly accounts for spatial and spectral endmember variabilities. The parameters of this model can be estimated using an optimization algorithm based on the alternating direction method of multipliers. The performance of the proposed unmixing method is evaluated on synthetic and real data. A comparison with state-of-the-art algorithms designed to model and estimate endmember variability allows the interest of the proposed unmixing solution to be appreciated.
\end{abstract}
\begin{IEEEkeywords}
Hyperspectral imagery, linear unmixing, endmember spatial and spectral variability, \emph{Alternating Direction Method of Multipliers} (ADMM).
\end{IEEEkeywords}
\section{Introduction} \label{sec:intro}

\IEEEPARstart{O}{ver} the past decades, hyperspectral imagery has been receiving an increasing interest. Whereas traditional red~/~green~/~blue or multispectral images are composed of a limited number of spectral channels (from three to tens), hyperspectral images are acquired in hundreds of contiguous spectral bands facilitating the analysis of the elements in the scene, e.g., determining their nature and relative proportions. However, the high spectral resolution of these images is mitigated by their lower spatial resolution, hence the need to unmix the data. Spectral unmixing is aimed at estimating the reference spectral signatures -- referred to as endmembers -- their abundance fractions and their number from which the $\nband $-multi-band observations are derived according to a predefined mixing model. Assuming the absence of any microscopic interaction between the materials of the imaged scene and a negligible declivity, a linear mixing model (LMM) is classically used to describe the structure of the collected data \cite{Bioucas2012jstars}.
However, the spectral signatures contained in a reference image can vary spectrally, spatially or temporally from an image to another due to varying acquisition conditions. This can result in significant estimation errors being propagated throughout the unmixing process. Various models either derived from a statistical or a deterministic point of view have been designed to address this issue \cite{Zare2014IEEESPMAG}. More precisely, the first class of methods assumes that the endmember spectra can be considered as realizations of multivariate distributions. The most popular models are the normal composition model \cite{Eches2010ip} and the beta compositional model \cite{Du2014}. The second class of methods considers the endmember signatures as members of spectral libraries associated with each material (bundles). Two methods using spectral libraries have been especially considered in the literature: the automated endmember bundles (AEB) \cite{Somers2012jstars2} and the Fisher discriminant null space (FDNS) \cite{Jing2010}. Whereas AEB enables the extraction of an endmember library to account for spectral variabilities, the aim of FDNS is to estimate a transformation projection matrix to project the hyperspectral data into a space minimizing the variability impact.

Since the identified endmembers can be considered as variable instances of reference endmembers, we introduce an extended version of the classical LMM to explicitly model the spectral variability. In \cite{Vengazones2014WIHSPERS}, the variability is assumed to only result from scaling factors. Conversely, in this paper, inspired by a model designed in \cite{Johnson2013}, each endmember is represented by a "pure" spectral signature corrupted by an additive perturbation accounting for its variability. The perturbation is allowed to vary from a pixel to another to represent spatial-spectral variabilities. As a result, the proposed perturbed LMM (PLMM) can capture endmember spatial and spectral variability within a given image. To the best of our knowledge, it is the first time endmember variability has been explicitly modeled as an additive perturbation.

The promising results obtained with the alternating direction method of multipliers (ADMM) in hyperspectral imagery \cite{Bioucas2009} and in image deblurring \cite{Alfonso2010,Alfonso2011,Almeida2013,Almeida_arXiv2013,Matakos2013} serve as an incentive to apply a similar framework to conduct PLMM-based unmixing. A key property of the ADMM framework lies in the introduction of appropriate splitting variables. Indeed, the specified constraints can be handled independently from the rest of the problem and often lead to analytical solutions when solving the resulting optimization problem. Using this fruitful principle, an ADMM-based algorithm for linear unmixing using a group lasso $\ell_{2,1}$-norm regularization was recently developed in \cite{Ammanouil2014,Ammanouil_12_2014}. Inspired by these examples, this paper proposes to exploit the advantages of an ADMM-based resolution of the linear unmixing problem to account for spatial and spectral endmember variabilities.

Throughout the article, the number of endmembers will be assumed to be \ap{} known or estimated by any state-of-the-art method (e.g., \cite{Bioucas2008}) since estimating the required number of endmembers to appropriately describe the data as well as endmember variability is a challenging task. Indeed, the choice of $\nendm$ drastically alters the representation of the imaged scene, and is thus a crucial step to the endmember identification and the subsequent abundance estimation \cite{Bioucas2008,Bioucas2012jstars,Halimi2015}.

The paper is organized as follows. The PLMM accounting for spectral and spatial variabilities is introduced in Section \ref{sec:LMM}. Section \ref{sec:ADMM} describes an ADMM-based algorithm to solve the resulting optimization problem. Experimental results obtained on synthetic and real data are reported in Section \ref{sec:simulations} and \ref{sec:experiments} respectively. The results obtained with the proposed algorithm are systematically compared to those of the vertex component analysis / fully constrained least squares (VCA/FCLS), the simplex identification via split augmented Lagrangian (SISAL) \cite{Bioucas2009} coupled with FCLS, AEB and FDNS. Section \ref{sec:conclusion} finally concludes this work.

\section{Problem statement} \label{sec:LMM}
\begin{table}
\centering
\caption{Table of notations.}
\begin{tabular}{|l|l|}
\hline
$\nbpix$ & number of pixels \\
$\nband $ & number of spectral bands \\
$\nendm $ & number of endmembers \\
$\mat{Y} \in \R[\nband ]{\nbpix}$ & lexicographically ordered pixels \\
$\mat{M} \in \R[\nband ]{\nendm }$ & endmember matrix\\
$\mat{dM}_n \in \R[\nband ]{\nendm }$ & $n$th variability matrix\\
$\mat{A} \in \R[\nendm ]{\nbpix}$ & abundance matrix\\
$\mat{V} \in \R[(\nendm-1)]{\nband}$ & projector on the space spanned by the \\
						 & $\nendm-1$ principal components of $\mat{Y}$\\
$\mat{U} \in \R[\nband]{(\nendm-1)}$ & inverse projector\\
$\mat{T} \in \R[(\nendm - 1)]{\nendm}$ & projection of $\mat{M}$ on the PCA space\\
$\mat{m}_{n,k}$, $\row{m}_{n,j}$ & column $k$, row $j$ of the matrix $ \mat{M}_n $ \\
$[\mat{AB}]_k,\{\mat{AB}\}_j$ & column $k$, row $j$ of the matrix $ \mat{AB} $ \\
$ \succeq $ & term-wise inequality \\  
$\mathcal{S}_{n,p}^+ $ &$ = \left\{ \mat{X} \in \R[n]{p} \middle| \mat{X} \succeq \Zero[n]{p}  \right\}$\\
$\mathbf{1}_n $& $ = [1,1, \ldots,1]^T \in \mathbb{R}^n$ \\
$\mathcal{I}_A(x)$& $= \left\{
\begin{array}{ll}
0 & \text{if} \, x \in A \\
+\infty & \text{otherwise.}
\end{array}
\right.$\\
\hline
\end{tabular}
\end{table}

	\subsection{Perturbed linear mixing model (PLMM)} \label{p:LMM}
In the absence of any specific prior knowledge on the variability nature (i.e., errors affecting the endmembers), we have chosen to explicitly represent the variability by a spatially varying additive endmember perturbation. This choice, inspired by a model designed in \cite{Johnson2013}, appears to be simple and flexible enough to account for the observed variability. %
Assuming that the number of endmembers $\nendm$ is known, the proposed PLMM differs from the classical LMM insofar as each pixel $\mathbf{y}_n$ is represented by a combination of the $\nendm$ endmembers -- denoted as $\mathbf{m}_k$ -- affected by a perturbation vector $\mathbf{dm}_{n,k}$ accounting for endmember variability. The resulting PLMM can be written
\begin{equation}
\label{eq:model}
\mathbf{y}_n  =  \sum_{k=1}^\nendm a_{kn}\Bigl(\mathbf{m}_k + \mathbf{dm}_{n,k} \Bigr) + \mathbf{b}_n \  \text{for} \  n = 1,\dotsc , \nbpix
\end{equation}%
where $\mat{y}_n$ denotes the $n$th image pixel, $\mat{m}_k$ is the $k$th endmember, $a_{kn}$ is the proportion of the $k$th endmember in the $n$th pixel, and $\mat{dm}_{n,k}$ denotes the perturbation of the $k$th endmember in the $n$th pixel. Finally, $\mathbf{b}_n$ models a zero-mean white Gaussian noise resulting from the data acquisition as well as modeling errors. We can note that the proposed PLMM reduces to the classical LMM in absence of variability. In matrix form, the PLMM \eqref{eq:model} can be written as follows
\begin{equation}
\mathbf{Y}  = \mathbf{MA} + %
\underbrace{\biggl[ \begin{array}{c|c|c} %
 \mathbf{dM}_1 \mathbf{a}_1 & \dotsc & \mathbf{dM}_\nbpix \mathbf{a}_\nbpix %
 \end{array} \biggr]}_{\Delt} + %
 \mathbf{B}
\end{equation}%
where $\Y = \left[ \mathbf{y}_1,\dotsc,\mathbf{y}_N \right]$ is an $\nband \times \nbpix$ matrix containing the image pixels, $\M$ is an $\nband \times \nendm$ matrix containing the endmembers, $\A$ is a $\nendm \times \nbpix$ matrix composed of the abundance vectors $\mathbf{a}_n$, $\dM_n$ is an $\nband \times \nendm$ matrix whose columns are the perturbation vectors associated with the $n$th pixel, and $\mathbf{B}$ is an $\nband \times \nbpix$ matrix accounting for the noise. The non-negativity and sum-to-one constraints usually considered to reflect physical considerations are
\begin{align} \label{eq:constraints}
\begin{split}
\mathbf{A} & \succeq \Zero[\nendm]{\nbpix}, \quad  \mathbf{A}^T \mathbf{1}_\nendm    = \mathbf{1}_\nbpix   \\%
\mathbf{M} & \succeq \Zero[\nband]{\nendm},\quad  \mathbf{M} + \mathbf{dM}_n  \succeq \Zero[\nband]{\nendm}, \, \forall n = 1,\dotsc , \nbpix.
\end{split}
\end{align}{}%
When compared to the underlying models proposed in the literature to mitigate variability \cite{Zare2014IEEESPMAG}, model \eqref{eq:model} presents the advantage to explicitly address the variability phenomenon in terms of an additive perturbation affecting each endmember. This perturbation accounts for any deviation from the linear mixing model (as will be illustrated in our experiments). The main contribution of this paper is to propose an unsupervised algorithm for estimating the endmembers contained in the image and the abundances and endmember variability for each pixel of this image.

\subsection{Problem formulation}
\label{subsec:problem_formulation}
As mentioned in Section \ref{sec:intro}, the PLMM \eqref{eq:model} and constraints \eqref{eq:constraints} can be combined to formulate a constrained optimization problem. An appropriate cost function is required to estimate the parameters $\mathbf{M,A,dM}$. Assuming the signal is corrupted by a zero-mean white Gaussian noise, we define the data fitting term as the Frobenius norm of the difference between the acquisitions $\mathbf{Y}$ and the reconstructed data $\mathbf{MA + \Delta}$. Since the problem is ill-posed, additional penalization terms are needed. In this paper, we propose to define penalization functions $\Phi,\Psi$ and $\Upsilon $ to reflect the available \emph{a priori} knowledge on $\mathbf{M,A}$ and $\mathbf{dM}$ respectively. As a result, the optimization problem is expressed as
\begin{equation}
\label{eq:general_problem}
(\mathbf{M}^*,\mathbf{dM}^*,\mathbf{A}^*) \in \argmin{\mathbf{M,dM,A}}%
\Bigl\{
	\mathcal{J}(\mathbf{M,dM,A}) \text{ s.t. \eqref{eq:constraints}}%
\Bigr\}%
\end{equation}{}%
with
\begin{equation}
\label{eq:objective}
\begin{split}
	\mathcal{J}(\mathbf{M,dM,A})  = & \frac{1}{2}\Norm_fro{\mathbf{Y} - \mathbf{MA} - \Delt } + %
	\alpha \Phi(\mathbf{A}) + \medskip\\ %
	& \beta\Psi(\mathbf{M}) + \gamma \Upsilon (\mathbf{dM})
\end{split}
\end{equation}{}%
where the penalization parameters $\alpha , \beta, \gamma$ control the trade-off between the data fitting term $ \frac{1}{2} \Norm_fro{\mathbf{Y} - \mathbf{MA-\Delta}}$ and the penalties $\Phi(\mathbf{A})$, $\Psi(\mathbf{M})$ and $\Upsilon (\mathbf{dM})$. In addition, we assume that the penalization functions are separable, leading to
\begin{align} \label{eq:assumption}
\Phi(\A) &= \sum_{n=1}^\nbpix \phi(\mathbf{a}_n) \\
\Psi(\M) &=\sum_{\ell=1}^\nband \psi(\row{m}_{\ell}) \\
\Upsilon (\dM) &= \sum_{n=1}^\nbpix \upsilon (\dM _n)
\end{align}%
where $\row{m}_{\ell}$ denotes the $\ell$th row of $\M$ and $\phi$, $\psi$ and $\upsilon$ are non-negative differentiable convex functions. This assumption is used to decompose \eqref{eq:general_problem} into a collection of simpler sub-problems described in Section \ref{sec:ADMM}. All these penalizations are described in the next paragraphs.

		\subsubsection{Abundance penalization} \label{subsec:A_penalization}

The abundance penalization $\Phi$ has been chosen to promote spatially smooth abundances as in \cite{Chen2014}. More precisely, the abundance spatial smoothness penalization is expressed in matrix form as
\begin{equation}
\Phi(\mathbf{A}) = \frac{1}{2}\Norm_fro{\mathbf{AH}}
\end{equation}
where $\mathbf{H}$ is a matrix computing the differences between the abundances of a given pixel and those of its $4$ nearest neighbors \cite{Chen2014}. The resulting expression of $\phi$ is detailed in Appendix \ref{app:constraints}.

		\subsubsection{Endmember penalization} \label{subsec:M_penalization}
		
As for $\Psi$, classical penalizations found in the literature consist of constraining the size of the simplex whose vertices are the endmember signatures. The volume criterion used in \cite{Miao2007,Chan2011tgrs} enables the volume exactly occupied by the $(\nendm-1)$-simplex formed by the endmembers to be penalized. The mutual distance between the endmembers introduced in \cite{Berman2004,Arngren2011} (which approximates the volume) has a similar purpose. Finally, if the endmembers are \emph{a priori} close from available reference spectral signatures, a penalization on the distance between the endmembers and these signatures can be implemented. The expression of the distance between the endmembers and some reference spectral signatures, the mutual distance between the endmembers and the volume penalization are recalled in the following lines. For each penalization type, the corresponding expression of $\psi$ is given in Appendix \ref{app:constraints}.

\begin{itemize}
\item The distance between the endmembers and some reference spectral signatures $\mathbf{M}_0$ is given by
\begin{equation}
\Psi(\mathbf{M}) = \frac{1}{2}\Norm_fro{\mathbf{M} - \mathbf{M}_0}.
\end{equation}
\item The mutual distance between the endmembers is expressed in matrix form as
\begin{equation}
\begin{split}
\Psi(\mathbf{M}) = \frac{1}{2}\sum_{i = 1}^\nendm  \Biggl( \underset{j \neq i}{\sum_{j = 1}^\nendm } \norm_2{\mathbf{m}_i - \mathbf{m}_j} \Biggr).
\end{split}
\end{equation}{}%
\item Under the pure pixel and linear mixture assumptions, the data points are enclosed in a $(\nendm -1)$-simplex whose vertices are the endmembers \cite{Chan2011tgrs}. Let $\mathbf{T}$ be the projection of $\mathbf{M}$ on the space spanned by the $\nendm -1$ principal components of $\mathbf{Y}$. The expression of the volume of this subspace is
\begin{equation*}
\mathcal{V}(\mathbf{T}) = \frac{1}{(\nendm -1)!} \left| \det %
\begin{pmatrix}
\mathbf{T} \\
\mat{1}_\nendm ^T
\end{pmatrix} \right| .
\end{equation*}{}%
To ensure the differentiability of the penalization \wrt{} $\mathbf{T}$, we propose to consider the following penalty
\begin{equation}
\Psi(\mathbf{M}) = \frac{1}{2} \mathcal{V}^2(\mat{T}).
\end{equation}
\end{itemize}

		\subsubsection{Variability penalization}
\label{subsec:dM_penalization}
The variability penalizing function $\Upsilon $ has been designed to limit the norm of the spectral variability. Indeed, it is legitimate to penalize the energy of the perturbation matrices $\mat{dM}_n$ in order to obtain a reasonable endmember variability. In this paper, we propose to consider the following penalty (having the advantage to be differentiable with respect to $\mat{dM}_n$)
\begin{equation}
\Upsilon  \left(\mat{dM} \right) =\frac{1}{2} \Norm_fro{\mat{dM}}  = \frac{1}{2} \sum_{n=1}^\nbpix \Norm_fro{\mat{dM}_n}.
\end{equation}{}
To the best of our knowledge, no specific information regarding the spatial distribution of the variability is available in the remote sensing literature so far. We have consequently preferred not to include any additional regularization on $\dM$. However, any spatial penalization satisfying the assumptions given in Paragraph \ref{subsec:problem_formulation} can be added when necessary (e.g., a group-Lasso $\ell_{2,1}$ penalization to promote spatial sparsity of the variability term $\dM$).

\begin{algorithm}[!t]
 \KwData{$\Y,\A^{(0)},\M^{(0)},\dM^{(0)}$} 
 \KwResult{$\A,\M,\dM$}
 \Begin{
  $k \leftarrow 1$\;
  \While{stopping criterion not satisfied}{
   \lnlset{alg1_A}{a}$\A^{(k)} \leftarrow \argmin{\A} \mathcal{J} \Bigl(\M^{(k-1)},\dM^{(k-1)},\A \Bigr)$ \medskip \;
    \lnlset{alg1_M}{b}$\M^{(k)} \leftarrow \argmin{\M} \mathcal{J} \Bigl(\M,\dM^{(k-1)},\A^{(k)} \Bigr)$ \medskip \;
    \lnlset{alg1_dM}{c}$\dM^{(k)} \leftarrow \argmin{\M} \mathcal{J} \Bigl(\M^{(k)},\dM,\A^{(k)}\Bigr)$ \medskip \;
   $k \leftarrow k+1$\;
   }
   $\A \leftarrow \A^{(k)}$\;
   $\M \leftarrow \M^{(k)}$\;
   $\dM \leftarrow \dM^{(k)}$\;
 }
 \caption{PLMM-unmixing: global algorithm.}
 \label{alg:global_optimization}
\end{algorithm}  

\section{An ADMM-based algorithm} \label{sec:ADMM}

Since the problem \eqref{eq:general_problem} is not convex, a minimization strategy similar to \cite{Almeida2013} has been adopted. Precisely, the cost function $\mathcal{J}$ is successively minimized \wrt{} each variable $\A,\M$ and $\dM$ until a stopping criterion is satisfied. The assumptions made on the penalization functions $\Phi,\Psi,\Upsilon $ in Section \ref{sec:LMM} allow the global optimization problem to be divided into a collection of strictly convex sub-problems. These sub-problems have the nice property to involve differentiable functions simplifying their resolution. Having introduced appropriate splitting variables to account for the constraints, these sub-problems are finally solved using ADMM steps admitting closed-form expressions due to the separability assumption. The three minimization steps considered in this algorithm present a highly similar structure. The details are reported in Appendix \ref{app:resolution} to facilitate the reading of this paper.

	\subsection{ADMM: general principle} \label{subsec:ADMM_gen}	
The ADMM is a technique combining the benefits of augmented Lagrangian and dual decomposition methods to solve constrained optimization problems \cite{Boyd2010}. More precisely, the method consists of solving the original optimization problem by successively minimizing the cost function of interest \wrt{} each variable. The following elements (extracted from \cite{Boyd2010}) recall a general formulation of the problem. Given $f : \mathbb{R}^p \rightarrow \mathbb{R}^+$, $g \in \mathbb{R}^m \rightarrow \mathbb{R}^+$, $\mathbf{A} \in \R[n]{p}$ and $\mathbf{B} \in \R[n]{m}$, consider the general optimization problem
\begin{equation}
\min_{\mathbf{x,z}} \biggl\{ f(\mathbf{x}) + g(\mathbf{z}) \biggl| \mathbf{Ax} + \mathbf{Bz} = \mathbf{c} \biggr\} \biggr. .
\end{equation}{}%
The scaled augmented Lagrangian associated with this problem can be written
\begin{equation*}
\mathcal{L}_\rho \left(\mathbf{x,z,u}\right) = f(\mathbf{x}) + g(\mathbf{z})  + \frac{\rho}{2} \norm_2{\mathbf{Ax} + \mathbf{Bz} -\mathbf{c} + \mathbf{u}}
\end{equation*}{}%
where $\rho > 0$. Denote as $\mathbf{x}^{(k+1)},\mathbf{z}^{(k+1)}$ and $\mathbf{u}^{(k+1)}$ the primal variables and the dual variable at iteration $k+1$ of the algorithm
\begin{eqnarray*}
\mathbf{x}^{(k+1)} & \in & \argmin{\mathbf{x}}\mathcal{L}_\rho\left(\mathbf{x},\mathbf{z}^{(k)},\mathbf{u}^{(k)}\right) \\
\mathbf{z}^{(k+1)} & \in & \argmin{\mathbf{z}}\mathcal{L}_\rho \left(\mathbf{x}^{(k+1)},\mathbf{z},\mathbf{u}^{(k)}\right) \\
\mathbf{u}^{(k+1)} & = &\mathbf{u}^{(k)} + \mathbf{Ax}^{(k+1)} + \mathbf{Bz}^{(k+1)} -\mathbf{c}.
\end{eqnarray*}
The ADMM consists in successively minimizing $\mathcal{L}_{\rho}$ \wrt{} $\mathbf{x}, \mathbf{z}$ and $\mathbf{u}$. A classical stopping criterion involves the primal and dual residuals at iteration $k+1$ (see \cite[p. 19]{Boyd2010}): the procedure is iterated until
\begin{equation} \label{eq:stop}
\norme_2{\mat{r}^{(k)}} \leq \varepsilon^{\text{pri}} \quad \text{and} \quad \norme_2{\mat{s}^{(k)}} \leq \varepsilon^{\text{dual}}
\end{equation}{}%
where the primal and dual residuals at iteration $k+1$ are respectively given by
\begin{align}
\mathbf{r}^{(k+1)} &= \mathbf{Ax}^{(k+1)} + \mathbf{Bz}^{(k+1)} - \mathbf{c} \\
\mathbf{s}^{(k+1)} &= \rho \mathbf{A}^T \mathbf{B} \left( \mathbf{z}^{(k+1)} - \mathbf{z}^{(k)} \right)
\end{align}{}%
and
\begin{align}
\varepsilon^{\text{pri}} & = \sqrt{p} \varepsilon^{\text{abs}} + \varepsilon^{\text{rel}} \max \Bigl\{ \norm_2{\mathbf{Ax}^{(k)}}, \norm_2{\mathbf{Bz}^{(k)}}, \norm_2{\mathbf{c}}  \Bigr\} \\
\varepsilon^{\text{dual}} & = \sqrt{n} \varepsilon^{\text{abs}} + \varepsilon^{\text{rel}} \norm_2{\mathbf{A}^T \mathbf{y}^{(k)}}.
\end{align}
Finally, the parameter $\rho$ can be adjusted using the rule described in \cite[p. 20]{Boyd2010}
\begin{equation}
\label{eq:update}
\rho^{(k+1)} = \left\{
\begin{array}{ll}
\tau^{\text{incr}} \rho^{(k)} & \text{if} \, \norme_2{\mat{r}^{(k)}} > \mu \norme_2{\mat{s}^{(k)}} \medskip\\
\rho^{(k)}/\tau^{\text{decr}} & \text{if} \, \norme_2{\mat{s}^{(k)}} > \mu \norme_2{\mat{r}^{(k)}} \medskip\\
\rho^{(k)} & \text{otherwise.}
\end{array}
\right.
\end{equation}
Note that this parameter adjustment does not alter the ADMM convergence as long as it is performed finitely many times.

\begin{algorithm}[!t]
 \KwData{$\mat{Y},\mat{A}^{(0)},\mat{M}^{(0)},\varepsilon_{\text{pri}},\varepsilon_{\text{dual}},\tau^{\text{incr}},\tau^{\text{decr}},\muAit{0}$}
 \KwResult{$\mat{A}$}
  \For{$n = 1$ \KwTo $N$}{
  $k \leftarrow 1$\;
  $\lagAit{0} =\mat{0}$\;
  $\splitAit{0}=\mat{0}$\;
    \While{stopping criterion not satisfied}{
   $\mat{a}_n^{(k)} \leftarrow \argmin{\mat{a}_n} \mathcal{L}_{\muAit{k-1}}\Bigl(\mat{a}_n,\splitAit{k-1},\lagAit{k-1} \Bigr)$\; 
   $\splitAit{k} \leftarrow \argmin{\splitA} \mathcal{L}_{\muAit{k-1}}\Bigl(\mat{a}_n^{(k)},\splitA,\lagAit{k-1}\Bigr)$\;
   $\lagAit{k} \leftarrow \lagAit{k-1} + \mat{Qa}_n^{(k)} + \splitAit{k} -\mat{s}$ \medskip \;
	$\muAit{k} \leftarrow \text{Update} \Bigl( \muAit{k-1} \Bigr)$ using \eqref{eq:update} \medskip \;
   $k \leftarrow k+1$ \;
    }
   $\mat{a}_n \leftarrow \mat{a}_n^{(k)}$\;
   }
 \caption{ADMM optimization w.r.t. $\mat{A}$ (step \eqref{alg1_A}).}
 \label{alg:opti_A}
\end{algorithm}{}

	\subsection{Optimization \wrt{} \textbf{A}} \label{p:opti_A}
	With the assumptions made in paragraph \ref{subsec:problem_formulation}, optimizing the cost function $\mathcal{J}$ \wrt{} $\mathbf{A}$ under the constraints \eqref{eq:constraints} is equivalent to solving the following problems
\begin{equation}
\mathbf{a}_n^* = \argmin{\mathbf{a}_n} \left\{ \begin{array}{c}
 	\frac{1}{2}   \norm_2{\mathbf{y}_n - (\mathbf{M+dM}_n) \mathbf{a}_n} + \alpha\phi(\mathbf{a}_n)\medskip \\
 	\text{s.t.} \quad
 		\mathbf{a}_n \succeq \mathbf{0}_\nendm ,  \quad
 		\mathbf{a}_n^T \mathbf{1}_\nendm    = 1
 \end{array} \right\}.
\end{equation}{}%
After introducing the splitting variables $\splitA \in \mathbb{R}^\nendm $ for $ n =~1, \dotsc ,\nbpix $ such that %
\begin{equation} \label{eq:split_A}
\underbrace{\begin{pmatrix}
\mathbf{I}_\nendm    \\
\mathbf{1}_\nendm ^T
\end{pmatrix}}_{\mathbf{Q}}\mathbf{a}_n + %
\underbrace{\begin{pmatrix}
-\mathbf{I}_\nendm    \\
\mathbf{0}_\nendm ^T
\end{pmatrix}}_{\mathbf{R}} \mathbf{w}_n = %
\underbrace{\begin{pmatrix}
\mathbf{0}_\nendm    \\
1
\end{pmatrix}}_{\mathbf{s}}
\end{equation}{}%
the resulting scaled augmented Lagrangian is expressed as
\begin{equation} \label{eq:lagrangian_A}
\begin{split}
\mathcal{L}_{\muA}&\Bigl(\a_n,\splitA,\lagA \Bigr) =  %
 \frac{1}{2}   \norm_2{\mathbf{y}_n - (\mathbf{M+dM}_n) \mathbf{a}_n} \\
& + \frac{\muA}{2}\norm_2{\mat{Qa}_n +%
 \mat{R}\splitA - \mat{s}+ \lagA} \\
& + \alpha \phi(\mathbf{a}_n) + \mathcal{I}_{\mathcal{S}_{\nendm ,1}^+}\Bigl( \splitA \Bigr)
\end{split}
\end{equation}{}%
with $\muA > 0$. The resulting algorithm (step \ref{alg1_A} of Algo. \ref{alg:global_optimization}) is detailed in Algo. \ref{alg:opti_A}, and the solution to each sub-problem is given in Appendix \ref{app:resolution}.

	\subsection{Optimization \wrt{} \textbf{M}} \label{p:opti_M}
	Similarly to Paragraph \ref{p:opti_A}, optimizing $\mathcal{J}$ \wrt{} $\mathbf{M}$ under the constraint \eqref{eq:constraints} is equivalent to solving
\begin{equation}
\tilde{\mathbf{m}}_\ell^* = %
\argmin{\row{m}_\ell}\left\{ \begin{array}{c}
	\frac{1}{2} \norm_2{\row{y}_\ell - \row{m}_\ell \A - \row{\boldsymbol{\delta}}_\ell} + \beta \psi(\row{m}_\ell) \medskip\\
	\text{s.t.} \quad \text{for} \quad n = 1, \dotsc,\nbpix \medskip \\
	\row{m}_\ell \succeq \mathbf{0}_\nendm ^T, \quad \row{m}_\ell + \row{dm}_{n,\ell} \succeq \mathbf{0}_\nendm ^T
\end{array} \right\}
\end{equation}{}%
where $\row{m}_{\ell}$ denotes the $\ell$th row of $\M$. 
Introducing the splitting variables $\splitM \in \R[(\nbpix +1)]{\nendm }$ for $ \ell =~1,\dotsc ,\nband $ such that
\begin{equation}
\label{eq:split_M}
\underbrace{\begin{pmatrix}
1 \\
\mathbf{1}_\nbpix
\end{pmatrix}}_{\mathbf{e}}\row{m}_\ell -
\splitM = -%
\underbrace{\begin{bmatrix}
\mathbf{0}_\nendm ^T \\
 \begin{pmatrix}
 \row{dm}_{1,\ell} \\
 \vdots \\
 \row{dm}_{\nbpix,\ell}
 \end{pmatrix}
\end{bmatrix}}_{\mathbf{F}_\ell}
\end{equation}{}%
the associated scaled augmented Lagrangian can be written
\begin{equation} \label{eq:lagrangian_M}
\begin{split}
\mathcal{L}_{\muM} & \Bigl(\row{m}_\ell,\splitM,\lagM \Bigr) = %
 \frac{1}{2} \norm_2{\row{y}_l - \row{m}_\ell\mathbf{A} - \row{\boldsymbol{\delta}}_\ell} \\
& + \frac{\muM}{2} \Norm_fro{\mathbf{e}\row{m}_\ell - \splitM + \mathbf{F}_\ell + \lagM} \\ %
& + \beta \psi(\row{m}_\ell) + \mathcal{I}_{\mathcal{S}_{\nbpix +1,\nendm }^+}\Bigl(\splitM \Bigr)
\end{split}
\end{equation}{}%
with $\muM > 0$. The resulting algorithm (step \ref{alg1_M} of Algo. \ref{alg:global_optimization}) is similar to Algo. \ref{alg:opti_A}. The solution to the optimization problems depends on the selected endmember penalizing function $\Psi$ chosen in paragraph \ref{subsec:M_penalization} (see Appendix \ref{app:resolution} for more details).
	
	\subsection{Optimization \wrt{} \textbf{dM}} \label{p:opti_dM}
	Finally, optimizing $\mathcal{J}$ \wrt{} $\mathbf{dM}$ under the constraint \eqref{eq:constraints} is equivalent to solving the sub-problems
\begin{equation}
 \dM_n^* = \argmin{\dM_n} %
 \left\{ \begin{array}{c}
 	\frac{1}{2} \norm_2{\mathbf{y}_n - (\mathbf{M+dM}_n) \mathbf{a}_n} \medskip \\ + %
 	\gamma \upsilon (\mathbf{dM}_n) \medskip \\
 	\text{s.t.} \quad \mathbf{M} + \mathbf{dM}_n \succeq \Zero[\nband]{\nendm}
\end{array} \right\}.
\end{equation}{}%
Introducing the splitting variables $\splitdM = \mathbf{M}+\mathbf{dM}_n$ for $n =~1,\dotsc ,\nbpix $, the resulting scaled augmented Lagrangian is given by
\begin{equation}
\label{eq:lagrangian_dM}
\begin{split}
\mathcal{L}_{\mudM} \Bigl( \mathbf{dM}_n,& \splitdM, \lagdM \Bigr) =  %
\frac{1}{2}\norm_2{\mathbf{y}_n -  (\mathbf{M}+\mathbf{dM}_n)\mathbf{a_n}}  \\%
& +\frac{\mudM}{2} \Norm_fro{\mathbf{dM}_n + \mathbf{M} - \splitdM + %
\lagdM} \\
& +\gamma \upsilon (\mathbf{dM}_n) + \mathcal{I}_{\mathcal{S}_{\nband ,\nendm }^+} \Bigl( \splitdM \Bigr)
\end{split}
\end{equation}{}%
with $\mudM > 0$. The resulting algorithm (step \ref{alg1_dM} of Algo. \ref{alg:global_optimization}) is similar to Algo. \ref{alg:opti_A}. The solution to these problems is given in Appendix \ref{app:resolution}.\\

The optimization procedures detailed above are performed sequentially until the stopping criterion is satisfied. The next sections evaluate the performance of the resulting unmixing strategy via several experiments conducted on synthetic and real data. 

	\subsection{Convergence and computational cost}
	The alternating scheme proposed in Alg. \ref{alg:global_optimization} is nothing but a block coordinate descent descent (BCD) which is guaranteed to converge to a stationary point of the objective function $\mathcal{J}$ as long as each sub-problem is exactly minimized \cite[Proposition 2.7.1]{Bertsekas1999}. Besides, the sub-problems tackled in \ref{p:opti_A}, \ref{p:opti_M} and \ref{p:opti_dM} are strongly convex, hence the convergence of the ADMM steps toward the unique minimum of each independent sub-problem when the augmented Lagrangian parameter has a constant value (see for instance \cite{Boyd2010}). The same convergence result applies to the ADMM with the parameter adjustment introduced in Paragraph \ref{subsec:ADMM_gen} as long as the parameter is updated finitely many times \cite{Boyd2010}. We may however mention that the proximal alternating linearized minimization (PALM) \cite{Bolte2013} could also be directly applied to the considered problem with a rigorous convergence proof based on the Kurdyka-{\L}ojasiewicz property. This alternative work has been presented in \cite{thouvenin2015gretsi}.

Considering the significant number of unknown parameters and the simple expression of the ADMM updates detailed in Appendix \ref{app:resolution}, we can note that the computational cost is dominated by matrix products, yielding an overall $\mathcal{O}(\nband \nendm^2 \nbpix)$ computational cost.

\section{Experiment with synthetic data} \label{sec:simulations}
\begin{figure}[!t]
\centering
\includegraphics[keepaspectratio,height=0.18\textheight, width=0.7\textwidth]{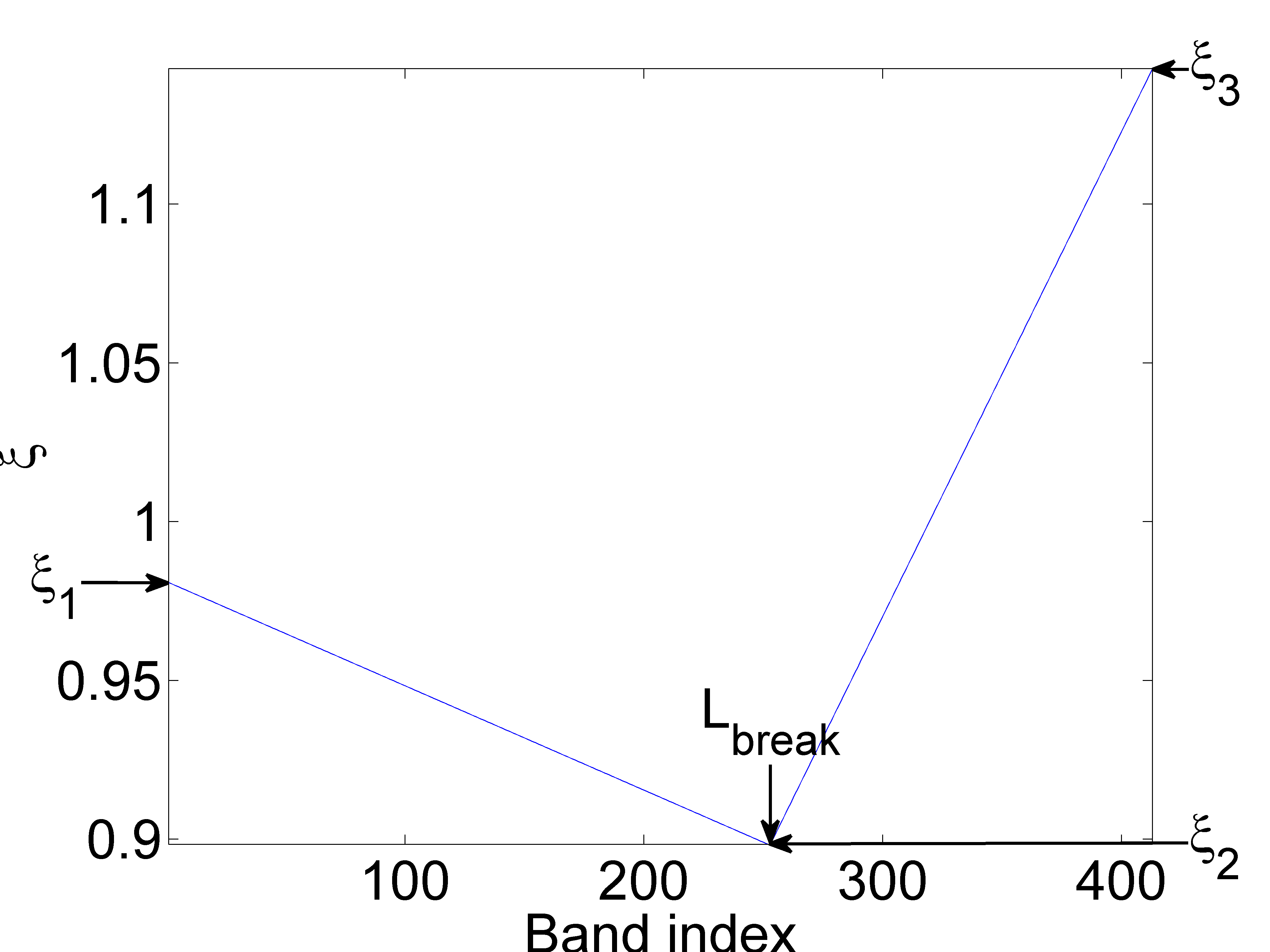}
\caption{Example of a randomly-generated affine function used to generate the synthetically perturbed endmembers.}
\label{fig:var_function}
\end{figure}

\begin{figure}[!t]
\centering
\subfloat{
\includegraphics[keepaspectratio,height=0.2\textheight , width=0.23\textwidth]{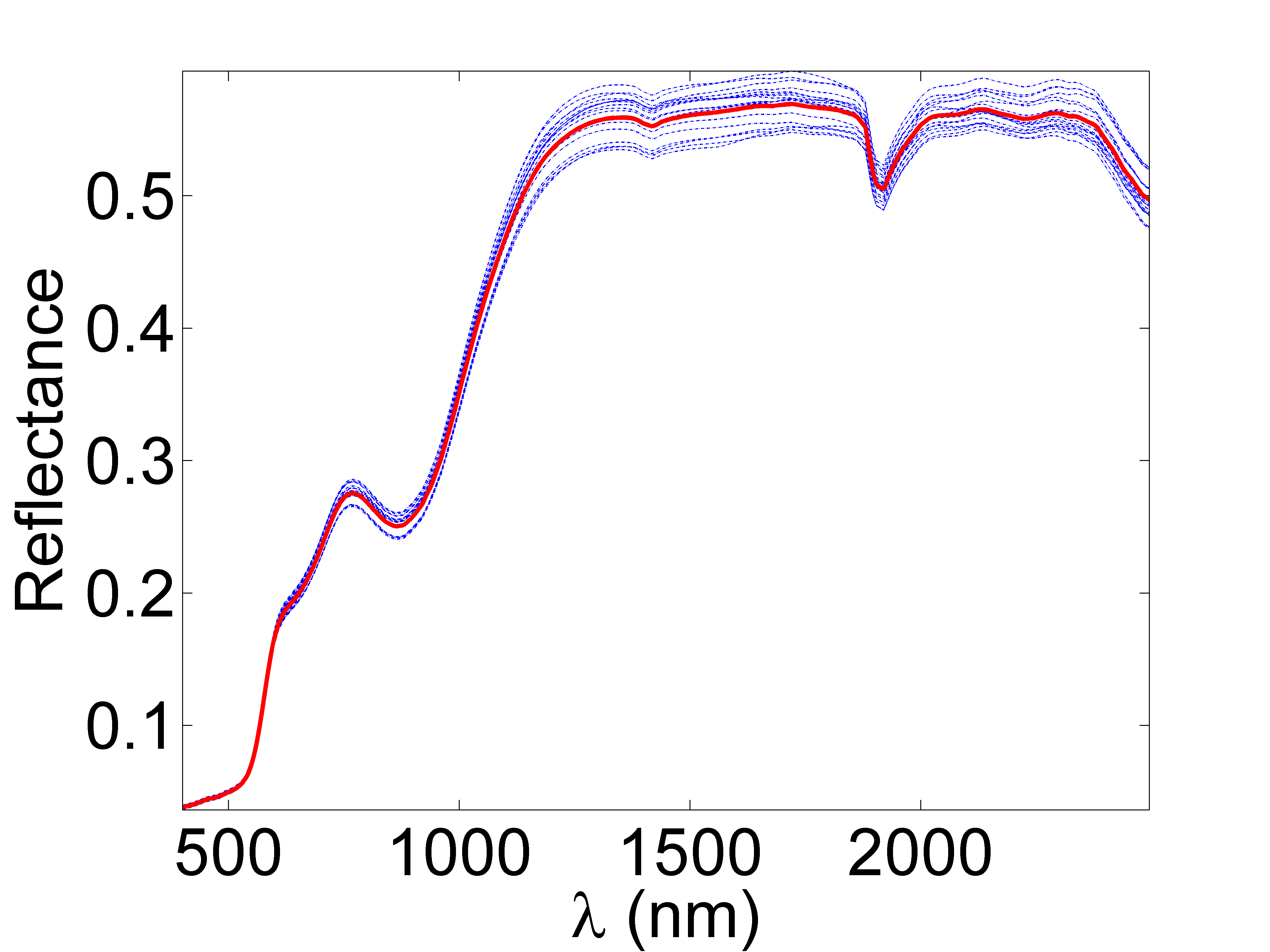}}
\subfloat{
\includegraphics[keepaspectratio,height=0.2\textheight , width=0.23\textwidth]{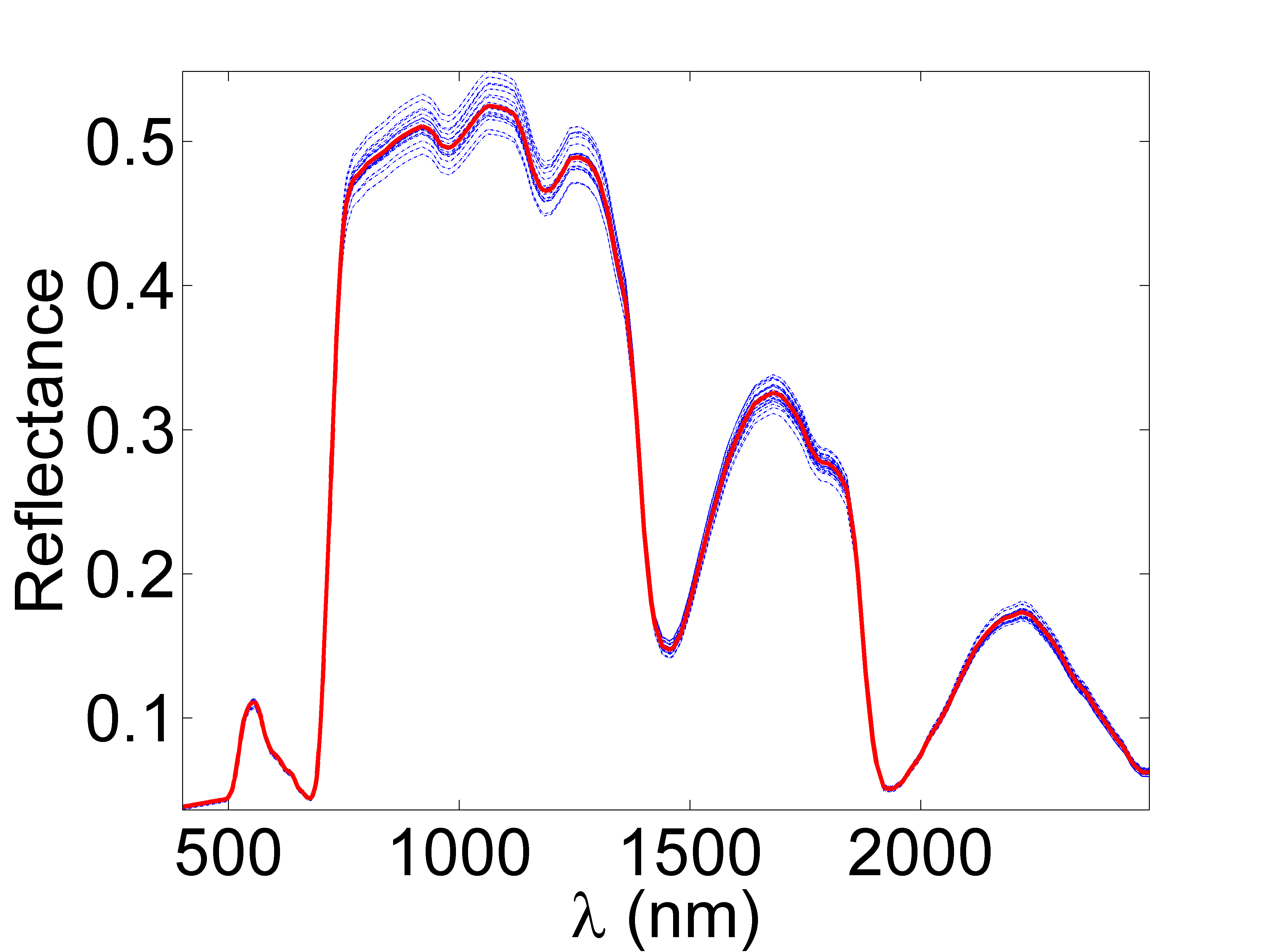}}
\\
\subfloat{
\includegraphics[keepaspectratio,height=0.2\textheight , width=0.23\textwidth]{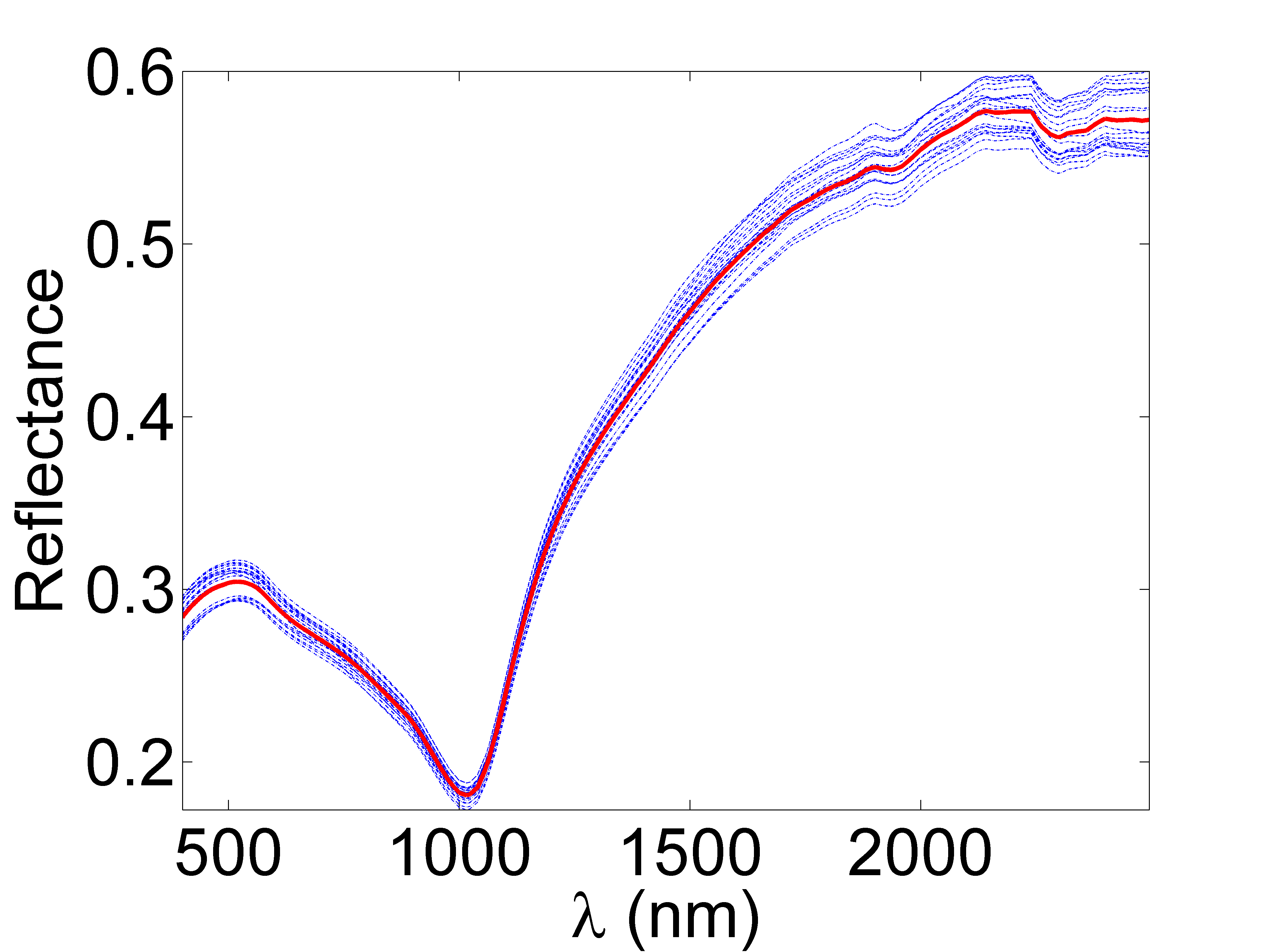}}
\subfloat{
\includegraphics[keepaspectratio,height=0.2\textheight , width=0.23\textwidth]{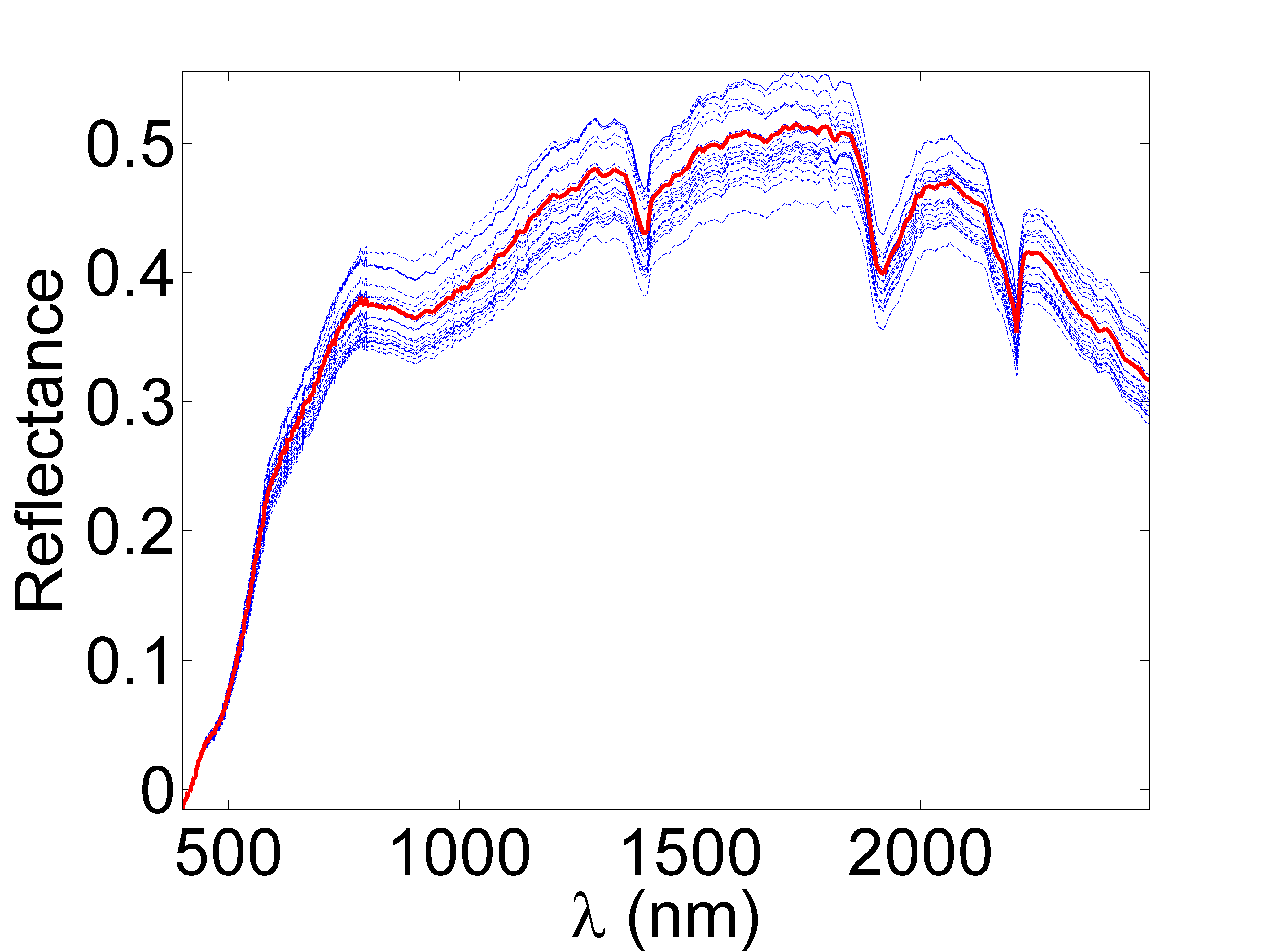}}
\\
\subfloat{
\includegraphics[keepaspectratio,height=0.2\textheight , width=0.23\textwidth]{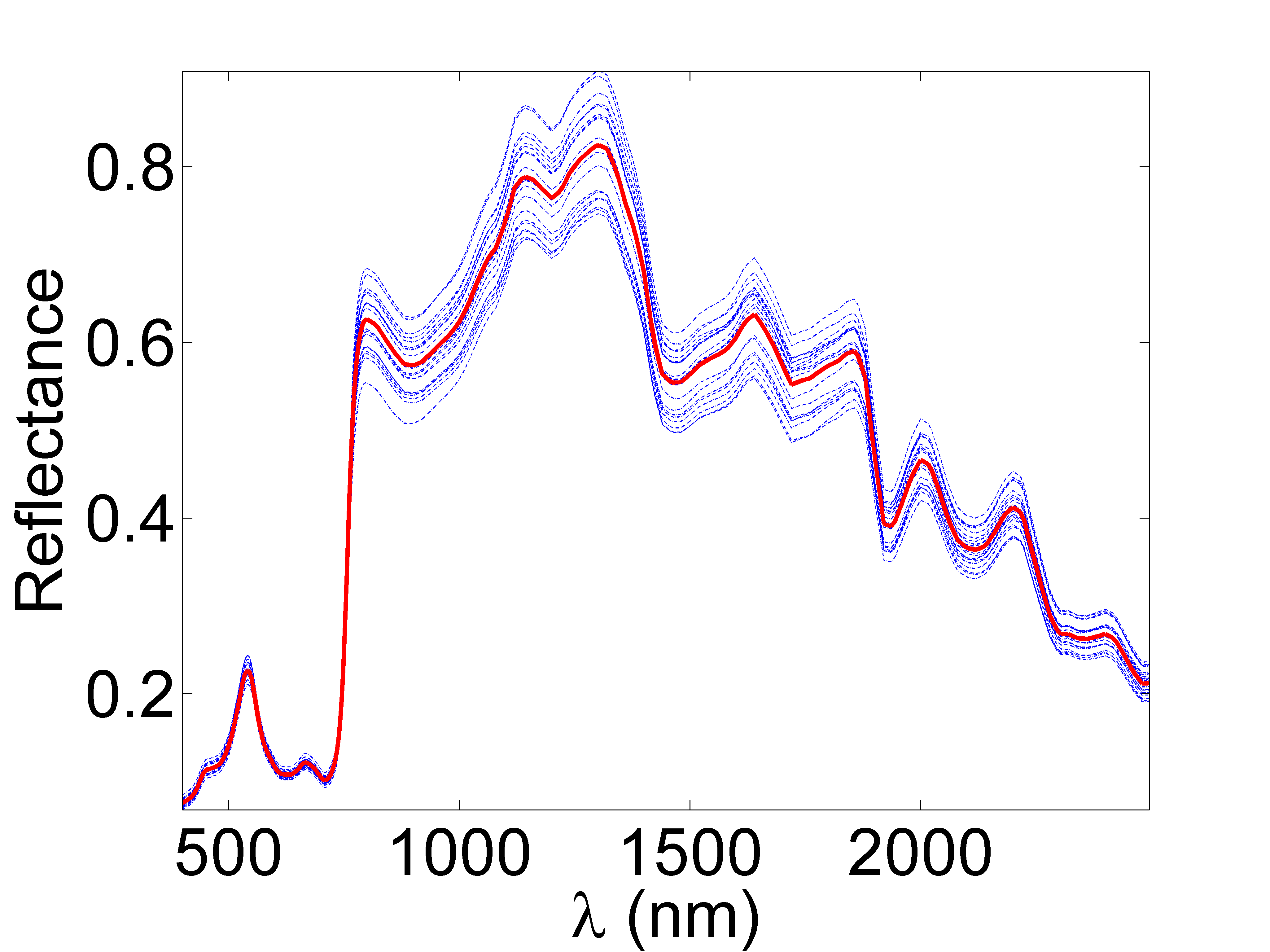}}
\subfloat{
\includegraphics[keepaspectratio,height=0.3\textheight , width=0.23\textwidth]{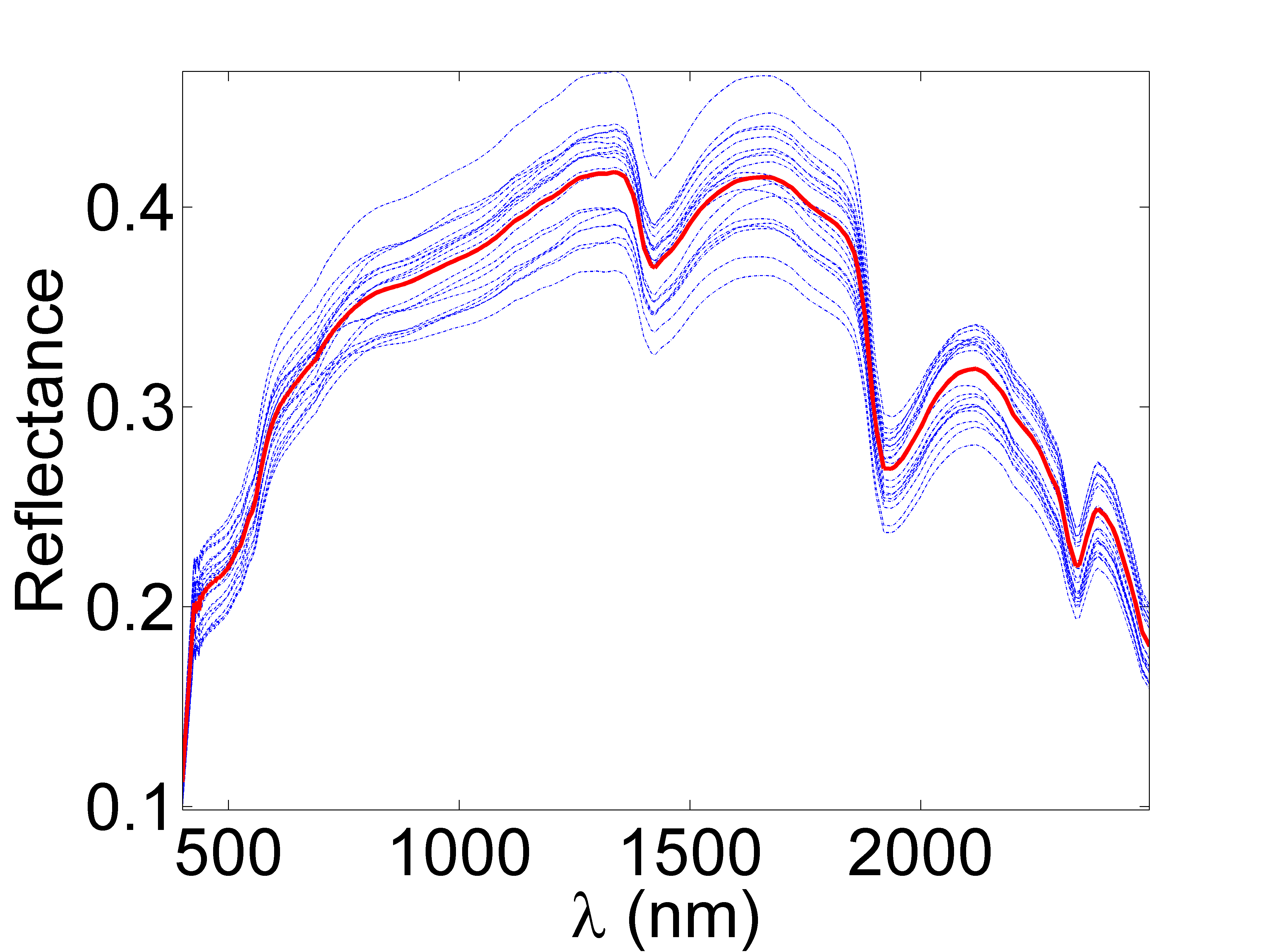}}
\caption{Reference endmembers (red lines) and $20$ corresponding instances under spectral variability (blue dotted lines) involved in the synthetic data experiments.}
\label{fig:endm_variability}	
\end{figure}

	This section considers four images of size $128~\times~64$ acquired in $413$ bands. Each image corresponds to a mixture of $\nendm$ endmembers with $\nendm \in \{ 3,6 \}$ in presence or absence of pure pixels (the absence of pure pixels is considered to evaluate the algorithm performance in a very challenging scenario). The synthetic linear mixtures have been corrupted by additive white Gaussian noise to ensure the signal-to-noise ratio is $\text{SNR} = 30\text{dB}$.
	Since no accepted variability model is available in the literature, we propose the following generation procedure to introduce controlled spectral variability. The corrupted endmembers involved in the mixture (see Fig. \ref{fig:endm_variability}) have been generated using the product of reference endmembers with randomly drawn piece-wise affine functions, providing realistic perturbed endmembers as represented in Fig. \ref{fig:var_function}. For a given variability coefficient $c_{\text{var}} > 0$ fixed by the user, the parameters $\xi_i$, $i \in \{1,2,3\}$ and $\nband_\text{break} \in \{1,\dotsc,\nband \}$ introduced in Fig. \ref{fig:var_function} have been generated as follows
\begin{align}
\xi_i \sim \mathcal{U}_{[1-c_{\text{var}}/2, 1+c_{\text{var}}/2]}, \: i \in \{1,2,3\} \\
\nband_{\text{break}} = \floor*{\nband / 2 + \floor*{\nband U / 3}}, \: U \sim \mathcal{N}(0,1)
\end{align}
where $\floor*{\cdot}$ denotes the floor operator. The synthetic data used in the proposed experiments have been generated with a value of $c_{\text{var}}$ that is lower in the upper half of the image ($c_{\text{var}} = 0.1$) than in the lower half ($c_{\text{var}}=0.25$). Some instances of the corresponding perturbed endmember spectra are depicted in Fig. \ref{fig:endm_variability}. Note that different affine functions have been considered for different endmembers and different pixels.

	\subsection{State-of-the-art methods}
The results of the proposed algorithm have been compared to those obtained with two classical linear unmixing methods (VCA/FCLS, SISAL/FCLS) and two variability accounting algorithms (AEB, FDNS). These methods are recalled below with their most relevant implementation details.
\begin{enumerate}
\item Classical unmixing methods (no variability)
	\begin{itemize}
	\item VCA/FCLS: the endmembers are first extracted using the vertex component analysis \cite{Nascimento2005}. The abundances are then estimated for each pixel using the fully constrained least squares (FCLS) algorithm \cite{Heinz2000});
	\item SISAL/FCLS: the endmembers are first extracted using the simplex identification via split augmented Lagrangian \cite{Bioucas2009}. The tolerance for the stopping rule has been set to $10^{-2}$ and VCA has been used as an initialization step. The abundances are then estimated for each pixel using FCLS.
	\end{itemize}
\item Variability accounting unmixing methods
	\begin{itemize}
	\item AEB \cite{Somers2012jstars2,Roberts1998,Goenaga2013}: the size of the bundles is equal to 25\% of the total pixel number. The endmembers and abundance are estimated using VCA/FCLS;
	\item FDNS \cite{Jing2010}: the endmembers and abundances are estimated by VCA/FCLS ;
	\item Proposed method (BCD/ADMM): endmembers and abundances have been initialized with VCA/FCLS estimates. Note that VCA/FCLS is a method assuming that there are pure pixels in the image, which can be problematic in case the imaged scene does not satisfy this assumption. The variability matrices have been initialized with all their entries equal to $eps$\footnote{\textsc{Matlab} constant $eps = 2.22 \times 10^{-16}$.}. The algorithm is stopped when the relative difference between two successive values of the objective function is less than $10^{-3}$.
	\end{itemize}
\end{enumerate}

\begin{table}[h!]
\caption{ADMM parameters.}
\label{tab:param}
	\begin{center}
		\begin{tabular}{@{}lcc@{}} \toprule
 		 & Synthetic data & Real data \\ \cmidrule{2-3} 		 
$\tau^{\text{incr}}$ 		&1.1	   &1.1	  	 \\
$\tau^{\text{decr}}$ 		&1.1	   &1.1	  	 \\
$\mu$ 						&$10$ & $10$	  	 \\
$\mu_n^{(\mathbf{A})(0)}$ 					&$10^{-4}$ & $10^{-4}$	  	 \\
$\mu_\ell^{(\mathbf{M})(0)}$ 					&$10^{-8}$ & $10^{-8}$	  	 \\
$\mu_n^{(\mathbf{dM})(0)}$ 				&$10^{-4}$ & $10^{-4}$	  	 \\
$\varepsilon^{\text{abs}}$ 	&$10^{-1}$ & $10^{-2}$	  	 \\
$\varepsilon^{\text{rel}}$ 	&$10^{-4}$ & $10^{-4}$	  	 \\ \bottomrule
		\end{tabular}
	\end{center}
\end{table}

\begin{figure}[t!]
\centering\subfloat[1][Th. Abundance 1]{
\includegraphics[keepaspectratio,height=0.3\textheight , width=0.13\textwidth]{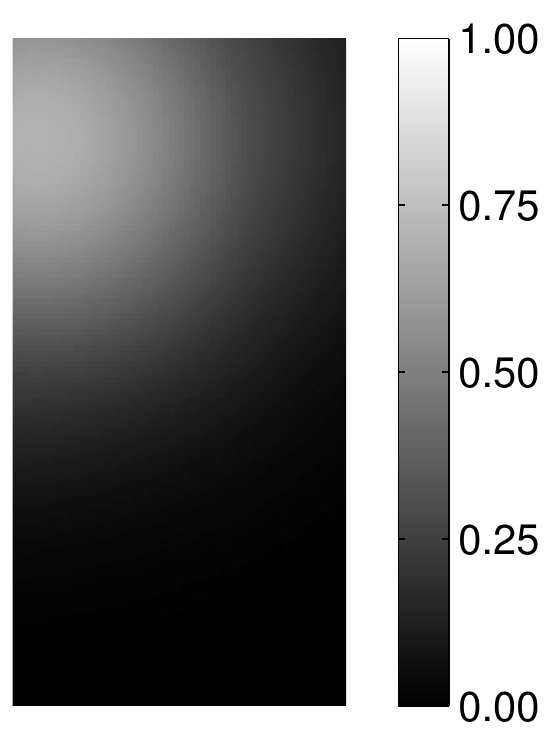}
\label{fig:smooth3_abundance_th1}}
\subfloat[2][Th. Abundance 2]{
\includegraphics[keepaspectratio,height=0.3\textheight , width=0.13\textwidth]{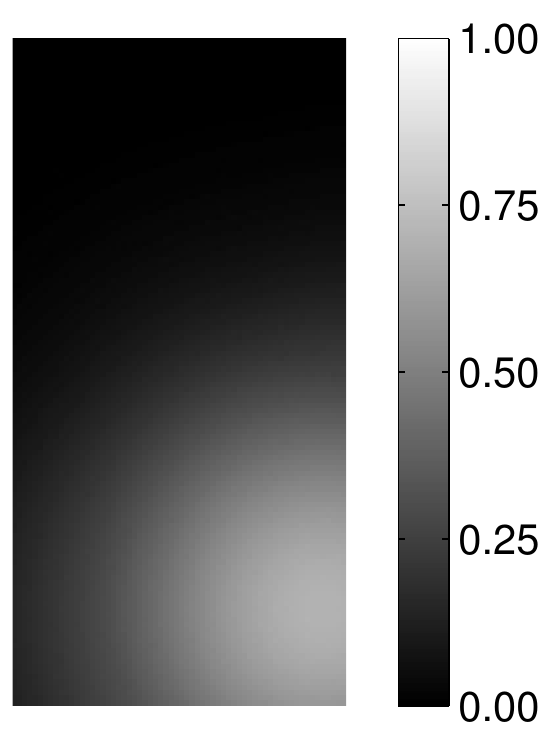}
\label{fig:smooth3_abundance_th2}}
\subfloat[3][Th. Abundance 3]{
\includegraphics[keepaspectratio,height=0.3\textheight , width=0.13\textwidth]{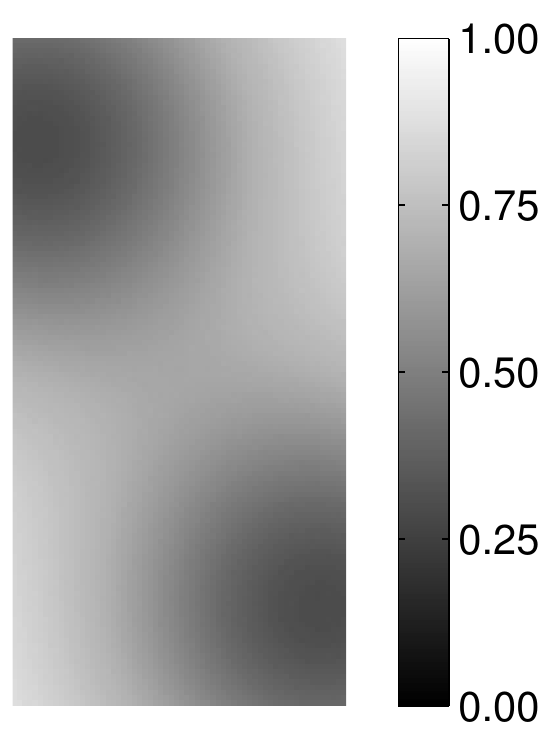}
\label{fig:smooth3_abundance_th3}}
\\
\subfloat[1][Abundance 1]{
\includegraphics[keepaspectratio,height=0.3\textheight , width=0.13\textwidth]{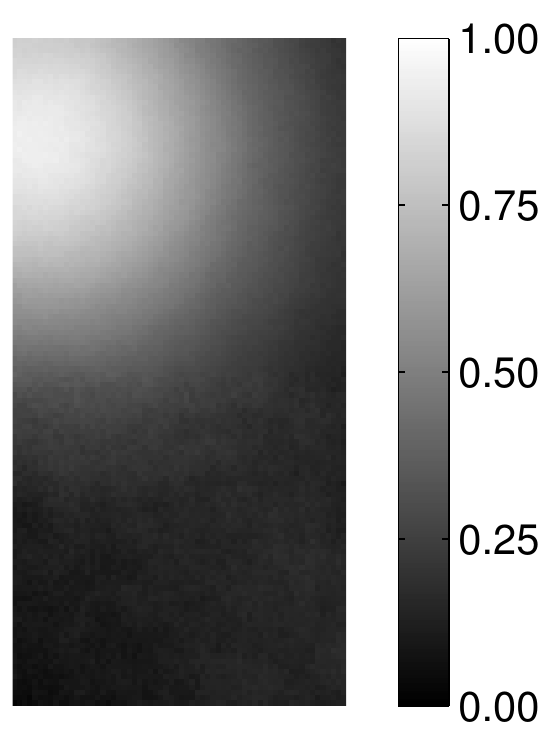}
\label{fig:smooth3_abundance1}}
\subfloat[2][Abundance 2]{
\includegraphics[keepaspectratio,height=0.3\textheight , width=0.13\textwidth]{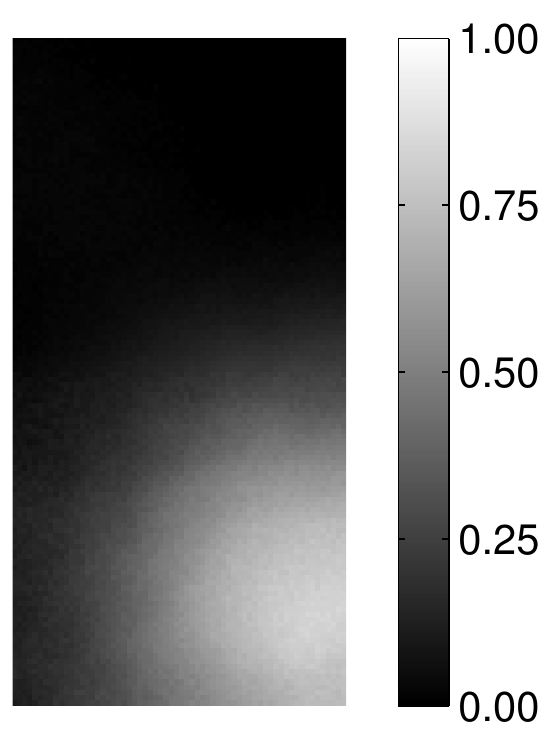}
\label{fig:smooth3_abundance2}}
\subfloat[3][Abundance 3]{
\includegraphics[keepaspectratio,height=0.3\textheight , width=0.13\textwidth]{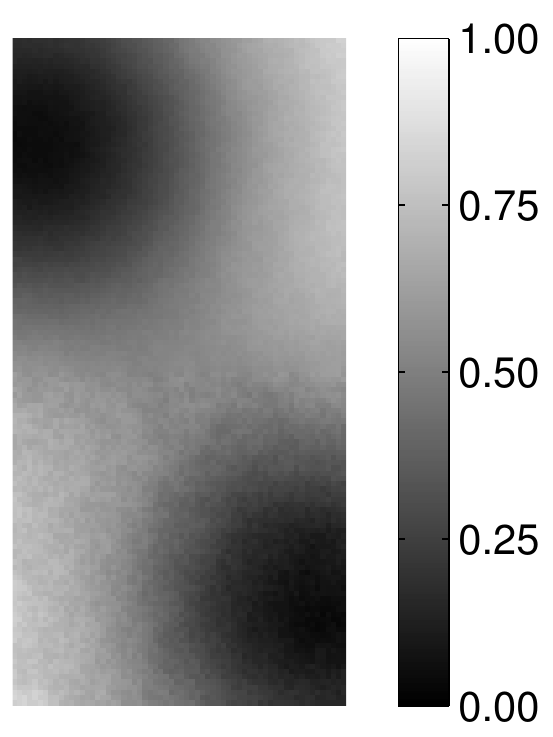}
\label{fig:smooth3_abundance3}}
\\
\subfloat[1][Variability 1]{
\includegraphics[keepaspectratio,height=0.3\textheight , width=0.13\textwidth]{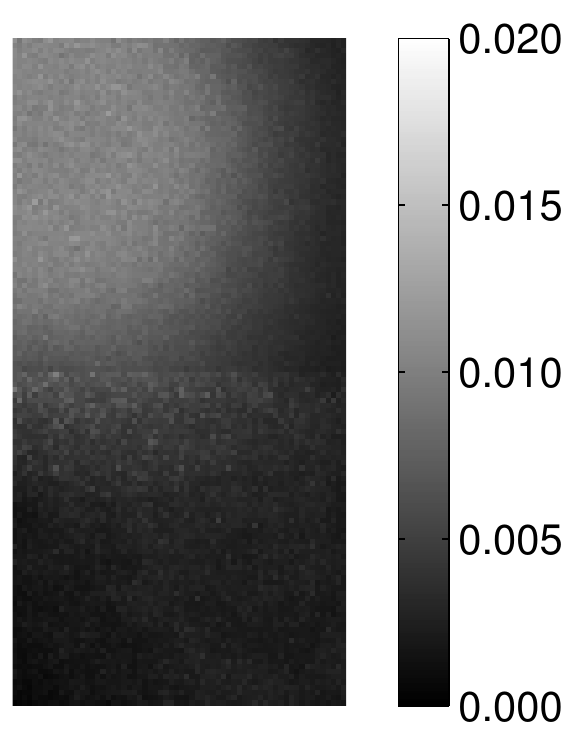}
\label{fig:smooth3_var_map1}}
\subfloat[2][Variability 2]{
\includegraphics[keepaspectratio,height=0.3\textheight , width=0.13\textwidth]{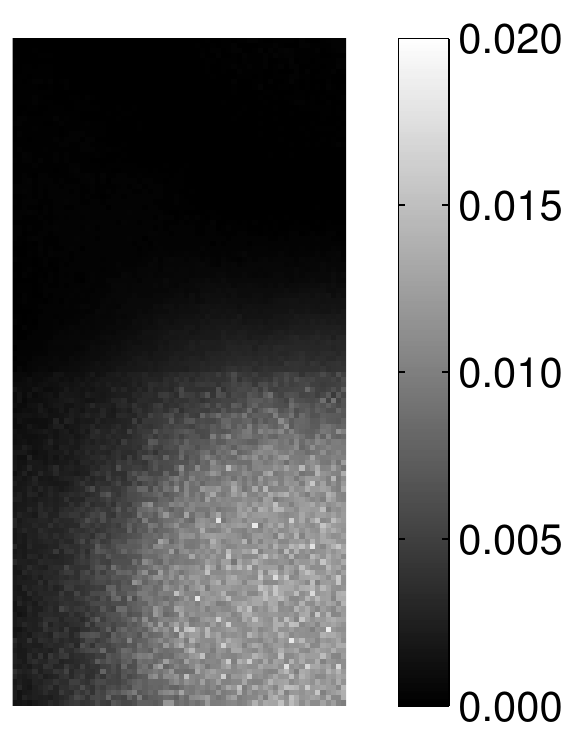}
\label{fig:smooth3_var_map2}}
\subfloat[3][Variability 3]{
\includegraphics[keepaspectratio,height=0.3\textheight , width=0.13\textwidth]{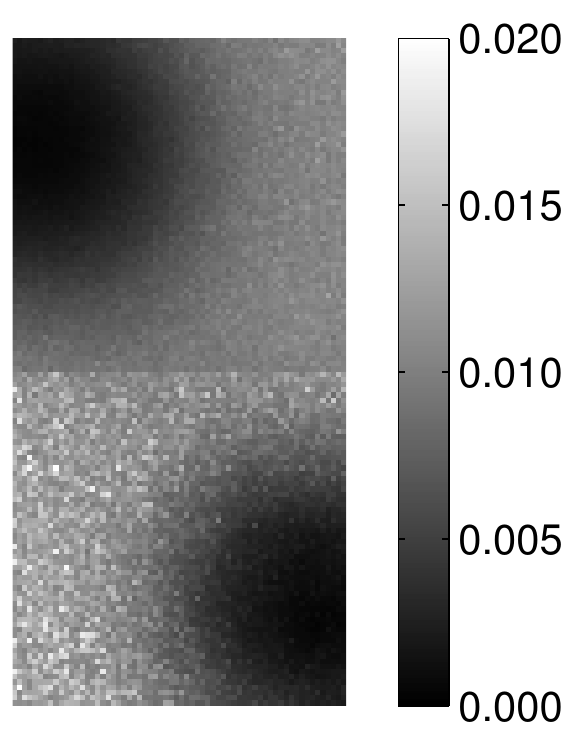}
\label{fig:smooth3_var_map3}}
\caption{True abundances (fig. \ref{fig:smooth3_abundance_th1} to \ref{fig:smooth3_abundance_th3}) and the ssmvBCD/ADMM-estimations (fig. \ref{fig:smooth3_abundance1} to \ref{fig:smooth3_endm3}) -- obtained with a synthetic dataset (no pure pixels, $\nendm = 3$). The spatial distribution of the variability \wrt{} each endmember is presented in terms of energy ($\frac{1}{\sqrt{\nband}} \lVert \mathbf{dm}_{n,k} \rVert_{2}$ for the $k$th endmember in the $n$th pixel) for visualization purpose in Figs. \ref{fig:smooth3_var_map1} to \ref{fig:smooth3_var_map3}.}
\label{fig:smooth3_abundance}
\end{figure}{}%

Different penalization combinations have been compared for the proposed method. The abbreviations \emph{ss}, \emph{mv} and \emph{vca} are used for \emph{spatially smooth}, \emph{minimum volume} and \emph{minimum distance to VCA} in the following. The absence of any additional abbreviation means that the method does not include any abundance or endmember penalization term.\\

The performance of the algorithm has been assessed in terms of endmember estimation using the average spectral angle mapper (aSAM)
\begin{equation*}
\aSAM(\M) = \frac{1}{\nendm } \sum_{k=1}^\nendm    \frac{ \left\langle \mathbf{m}_k \middle|  \widehat{\mathbf{m}}_k \right\rangle } { \lVert \mathbf{m}_k \rVert_2 \lVert \widehat{\mathbf{m}}_k \rVert_2   }
\end{equation*}{}%
as well as in terms of abundance and perturbation estimation by global mean square errors (GMSEs)
\begin{align*}
\GMSE(\A)  & = \frac{1}{\nendm \nbpix} \lVert \A - \widehat{\A}
\rVert_{\text{F}}^2 \\
\GMSE(\dM) & = \frac{1}{\nbpix \nband \nendm } \sum_{n=1}^\nbpix \lVert \dM_n - \widehat{\dM}_n \rVert_{\text{F}}^2 .
\end{align*}{}
As a measure of fit, the following reconstruction error (RE) has been also considered
\begin{align*}
\RE &= \frac{1}{\nband \nbpix}\Norm_fro{\Y - \widehat{\Y}}
\end{align*}{}%
where $\widehat{\Y}$ is the matrix formed of the pixels reconstructed using the parameters estimated by the algorithm. 

\begin{figure}[!t]
\centering
\subfloat[Endmember 1]{
\includegraphics[keepaspectratio,height=0.2\textheight , width=0.24\textwidth]{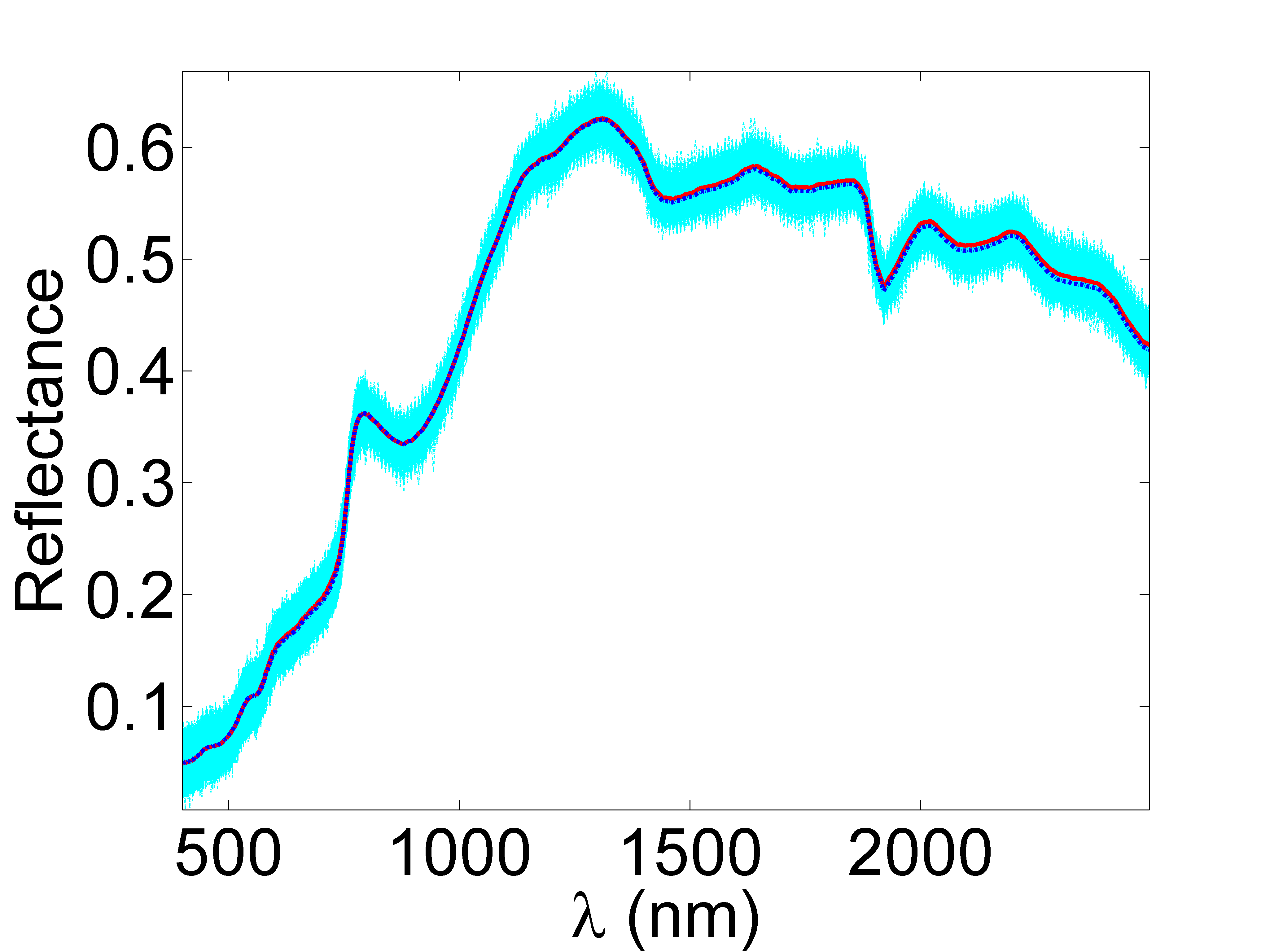}
\label{fig:smooth3_endm1}}
\subfloat[Endmember 2]{
\includegraphics[keepaspectratio,height=0.2\textheight , width=0.24\textwidth]{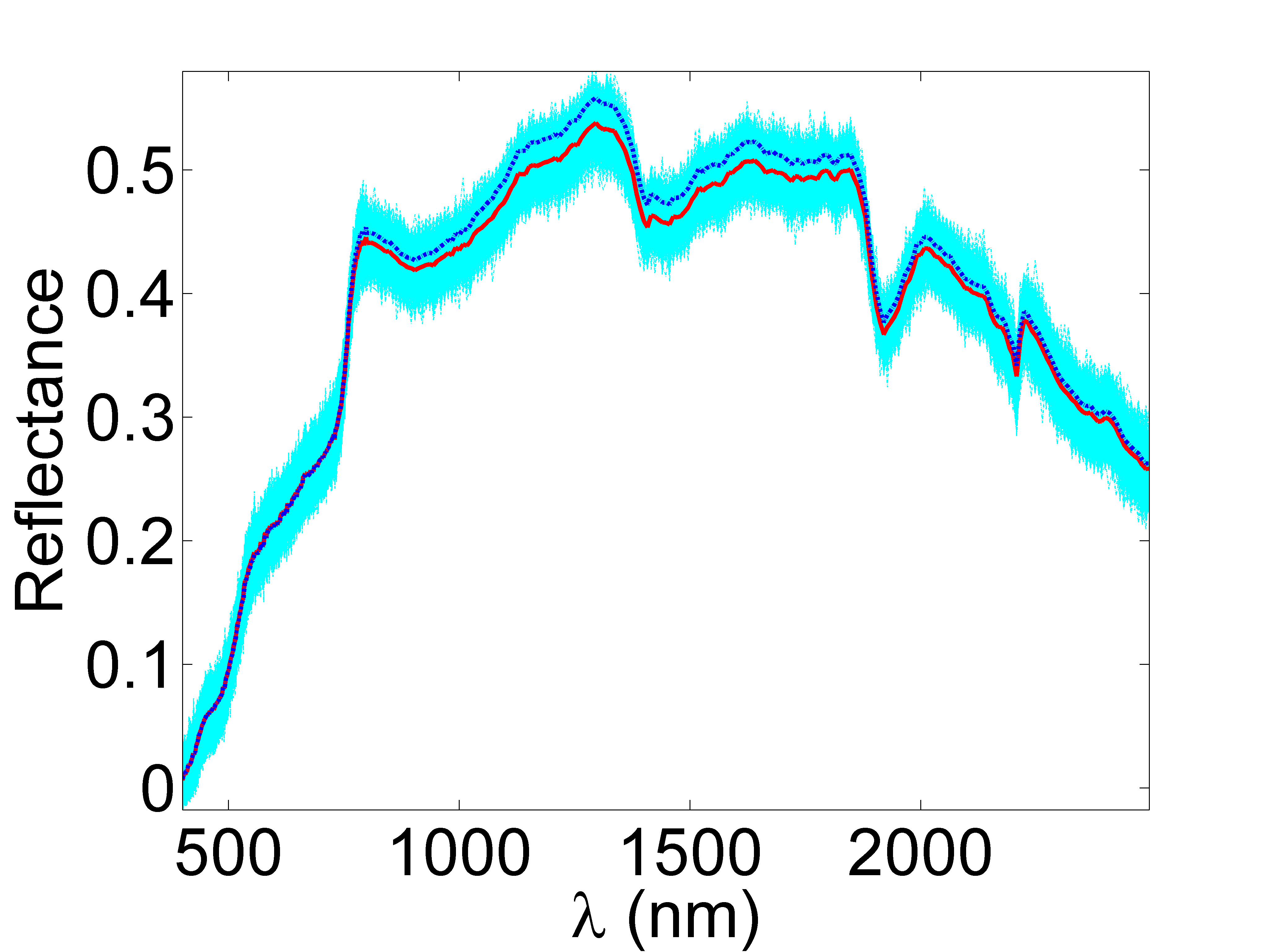}
\label{fig:smooth3_endm2}}
\\
\subfloat[Endmember 3]{
\includegraphics[keepaspectratio,height=0.2\textheight , width=0.24\textwidth]{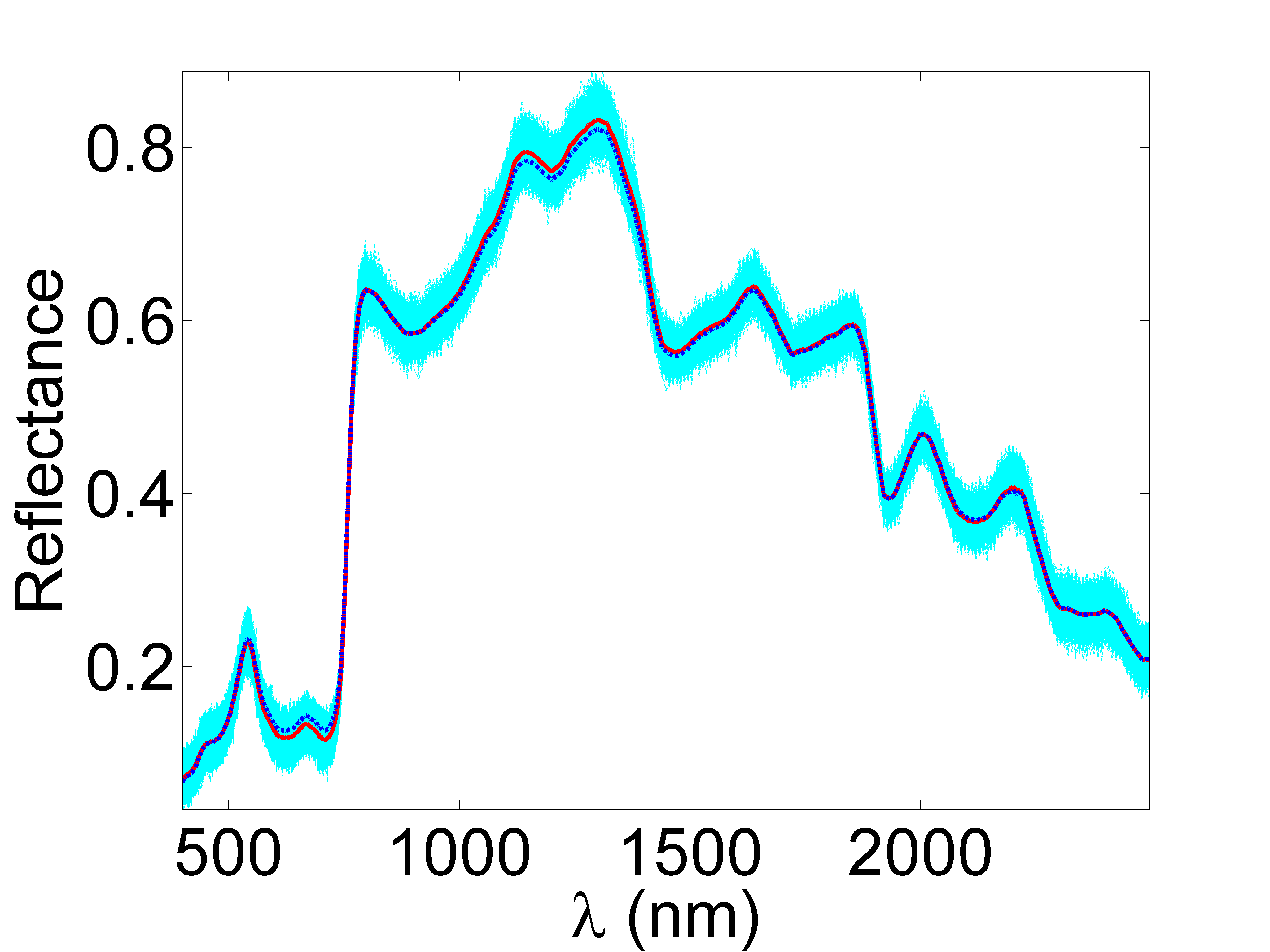}
\label{fig:smooth3_endm3}}
\caption{Endmember estimations obtained on synthetic data in absence of pure pixels (cf. Figs. \ref{fig:smooth3_abundance} for the abundance estimations). The ssmvBCD/ADMM-estimated endmembers (red lines) are given with typical examples of the estimated variability (cyan dotted lines). The VCA endmembers are given in blue dotted lines for comparison.}
\label{fig:smooth3_endm}
\end{figure}{}%
\begin{table*}
\caption{Simulation results for synthetic data in presence of pure pixels (GMSE($\mathbf{A}$)$\times 10^{-2}$, GMSE($\mathbf{dM}$)$\times 10^{-4}$, RE $\times 10^{-4}$, $\gamma = 1$).}
	\begin{center}
		\begin{tabular}{@{}lccccc|ccccc@{}} \toprule
			  & \multicolumn{5}{c}{$\nendm = 3$, $(\alpha,\beta) = (1.4,2.5 \times 10^{-5})$}& \multicolumn{5}{c}{$\nendm = 6$, $(\alpha,\beta) = (0.37,5.1 \times 10^{-4})$} \\ \cmidrule{2-11}
		      & aSAM($\mathbf{M}$) (\textdegree)& GMSE($\mathbf{A}$) & GMSE($\mathbf{dM}$) & RE & time (s) 
		      & aSAM($\mathbf{M}$) (\textdegree)& GMSE($\mathbf{A}$) & GMSE($\mathbf{dM}$) & RE & time (s) \\ \cmidrule{2-11}
VCA/FCLS	  & 6.0038 & 3.80  &/		& 7.56 & \textbf{1}   %
              &	6.3313 & 2.24  &/		& 2.92 & \textbf{1} 	\\
SISAL		  & \textbf{5.2665} & 3.08  &/		& 3.35 & 2	%
			  &	\textbf{3.8365} & 3.05  &/		& 2.25 & 3		\\
FDNS		  & 6.0038 & 3.79  &/		& 7.56 & 4   %
              &	6.3313 & 2.22  &/		& 2.92 & 5	\\
AEB			  & 5.6971 & \textbf{2.07}  &/		& 3.50 & 52  %
			  &	5.7017 & \textbf{1.31}  &/		& 2.40 & 142	\\
BCD/ADMM		  & 5.9910 & 3.51  &\textbf{4.00}   & \textbf{0.20} & 92 	 %
			  &	6.2965 & 1.59  &\textbf{2.93} 	& \textbf{0.05} & 230 	\\
ssBCD/ADMM	  & 5.7765 & 3.15  & 4.25 	& 0.23 & 422 	 %
			  &	6.0304 & 1.44  & 2.97 	& 0.07 & 848 	\\
ssmvBCD/ADMM  & 5.4390 & 3.01  & 4.25 	& 0.25 & 530%
			  &	6.3397 & 1.42  & 2.97 	& 0.07 & 603 	\\ \bottomrule
		\end{tabular}
	\end{center}
\label{tab:results_synth_smooth_pp}
\end{table*}

\begin{table*}
\caption{Simulation results for synthetic data in absence of pure pixels (GMSE($\mathbf{A}$)$\times 10^{-2}$, GMSE($\mathbf{dM}$)$\times 10^{-4}$, RE $\times 10^{-4}$, $\gamma = 1$).}
	\begin{center}
		\begin{tabular}{@{}lccccc|ccccc@{}} \toprule
			  & \multicolumn{5}{c}{$\nendm = 3$, $(\alpha,\beta) = (24.5,4.2 \times 10^{-9})$}& \multicolumn{5}{c}{$\nendm = 6$, $(\alpha,\beta) = (0.71,4.8 \times 10^{-4})$} \\ \cmidrule{2-11}
		      & aSAM($\mathbf{M}$) (\textdegree)& GMSE($\mathbf{A}$) & GMSE($\mathbf{dM}$) & RE & time (s) & aSAM($\mathbf{M}$) (\textdegree)& GMSE($\mathbf{A}$) & GMSE($\mathbf{dM}$) & RE & time (s) \\ \cmidrule{2-11}
VCA/FCLS	  & 5.0639 & 2.07   &/		& 2.66   & \textbf{1}   %
              &	6.5530 & 2.52   &/		& 2.82   & \textbf{4} 	\\
SISAL		  & 4.4318 & 2.16 &/		& 2.56 & 2	 %
			  &	6.0431 & \textbf{1.63} &/		& 2.02 & 5   	\\
FDNS		  & 5.0639 & 2.06 &/		& 2.66 & 3   %
              &	6.5530 & 2.53 &/		& 2.82 & 7	\\
AEB			  & 5.1104 & 2.11 &/		& 2.66 & 33  %
			  &	\textbf{6.0016} & 1.78 &/		& 1.85 & 208\\
BCD/ADMM	  & 5.2480 & 2.13 &\textbf{3.81}	& \textbf{0.25} & 140 %
			  &	6.2785 & 2.14	& 3.33	& 0.30 & 3041 \\
ssBCD/ADMM	  & \textbf{4.1549} & \textbf{1.44} & 4.36	& 0.38 & 1263 %
			  &	6.2763 & 1.74 & \textbf{3.04}	& \textbf{0.076} & 1527 \\
ssmvBCD/ADMM  & 5.0584 & 1.94 & 4.59	& 0.47 & 1667 %
			  &	6.3207 & 1.67 & 3.05	& 0.08 & 795 \\ \bottomrule
		\end{tabular}
	\end{center}
\label{tab:results_synth_smooth}
\end{table*}

	\subsection{Results}

The parameters used for the ADMM algorithms are detailed in Table \ref{tab:param}, and the values chosen by cross-validation for $\alpha$, $\beta$ and $\gamma$ are reported in Table \ref{tab:results_synth_smooth_pp} and \ref{tab:results_synth_smooth}. The performance measures returned by the unmixing methods are provided in Table \ref{tab:results_synth_smooth_pp} for the datasets containing pure pixels, and in Table \ref{tab:results_synth_smooth} for images without pure pixels, leading to the following conclusions.

\begin{itemize}
\item The proposed method is robust to the absence of pure pixels;
\item The proposed method provides competitive results in terms of $\aSAM$ while allowing endmember variability to be estimated for each endmember in each pixel;
\item For most simulation scenarios, the abundance MSEs and the REs are lower than the MSEs and REs resulting from state-of-the-art methods;
\item The proposed method is computationally more expensive than existing algorithms.
\end{itemize}

We can note that the smoothness penalization on the abundances proves to be particularly appropriate in this experiment. Moreover, an increasing number of endmembers implies a loss of estimation performance. This result can be expected since VCA/FCLS algorithm is used as an initialization step.

 Finally, the variability captured by the proposed model is presented in Figs. \ref{fig:smooth3_abundance} and \ref{fig:smooth3_endm} for three endmembers: the difference between the variability intensities detected in the upper and the lower part of the scene is due to the different variability coefficients applied to these areas, thus illustrating the consistency of the proposed method.

\section{Experiment with real data} \label{sec:experiments}

	\subsection{Description of the datasets}

The proposed algorithm has been applied to real hyperspectral datasets obtained by the Airborne Visible Infrared Imaging Spectrometer (AVIRIS).
The first scene was acquired over Moffett Field, CA, in 1997. Water absorption bands were removed from the 224 spectral bands, leaving 189 exploitable spectral bands. The scene of interest (50 $\times$ 50) is partly composed of a lake and a coastal area.

The second scene is a $190 \times 250$ image extracted from the well-known Cuprite dataset\footnote{The Moffett and Cuprite images are available at \url{http://www.ehu.es/ccwintco/index.php?title=Hyperspectral_Remote_Sensing_Scenes}, and \url{http://aviris.jpl.nasa.gov/}}. The number of spectral bands is 189 after removing the water-absorption and low SNR bands. Many works previously conducted on this image provide reference abundance estimation maps.

	The parameters used for the proposed approach are identical to those used for the experiments with synthetic data (see Table \ref{tab:param}). The only difference is that the algorithm has been stopped when the relative difference between two successive values of the objective function is less than $10^{-2}$. This value has been chosen to obtain a compromise between the estimation accuracy and the computational cost implied. The values selected by cross-validation for $\alpha$, $\beta$ and $\gamma$ are given in Table \ref{tab:results_real}.

\begin{table}[h!]
\caption{Experiment results conducted on real data [ssBCD/ADMM for Moffett with $(\alpha,\beta) = (0.05,0)$, ssvcaBCD/ADMM for Cuprite with $(\alpha,\beta) = (0.014,404)$, RE $\times 10^{-4}$,$\gamma = 1$].}
\label{tab:results_real}
	\begin{center}
		\begin{tabular}{@{}lcc|cc@{}} \toprule
			  & \multicolumn{2}{c}{Moffett} & \multicolumn{2}{c}{Cuprite}\\ \cmidrule{2-5}
		      & RE	& time (s)  & RE   & time (s) \\ \cmidrule{2-5}
VCA/FCLS	  & 2.50	& \textbf{0.4}		& 3.69 & \textbf{9.9}	    \\ 
SISAL		  & 1.12	& 30		& 2.16 & 15    \\ 
FDNS		  & 2.69	& 1 		& 3.69 & 11        \\
AEB			  & 6.25	& 10		& 0.40 & 615       \\
BCD/ADMM	  & \textbf{0.18}	& 144 		& \textbf{0.23} & 1.9e4      \\ \bottomrule
		\end{tabular}
	\end{center}
\end{table}	


	\subsection{Results}
	
\begin{figure}[b!]
\centering
\subfloat[1][Abundance 1]{
\includegraphics[keepaspectratio,height=0.3\textheight , width=0.13\textwidth]{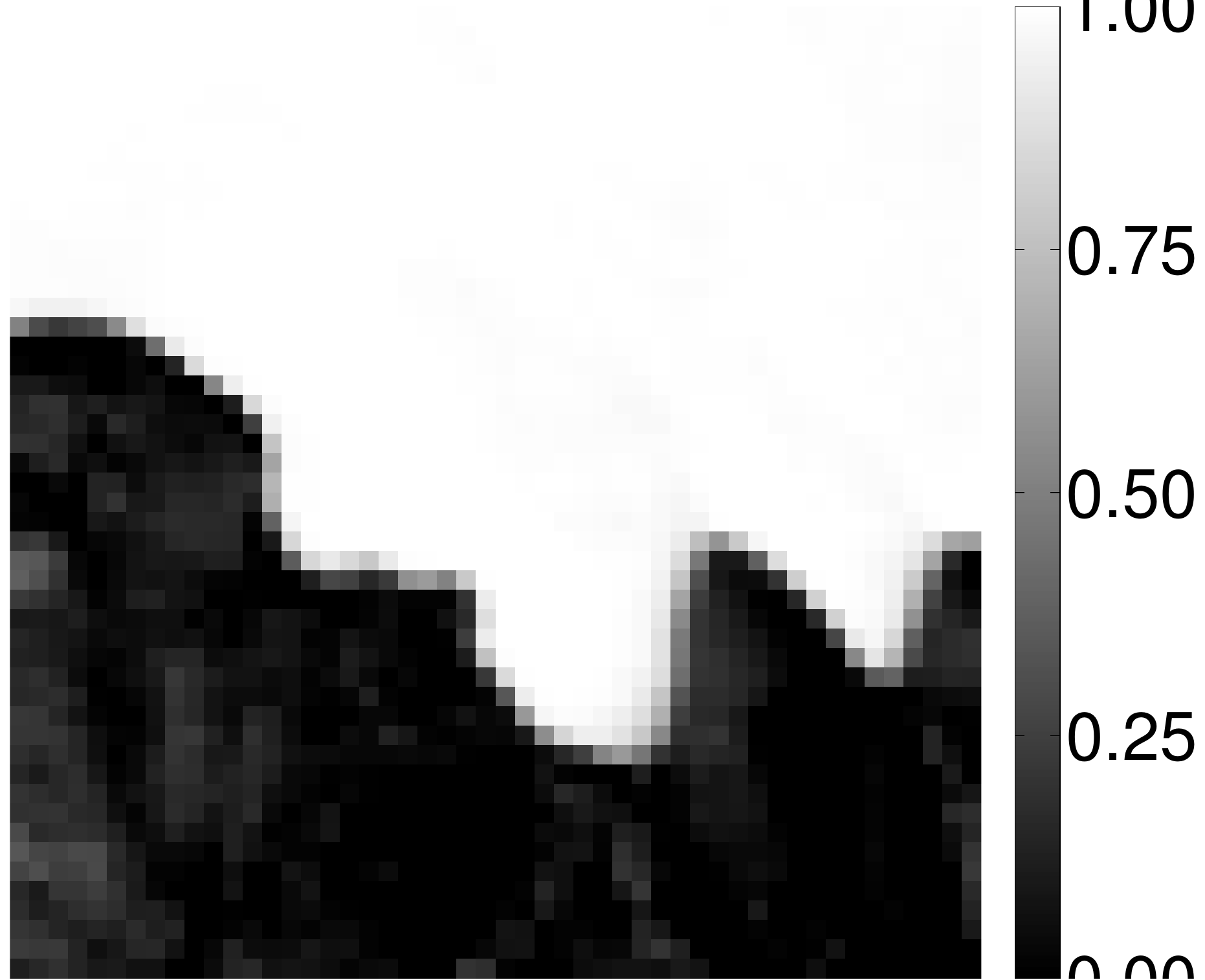}
\label{fig:moffett_abundance1_vca}}
\subfloat[2][Abundance 2]{
\includegraphics[keepaspectratio,height=0.3\textheight , width=0.13\textwidth]{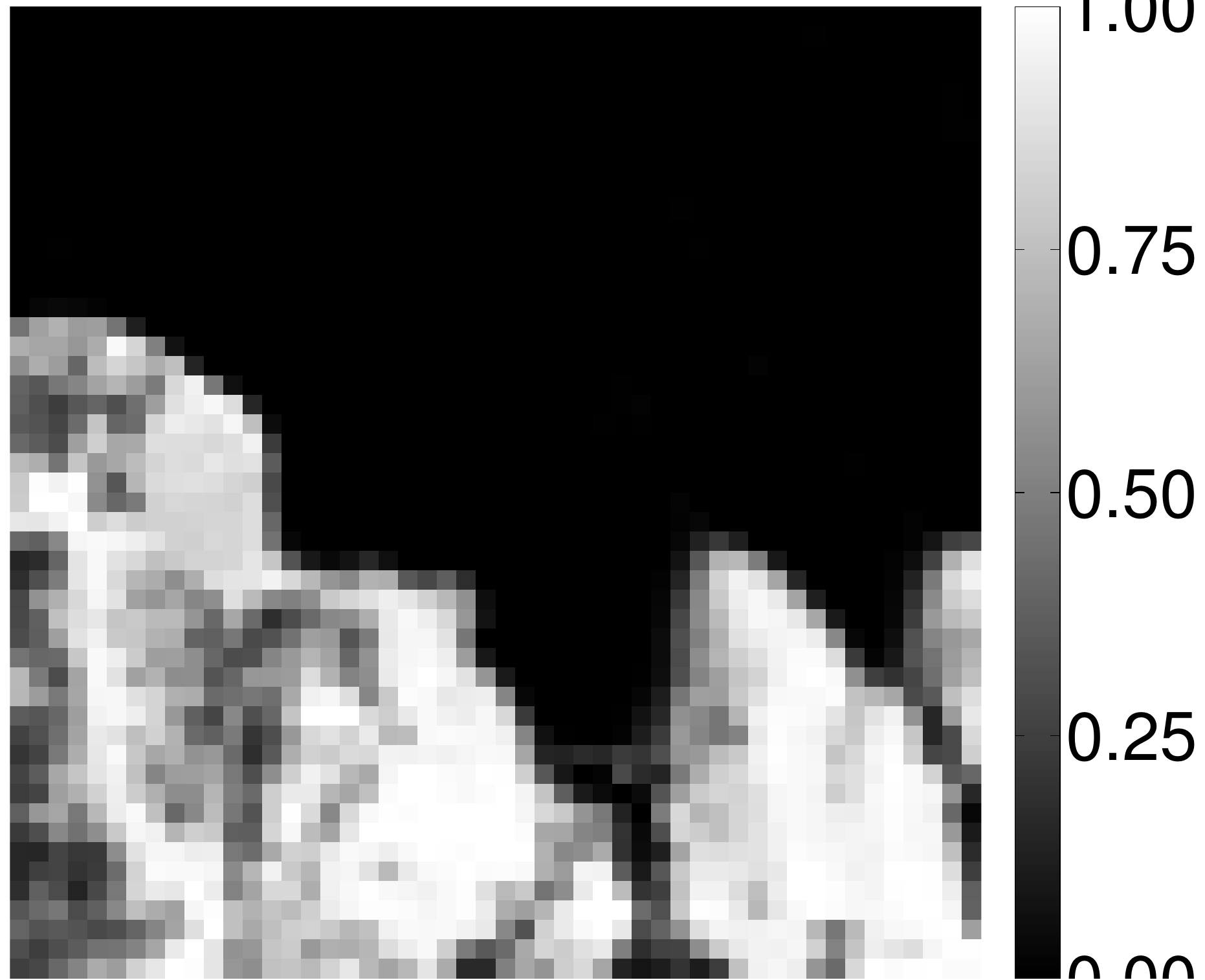}
\label{fig:moffett_abundance2_vca}}
\subfloat[3][Abundance 3]{
\includegraphics[keepaspectratio,height=0.3\textheight , width=0.13\textwidth]{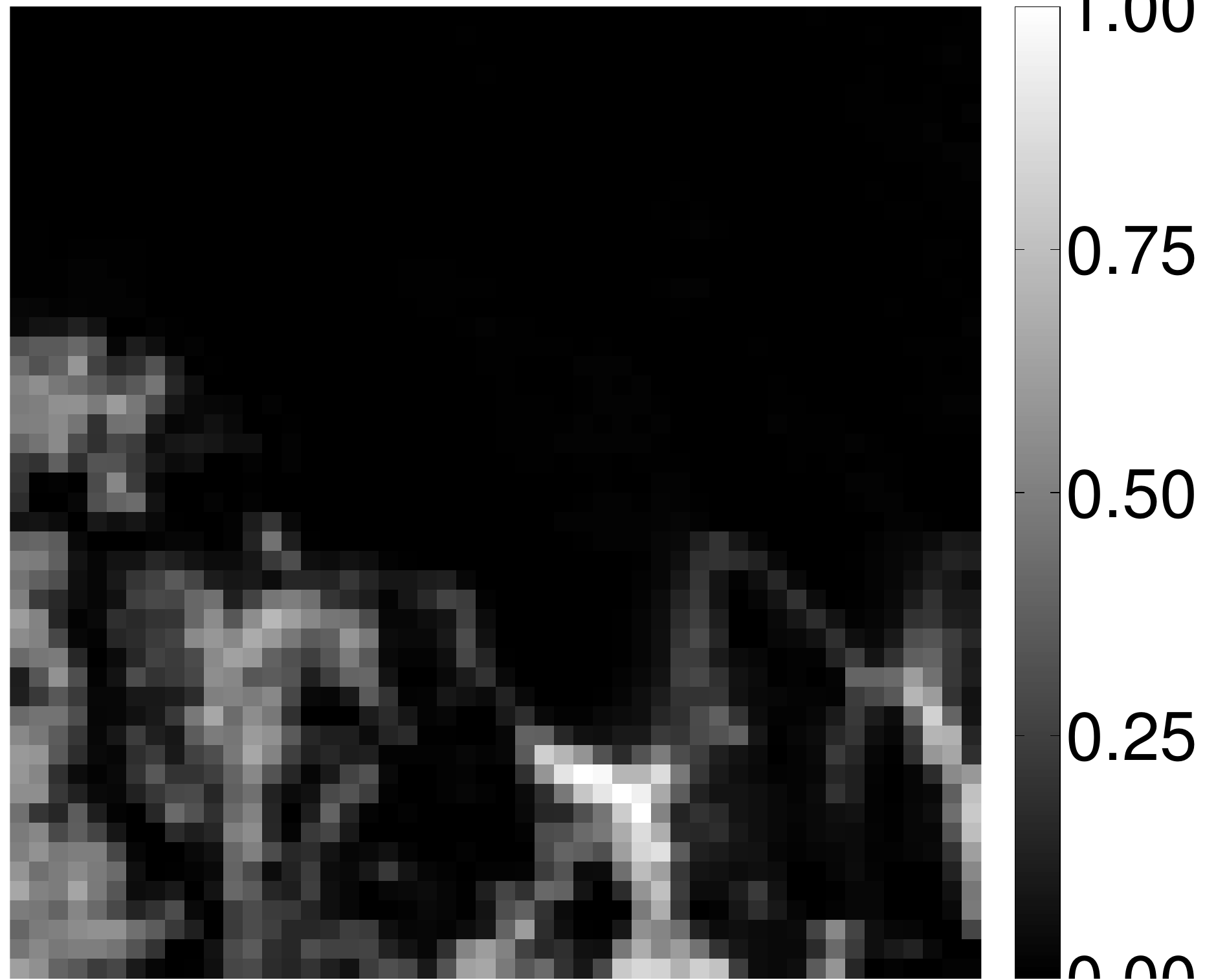}
\label{fig:moffett_abundance3_vca}}
\\
\subfloat[1][Abundance 1]{
\includegraphics[keepaspectratio,height=0.3\textheight , width=0.13\textwidth]{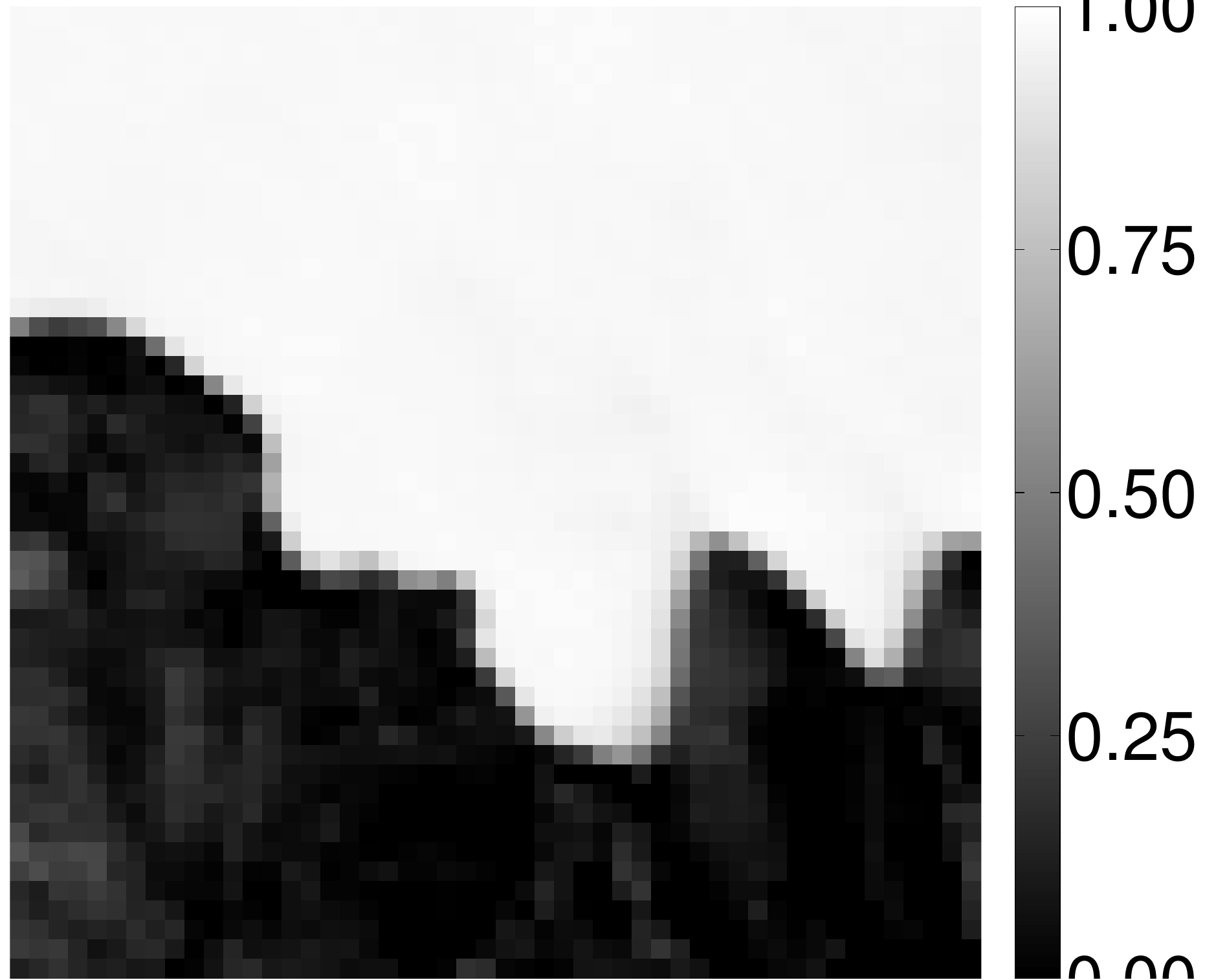}
\label{fig:moffett_abundance1}}
\subfloat[2][Abundance 2]{
\includegraphics[keepaspectratio,height=0.3\textheight , width=0.13\textwidth]{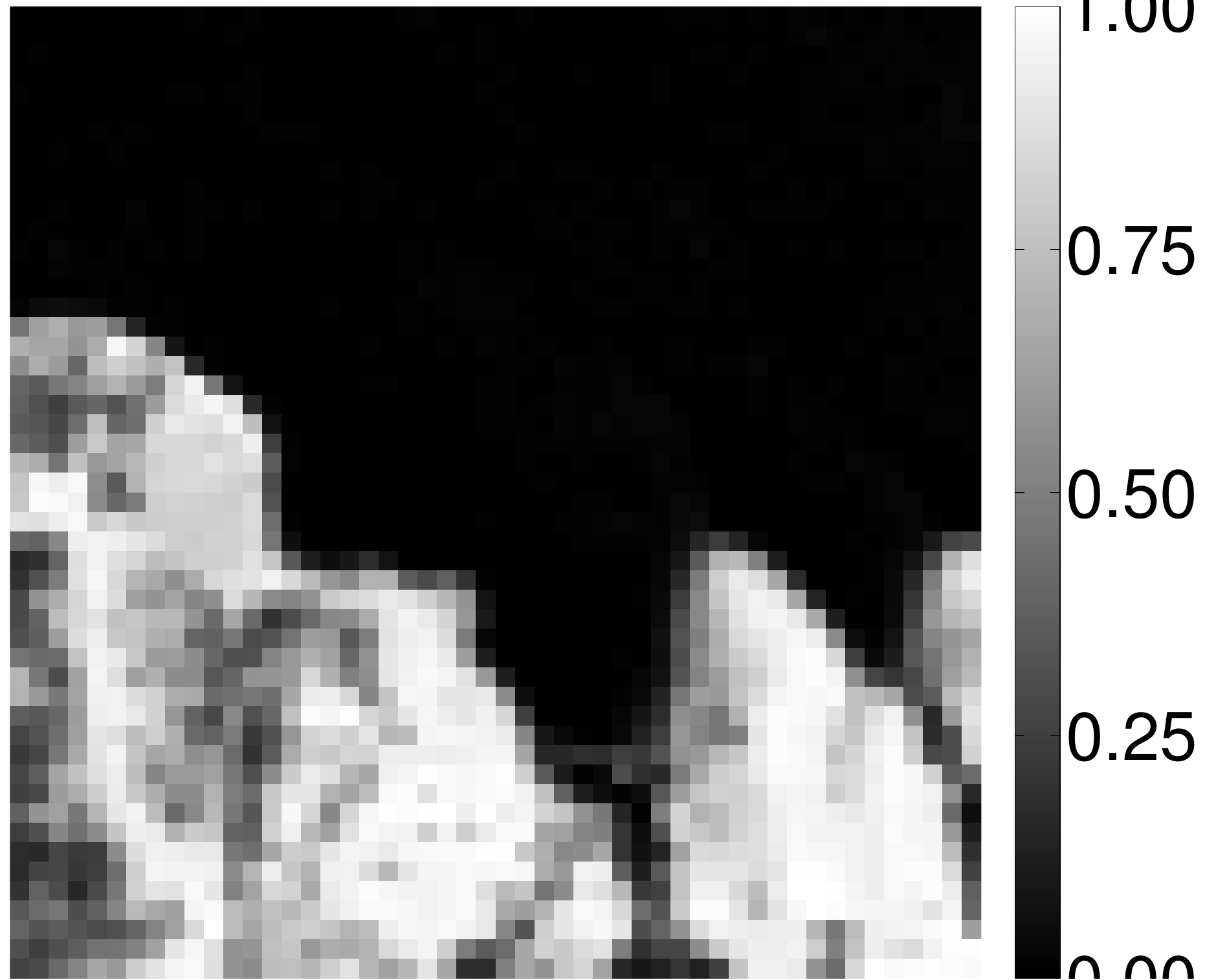}
\label{fig:moffett_abundance2}}
\subfloat[3][Abundance 3]{
\includegraphics[keepaspectratio,height=0.3\textheight , width=0.13\textwidth]{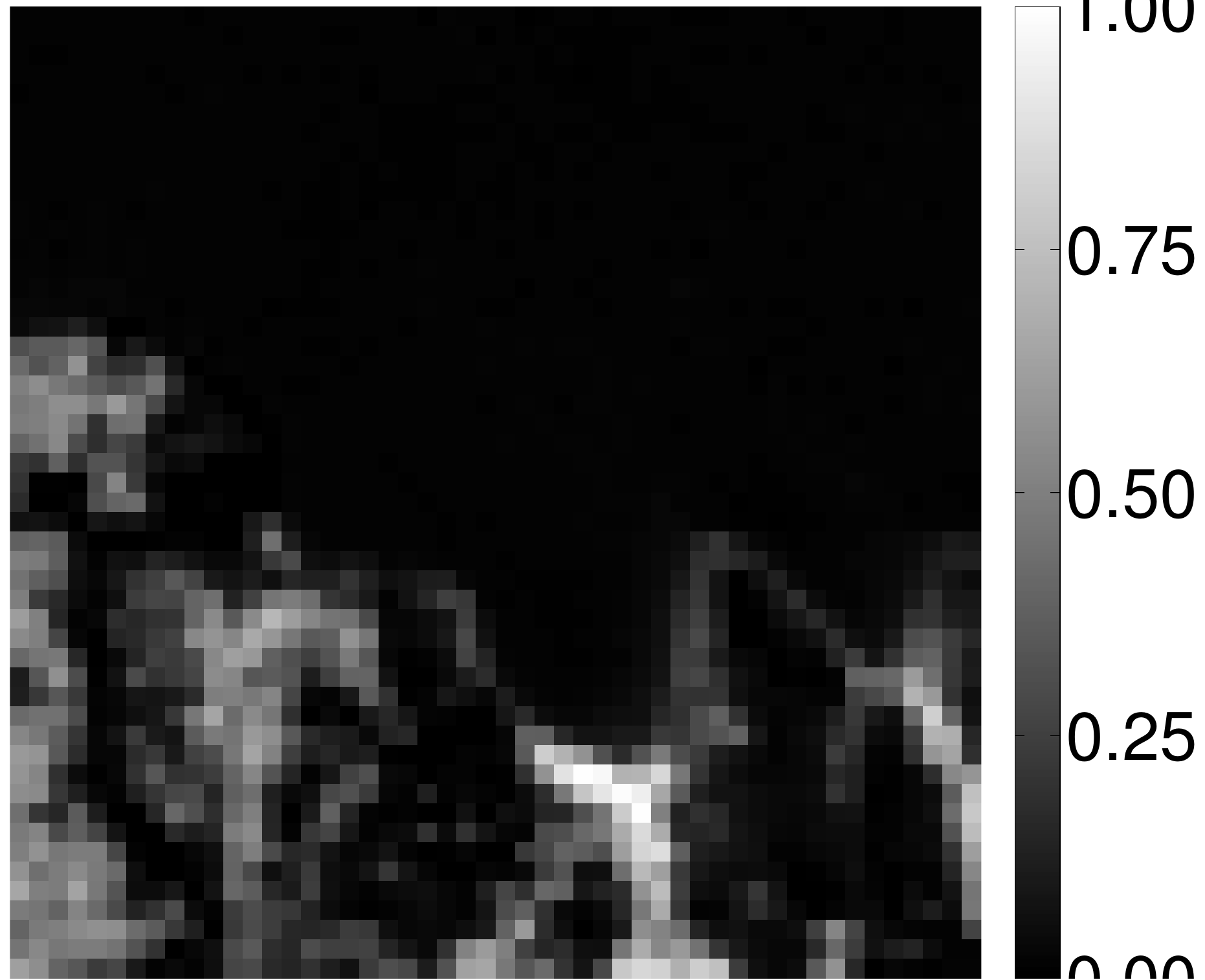}
\label{fig:moffett_abundance3}}
\\
\subfloat[7][Variability 1]{
\includegraphics[keepaspectratio,height=0.3\textheight , width=0.13\textwidth]{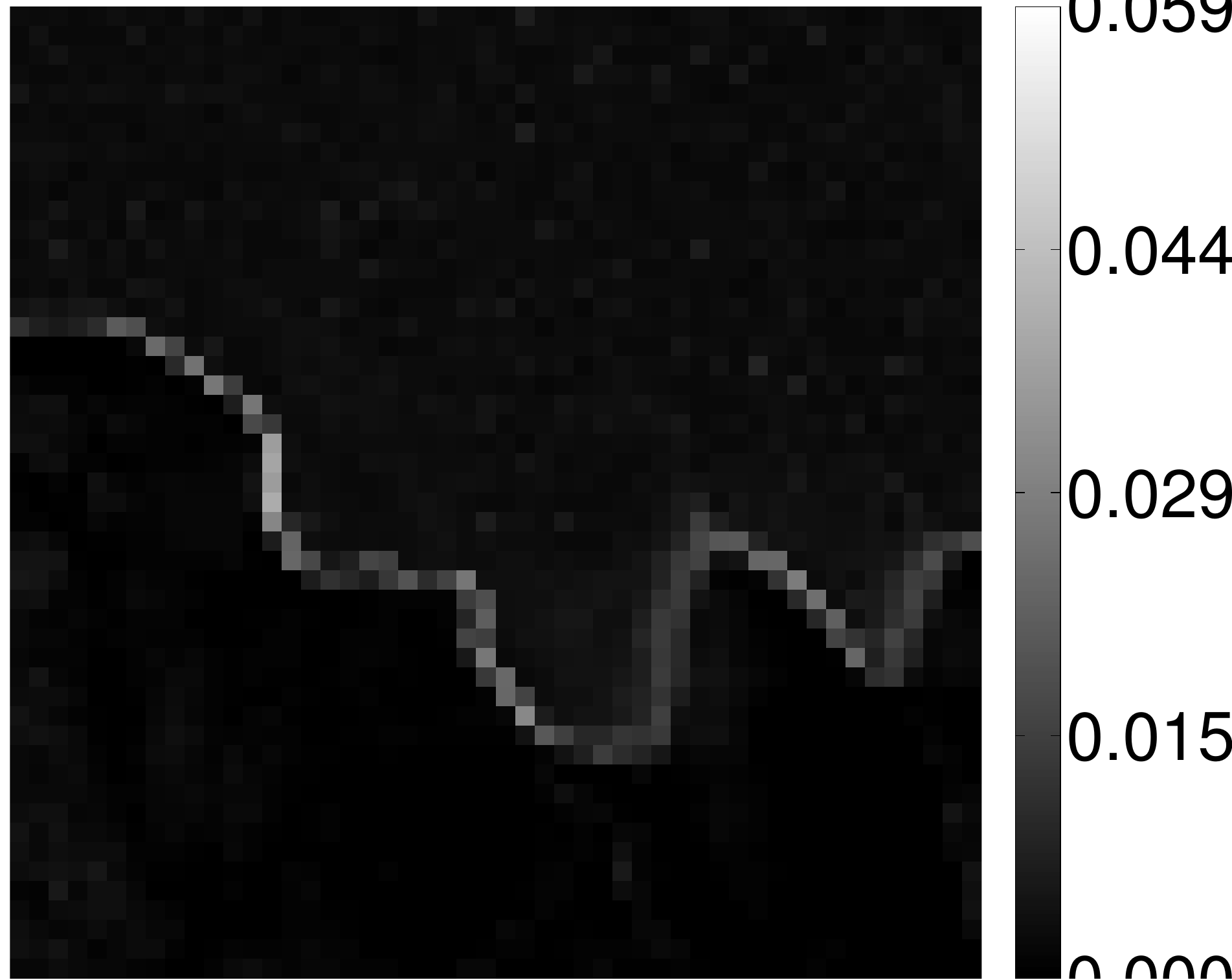}
\label{fig:moffett_var_map1}}
\subfloat[8][Variability 2]{
\includegraphics[keepaspectratio,height=0.3\textheight , width=0.13\textwidth]{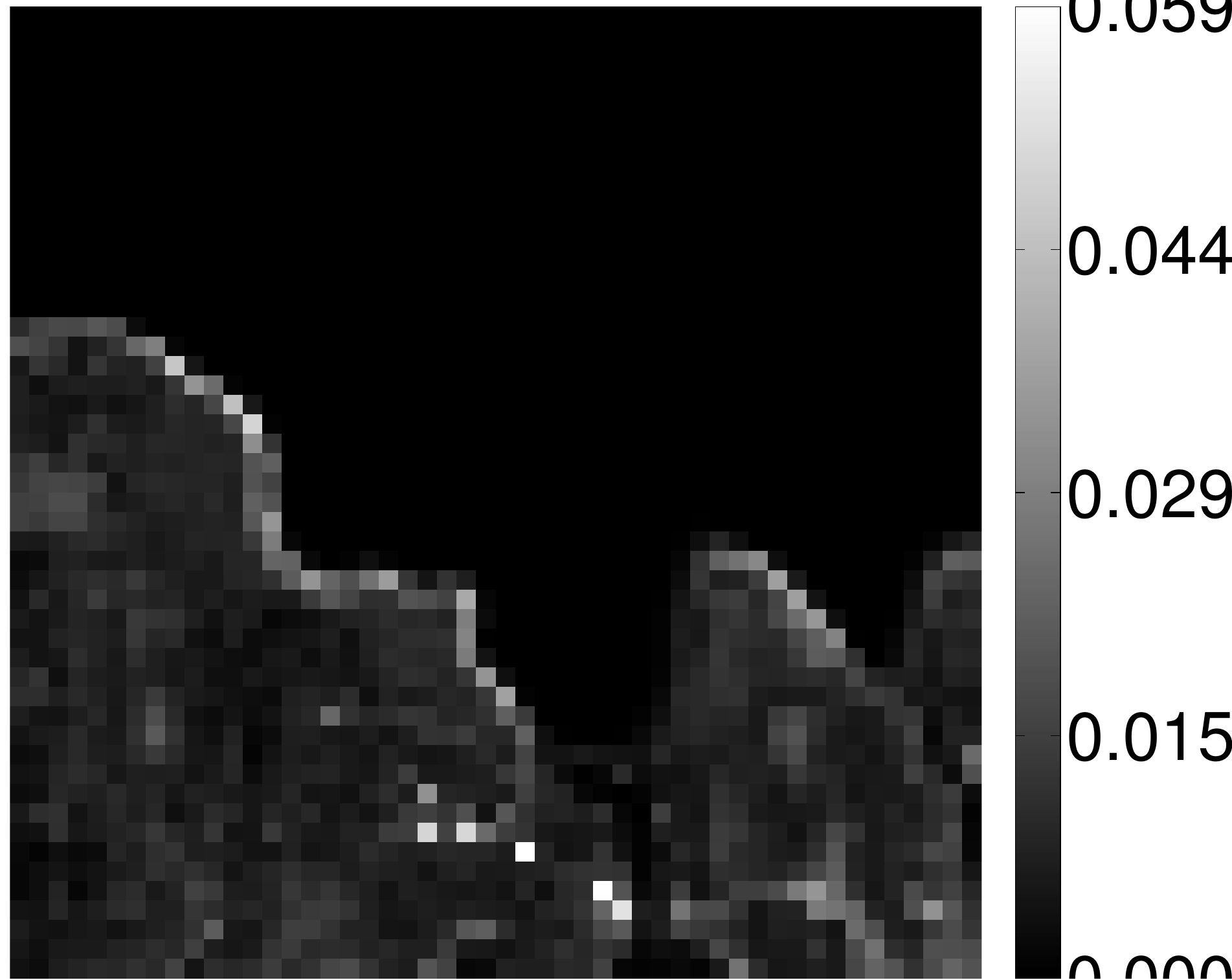}
\label{fig:moffett_var_map2}}
\subfloat[9][Variability 3]{
\includegraphics[keepaspectratio,height=0.3\textheight , width=0.13\textwidth]{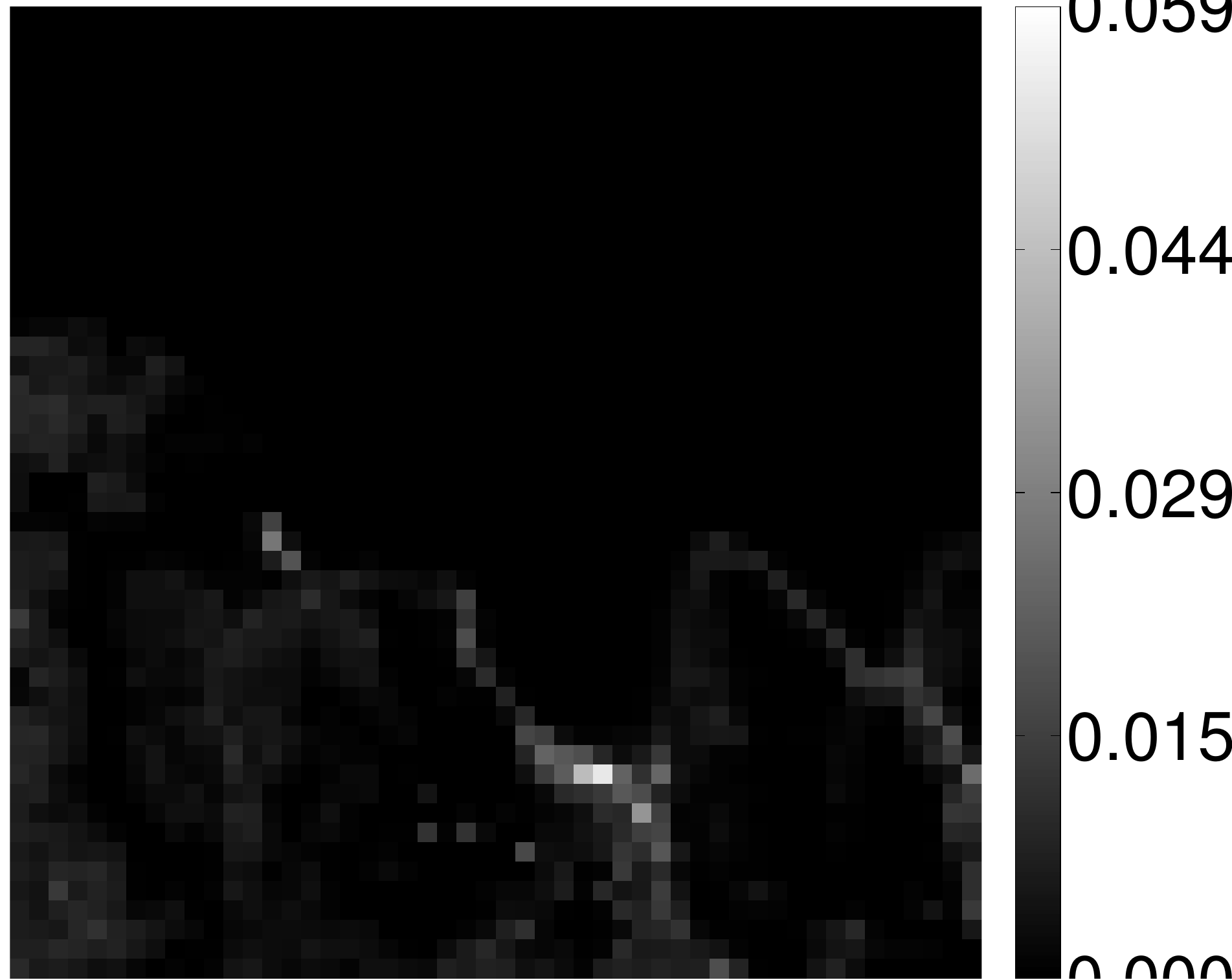}
\label{fig:moffett_var_map3}}
\caption{ssmvBCD/ADMM abundance estimations (Moffett scene). VCA/FCLS results are shown in Figs. \ref{fig:moffett_abundance1_vca} to \ref{fig:moffett_abundance3_vca} whereas Figs. \ref{fig:moffett_abundance1} to \ref{fig:moffett_endm3} are for the ssmvBCD/ADMM. The spatial distribution of the variability \wrt{} each endmember is presented in terms of energy ($\frac{1}{\sqrt{\nband}} \lVert \mathbf{dm}_{n,k} \rVert_{2}$ for the $k$th endmember in the $n$th pixel) for visualization purpose.}
\label{fig:moffett_abundance}
\end{figure}

\begin{figure}[hb!]
\centering
\subfloat[4][Endmember 1]{
\includegraphics[keepaspectratio,height=0.2\textheight , width=0.23\textwidth]{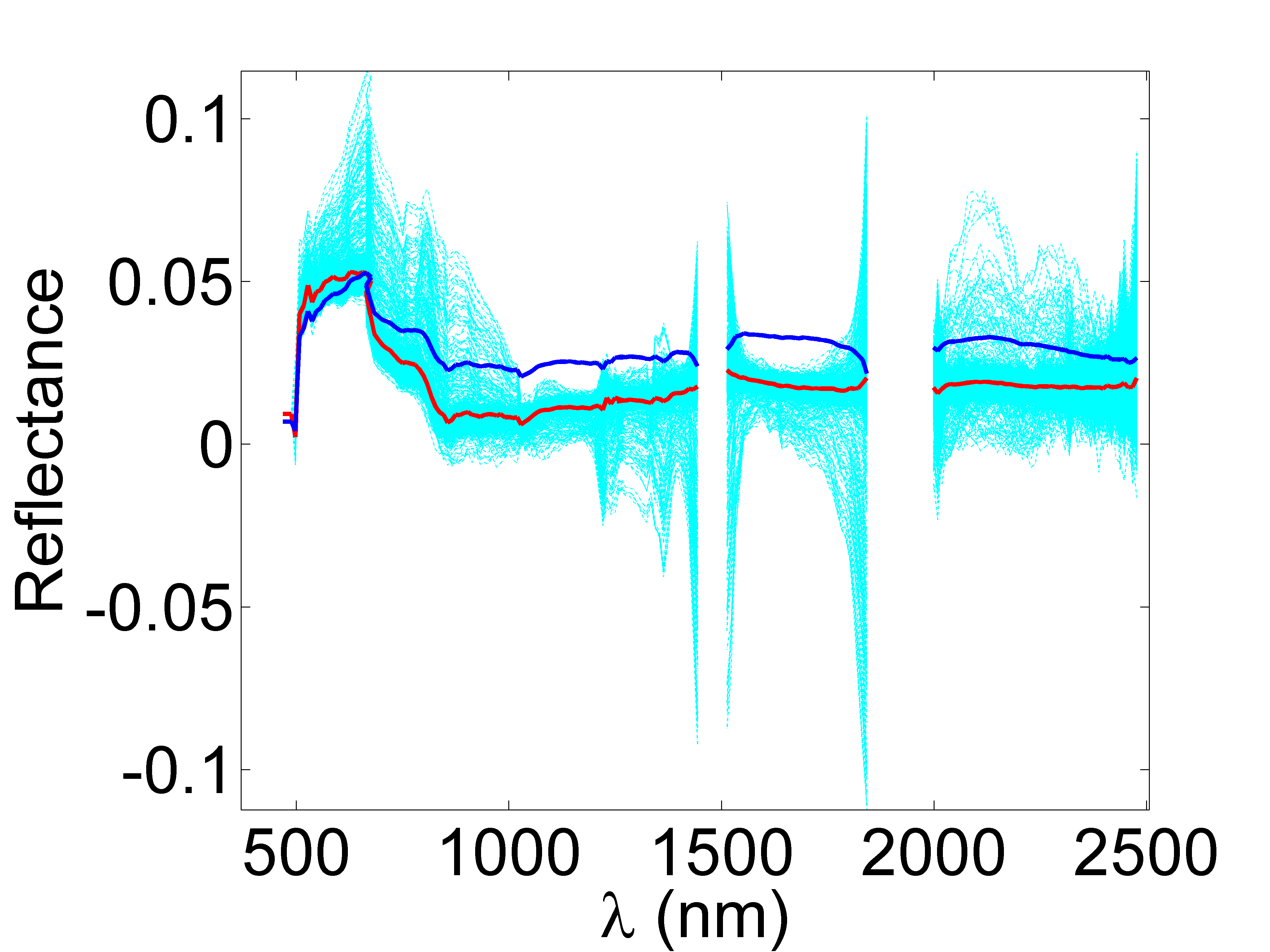}
\label{fig:moffett_endm1}}
\subfloat[5][Endmember 2]{
\includegraphics[keepaspectratio,height=0.2\textheight , width=0.23\textwidth]{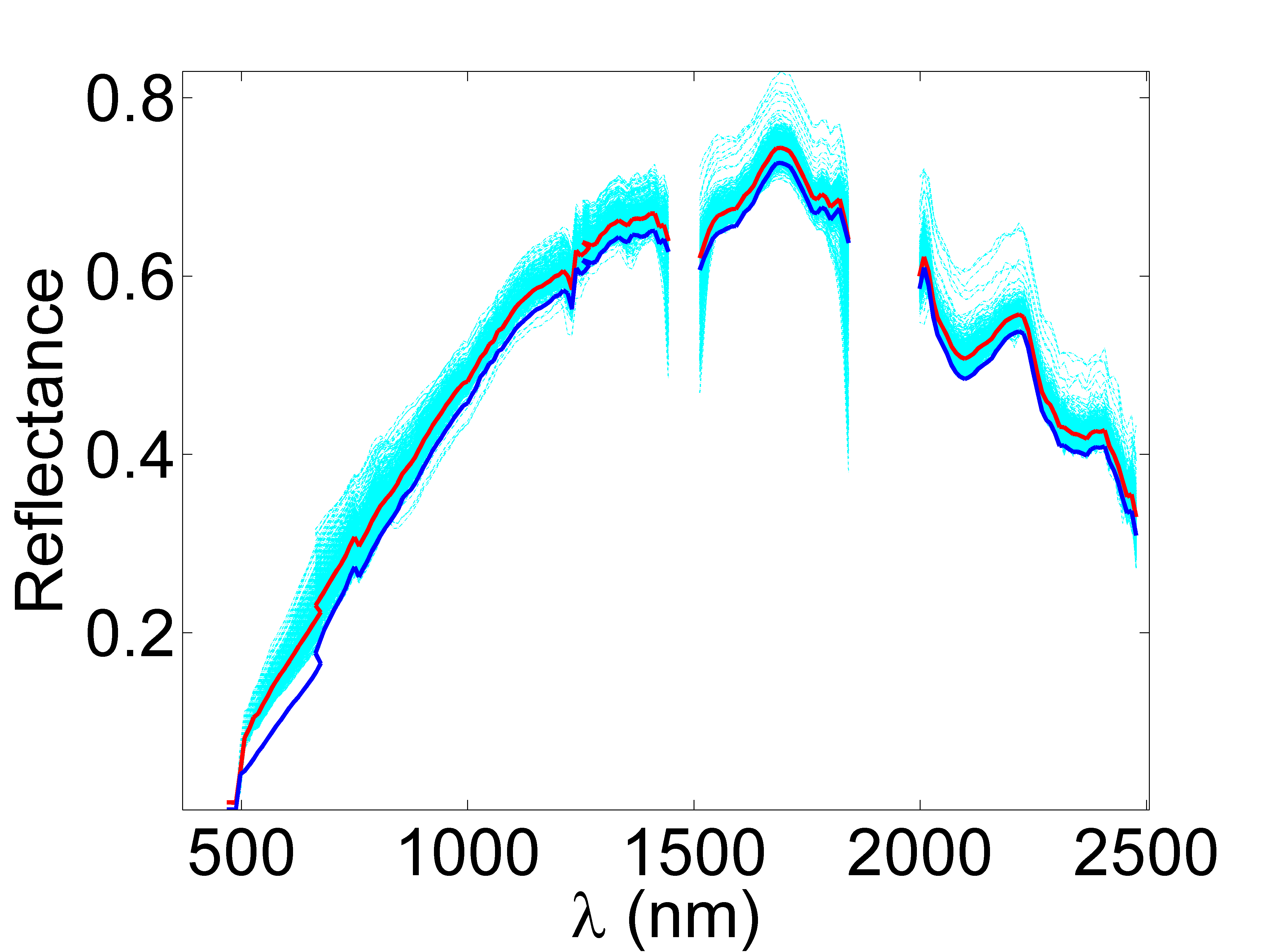}
\label{fig:moffett_endm2}}
\\
\subfloat[6][Endmember 3]{
\includegraphics[keepaspectratio,height=0.2\textheight , width=0.24\textwidth]{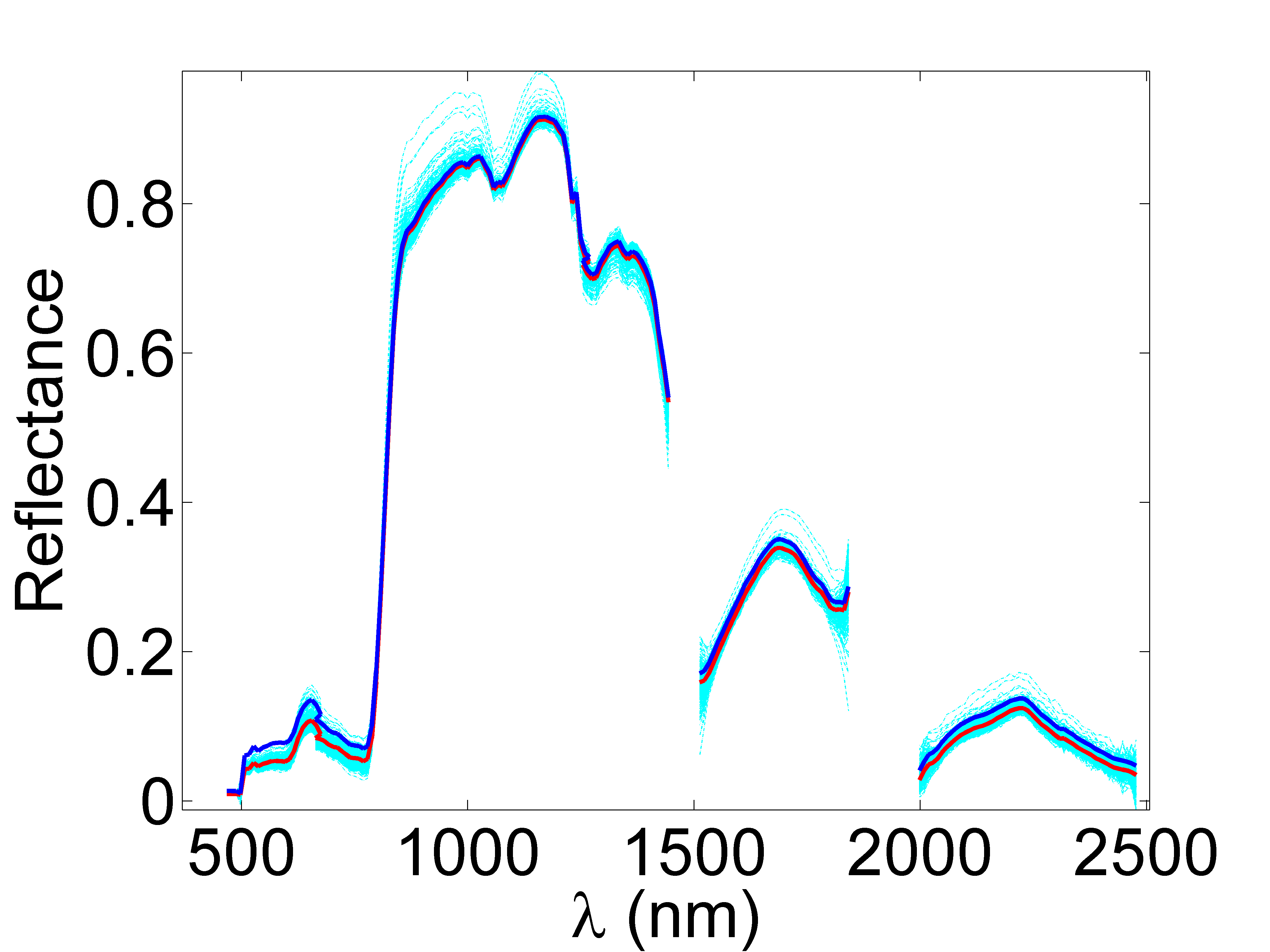}
\label{fig:moffett_endm3}}
\caption{ssmvBCD/ADMM endmembers (Moffett scene). The ssmvBCD/ADMM-estimated endmembers (red lines) are plotted with the VCA endmembers (blue lines) for comparison, and typical examples of the estimated variability are given in cyan dotted lines.}
\label{fig:moffett_endm}
\end{figure}

	The unmixing performance are reported in Table \ref{tab:results_real}. For the Moffett image, the variability detected by the proposed algorithm is displayed in Figs. \ref{fig:moffett_abundance} and \ref{fig:moffett_endm}. The variability seems to be more significant on the coastal area where the mixture is not appropriately described by a linear model. The potential non-linearities usually observed close to the coastal areas \cite{Altmann2011whispers,Fevotte2015,Halimi2011} are interpreted as variability in the proposed method, which tends to corroborate its consistency. Note that the advantage of the proposed method is that it does not require to consider a sophisticated non-linear model accounting for interactions between the different endmembers as in \cite{Altmann2011whispers,Halimi2011,Dobigeon2014}. Conversely, all deviations from the LMM are contained in the variability components $\dM_{n,k}$. We can also note that the variability peaks observed in Fig. \ref{fig:moffett_endm} are a clear indication that several corrupted spectral bands have not been removed prior to the unmixing process.

	The results obtained for the Cuprite scene are reported in Figs. \ref{fig:cuprite}, \ref{fig:cuprite_var} and \ref{fig:cuprite_endm}. Comparing our results with those of \cite{Nascimento2005}, we visually found out that some similar endmembers that were identified as different signatures by VCA for $K = 14$ \cite{Nascimento2005} are interpreted as multiple instances of single endmembers in our setting ($K = 10$). The identification is given in Figs. \ref{fig:cuprite} and \ref{fig:cuprite_endm}. Fig. \ref{fig:cuprite_var} shows that the algorithm captured a significant variability level in the pixels where many different endmembers are detected, which reveals that the spectral mixture may not be strictly linear in these pixels.


{}%

{}%

\section{Conclusion and future work} \label{sec:conclusion}
This paper introduced a new linear mixing model accounting for spatial and spectral endmember variabilities. The proposed model extended the classical LMM by including an additive spatially varying perturbation matrix that can capture endmember variabilities.
The resulting unmixing problem was solved by alternating marginal minimizations of an appropriately regularized cost function, each minimization being performed by an ADMM algorithm. Simulations conducted on synthetic and real data enabled the interest of the proposed solution to be appreciated. Indeed, the proposed method compared favorably with state-of-the-art approaches while providing a relevant variability estimation.  
The choice of the penalization parameters $\alpha,\beta$ and $\gamma$ was performed by cross validation. We think that it would be interesting to develop automatic strategies for estimating these parameters. 
Finally, due to the  significant number of unknown parameters, the proposed method is not intended to be applied to very large images. The proposed approach can be applied as a complementary tool when analyzing small hyperspectral images \ap{} believed to be affected by a non-negligible variability level. 
Decreasing the computational complexity of the algorithm introduced in this work is clearly an interesting prospect.

\begin{appendices}
\section{Constraints and Penalization Terms \label{app:constraints}}
\subsection{Abundance penalization: spatial smoothness}

	The abundance smoothness is expressed in matrix form as
\begin{equation}
\Phi(\mathbf{A}) = \frac{1}{2}\Norm_fro{\mathbf{AH}}
\end{equation}
where $\mathbf{H}$ denotes the matrix computing the differences between the abundances of a given pixel and the respective abundances of its 4 neighbors
\begin{equation*}
\mathbf{H} = \biggl[ \begin{array}{c|c|c|c} %
 \mathbf{H}_{\leftarrow} & \mathbf{H}_{\rightarrow} & \mathbf{H}_{\uparrow} &  \mathbf{H}_{\downarrow} %
 \end{array} \biggr] \in \mathbb{R}^{\nbpix  \times 4\nbpix }.
\end{equation*}{}%
For $h = 1, \dotsc , H $, we introduce %
\begin{align*}
\mathbf{H}_h &= \begin{pmatrix}
0         	& -1       	& 0        	& \cdots 	&   0      \\
0         	&  1        	& \ddots 	& \ddots 	& \vdots \\
\vdots  	& \ddots 	& \ddots 	& \ddots 	&   0      \\
\vdots 	& 		  	& \ddots  	& 1     		&   -1     \\
0        	& \dotsc  	& \dotsc  	& 0    		&    1      \\
\end{pmatrix} \in \mathbb{R}^{W \times W} \\ %
\tilde{\mathbf{H}}_h &=  \begin{pmatrix}
1         	& 0       	& \cdots	& \cdots 	&  0 \\
-1       	&  1        	& \ddots 	&  			& \vdots \\
0 			& \ddots 	& \ddots 	& \ddots 	& \vdots \\
\vdots 	& \ddots  	& \ddots  	& 1     		&  0 \\
0         	& \cdots  	& 0        	& -1   		&  0 \\
\end{pmatrix} \in \mathbb{R}^{W \times W}.
\end{align*}{}%
Hence
\begin{equation*}
\mathbf{H}_{\leftarrow} = \text{Diag}(\mathbf{H}_1, \dotsc, \mathbf{H}_H) \quad \text{and} \quad \mathbf{H}_{\rightarrow} = \text{Diag}(\tilde{\mathbf{H}}_1, \dotsc, \tilde{\mathbf{H}}_H).
\end{equation*}{}%
In addition
\begin{align*}
\mathbf{H}_{\uparrow} = 
	\begin{bmatrix}
		\Zero[\nbpix]{W}, \mathbf{H}_{\uparrow}^1
	\end{bmatrix} \quad \text{and} \quad
\mathbf{H}_{\downarrow} = 
	\begin{bmatrix}
		\mathbf{H}_{\downarrow}^1, \Zero[\nbpix]{W}
	\end{bmatrix}
\end{align*}
with
%
%
\begin{align*}
\mathbf{H}_{\uparrow}^1 &= \left( \begin{array}{cc}
W & \left\updownarrow  \begin{array}{cccc}
  -1     	&     0      & \dotsc 	&   0      \\
        	&  \ddots 	& \ddots 	& \vdots \\
 1     	&     		& \ddots 	&   0      
 \end{array} \right. \\
N-W&  \left\updownarrow  \begin{array}{cccc}
  0     	&  \ddots &				&   -1    \\ 
\vdots	&  \ddots	&	\ddots 	&          \\
  0       	&  \dotsc 	& 0		 	&  1  
\end{array}  \right.     
\end{array} \right) \in \mathbb{R}^{\nbpix \times (\nbpix - W)} \\
\mathbf{H}_{\downarrow}^1 &= -\mathbf{H}_{\uparrow}^1.
\end{align*}
%
%
The only terms in $\frac{1}{2} \Norm_fro{\mathbf{AH}}$ related to $\mathbf{a}_n$ are
\begin{equation}
\label{eq:penalizationA}
\begin{split}
\phi(\mat{a}_n) = & \frac{1}{2} \underbrace{\biggl(\sum_{k=0}^3 h_{n,n+k\nbpix }^2 \biggr)}_{cA_n}\norm_2{\mathbf{a}_n} \\ %
& + \underbrace{\biggl(\underset{i \neq n}{\sum_{i = 1}^\nbpix } \sum_{k = 0}^3 h_{n,n+k\nbpix }h_{i,n+k\nbpix }\mathbf{a}_i^T\biggr)}_{\mathbf{c}_n^T} \mathbf{a}_n.
\end{split}
\end{equation}

\subsection{Endmember penalization}
	\subsubsection{Distance between the endmembers and reference signatures}
The distance between the endmembers and the available reference signatures is
\begin{equation}
\Psi(\mathbf{M}) = \frac{1}{2}\Norm_fro{\mathbf{M} - \mathbf{M}_0} = \frac{1}{2} \sum_{\ell=1}^\nband  \norm_2{\row{m}_\ell - \row{m}_{0,\ell}}.
\end{equation}{}%
As a consequence, the penalty for the $\ell$th band is
\begin{equation}
\label{eq:penalization_vca}
\psi (\row{m}_\ell ) = \frac{1}{2}\norm_2{\row{m}_\ell  - \row{m}_{0,\ell}}.
\end{equation}
	\subsubsection{Mutual distance between the endmembers}
The distance between the different endmembers can be expressed as follows 
\begin{equation}
\begin{split}
\Psi(\mathbf{M}) & = \frac{1}{2}\sum_{i = 1}^\nendm  \Biggl( \underset{j \neq i}{\sum_{j = 1}^\nendm } \norm_2{\mathbf{m}_i - \mathbf{m}_j} \Biggr) \\
& = \frac{1}{2} \sum_{k = 1}^\nendm  \Norm_fro{\mathbf{MG}_k} = \frac{1}{2} \Norm_fro{\mathbf{MG}}
\end{split}
\end{equation}{}%
with
\begin{equation*}
\mathbf{G} = \biggl[ \begin{array}{c|c|c} %
 \mathbf{G}_1&\cdots &  \mathbf{G}_\nendm  %
 \end{array} \biggr] \in \R[\nendm ]{\nendm ^2}
 \end{equation*}{}%
and for $k = 1,\dotsc , \nendm $
%
\begin{equation*}
\mathbf{G}_k = - \mathbf{I}_K + \mathbf{e}_k \mathbf{1}_K^T
\end{equation*}
where $\mathbf{e}_k$ denotes the $k$th vector in the canonical basis of $\mathbb{R}^K$.
Hence%
\begin{equation}
\label{eq:penalization_dendm}
\psi (\row{m}_\ell ) = \frac{1}{2} \sum_{k=1}^\nendm  \norm_2{\row{m}_\ell  \mat{G}_k}.
\end{equation}

	
	\subsubsection{Volume and endmember positivity constraint}
The volume penalization is expressed using $\mathbf{T}$, hence the need to find a condition equivalent to the positivity of $\mathbf{M}$ (see \cite{Dobigeon2009}). We will first analyze the general expression of the volume penalization \wrt{} $\row{t}_k$, and then give a condition on $\mat{T}$ ensuring the positivity of $\mat{M}$ (respectively $\mat{M} + \mat{dM}_n$ when endmember variability is considered).

		\paragraph{Volume}

The determinant of a matrix $\mat{X} \in~\R[\nendm ]{\nendm }$ can be developed along its $i$th row yielding%
\begin{equation*}
\det (\mat{X}) = \sum_j (-1)^{i+j} x_{ij}\det(\mat{X_{ij}}) = \row{x}_i\mathbf{f}_i
\end{equation*}
with
\begin{equation*}
\mathbf{f}_i = \left[(-1)^{i+j} \det(\mat{X_{ij}})  \right]_{j=1}^\nendm  \in \mathbb{R}^\nendm .
\end{equation*}
Consequently, for $k = 1,\dotsc , \nendm -1$ 
\begin{equation*}
 \det %
\begin{pmatrix}
\mathbf{T} \\
\mat{1}_\nendm ^T
\end{pmatrix} =  %
\row{t}_k \mathbf{f}_k .
\end{equation*}{}%
Using previous developments
\begin{equation}
\label{eq:penalization_volume}
\psi ( \row{t}_k  ) = \frac{1}{2(\nendm -1)!^2} ( \row{t}_k  \mat{f}_k )^2.
\end{equation}
%

		\paragraph{Positivity constraint on \textbf{M}}
		Using the following notations
\begin{align*}
\mathbf{Y} & = \mathbf{UY}_{\text{proj.}} + \mathbf{\bar{Y}}_1, \quad %
&\mathbf{\bar{Y}}_1 &= [\bar{\mathbf{y}} | \dotsc| \bar{\mathbf{y}}] \in \R[\nband ]{\nbpix } \bigskip \\
\mathbf{M} & = \mathbf{UT} + \mathbf{\bar{Y}}_2, \quad %
&\mathbf{\bar{Y}}_2 &= [\bar{\mathbf{y}} | \dotsc| \bar{\mathbf{y}}] \in \R[\nband ]{\nendm }
\end{align*}{}%
one has
\begin{equation*}
m_{\ell r} = \sum_j u_{\ell j}t_{jr} + \bar{y}_\ell  = \sum_{j \neq k }u_{\ell j}t_{jr} + u_{\ell k}t_{kr} + \bar{y}_\ell .
\end{equation*}{}%
The positivity constraint for $m_{\ell r}$ can then be expressed as
\begin{equation*}
t_{kr} \geq -\frac{\bar{y}_\ell  + \sum_{j \neq k }u_{\ell j}t_{jr}}{u_{\ell k}}.
\end{equation*}
Introducing the two sets of integers
\begin{align*}
\mathcal{U}_k^+ & = \{\ell | u_{\ell k} > 0 \} \\
\mathcal{U}_k^- & = \{\ell | u_{\ell k} < 0 \}
\end{align*}
the previous equation implies that $t_{kr}~\in~[t_{kr}^- , t_{kr}^+]$, with 
\begin{eqnarray}
\label{eq:cT}
t_{kr}^- &=& \max_{\ell \in \mathcal{U}_k^+} \Biggl( -\frac{\bar{y}_\ell  + \sum_{j \neq k }u_{\ell j}t_{jr}}{u_{\ell k}} \Biggr) \bigskip \\
t_{kr}^+ &=& \min_{\ell \in \mathcal{U}_k^-} \Biggl(  -\frac{\bar{y}_\ell  + \sum_{j \neq k }u_{\ell j}t_{jr}}{u_{\ell k}} \Biggr).
\end{eqnarray}

	\subsubsection{Positivity constraint on \textbf{M} and \textbf{dM}}
This case differs from the previous one as the positivity constraint must be verified simultaneously by $\mat{M}$ and $\mat{M}_n = \mat{M} + \mat{dM}_n$. We will consequently derive a condition similar to \eqref{eq:cT}. Let $\mat{T}_n$ be the projection of $\mat{M}_n$ in the PCA subspace
\begin{equation*}
\mat{M}_n = \mat{UT}_n + \mat{\bar{Y}}_2.
\end{equation*}{}%
Since
\[
\mat{T}_n = \mat{V} \left( \mat{M}_n - \mat{\bar{Y}}_2 \right) = \left( \mat{T + dT}_n \right) + \underbrace{\mat{V}\mat{\bar{Y}}_2}_{\mathbf{Z}}
\]{}%
with
\[
\left\{\begin{array}{ll}
\mat{T} &= \mat{V} \left( \mat{M} - \mat{\bar{Y}}_2 \right) \\
\mat{dT}_n &= \mat{V} \left( \mat{dM}_n - \mat{\bar{Y}}_2 \right)
\end{array} \right.
\]
the positivity constraint can be written
\begin{eqnarray*}
m_{\ell r}^n \geq 0 &\Leftrightarrow & t_{kr}^n \geq -\frac{\bar{y}_\ell  + \sum_{j \neq k }u_{\ell j}t_{jr}^n}{u_{\ell k}} \\%
&\Leftrightarrow & t_{kr}^n \in \left[t_{kr}^{n-},t_{kr}^{n+}  \right]
\end{eqnarray*}{}
with
\begin{eqnarray}
\label{eq:cTn}
t_{kr}^n &=& t_{kr} + dt_{kr}^n + z_{kr} \\%
t_{kr}^{n-} &=& \max_{\ell \in \mathcal{U}_k^+} \Biggl( - \frac{\bar{y}_\ell  + \sum_{j \neq k }u_{\ell j} t_{jr}^n}{u_{\ell k}} \Biggr) \bigskip \\%
t_{kr}^{n+} &=& \min_{\ell \in \mathcal{U}_k^-} \Biggl( - \frac{\bar{y}_\ell  + \sum_{j \neq k }u_{\ell j} t_{jr}^n}{u_{\ell k}} \Biggr).
\end{eqnarray}{}%
We introduce the functions $g_k $ defined by
\begin{equation}
\begin{array}{cccc}
\label{constraintTvar}
g_k  : &\R[1]{\nendm }& \rightarrow &\R[2(\nbpix +1)]{\nendm } \\
	  &  \row{x}	 & \mapsto            & \begin{bmatrix}
	 \row{x} - \row{t}_k ^- \medskip\\
	 -\row{x} + \row{t}_k ^+ \bigskip \\
	 \mat{1}_\nbpix (\row{x} + \row{z}_k  ) +%
	   \begin{pmatrix}
	    \row{dt}_{1,k}  - \row{t}_{1,k}^- \\
	    \vdots \\
	    \row{dt}_{\nbpix,k}  - \row{t}_{\nbpix,k}^-
	   \end{pmatrix}\bigskip \\
	   -\mat{1}_\nbpix (\row{x} + \row{z}_k  ) +%
	   \begin{pmatrix}
	    \row{t}_{1,k}^+ - \row{dt}_{1,k}  \\
	    \vdots \\
	    \row{t}_{\nbpix,k}^+ - \row{dt}_{\nbpix,k} 
	   \end{pmatrix}
	 \end{bmatrix},
\end{array}
\end{equation}
where
\begin{align*}
\row{t}_k^+ &= [t_{k1}^+, \cdots , t_{k\nendm}^+] \\
\row{t}_k^- &= [t_{k1}^-, \cdots , t_{k\nendm}^-].
\end{align*}
%
Finally the positivity constraint on the sum of the endmembers and their variability can be written
\begin{align}
& \row{m}_\ell  + \row{dm}_\ell ^n \succeq \mat{0}_\nendm ^T \quad \forall \ell , \quad \forall n \\ %
& \Leftrightarrow g_k(\row{t}_k) \succeq \Zero[2(\nbpix +1)]{\nendm} \quad \forall k = 1,\dotsc ,\nendm -1.
\end{align}{}%
\subsection{Variability penalization}

The variability energy penalty is%
\begin{equation}
\label{eq:penalization_dM}
\Upsilon \left(\mat{dM} \right) =\frac{1}{2} \Norm_fro{\mat{dM}} \Rightarrow %
\upsilon  \left( \mat{dM}_n \right) = \frac{1}{2} \Norm_fro{\mat{dM}_n}. 
\end{equation}

\section{Solutions to the optimization sub-problems} \label{app:resolution}
\subsection{Resolution \wrt{} A}

Using \eqref{eq:penalizationA}, the scaled augmented Lagrangian \eqref{eq:lagrangian_A} becomes
\begin{equation*}
\begin{split}
\mathcal{L}_{\muA}& \Bigl(\mat{a}_n,\splitA,\lagA \Bigr) = %
 \frac{1}{2}   \norm_2{\mathbf{y}_n - (\mathbf{M+dM}_n) \mathbf{a}_n} \\ %
 & +\frac{\alpha}{2} \left(\csteA \norm_2{\mathbf{a}_n} + 2 \mathbf{c}_n^T \mathbf{a}_n \right)  + \mathcal{I}_{\mathcal{S}_\nendm^+} \Bigl(\splitA \Bigr) \\
 & + \frac{\muA}{2}\norm_2{\mat{Qa}_n %
 + \mat{R}\splitA-\mat{s}+\lagA}.
 \end{split}
\end{equation*}
Thus, for $n = 1,\dotsc , \nbpix $
\begin{equation}
 \begin{split}
& \mat{a}_n^* = \left[ (\mat{M+dM}_n)^T(\mat{M+dM}_n) + \muA \mat{Q}^T\mat{Q} + \alpha \csteA\mathbf{I}_\nendm \right ]^{-1} \\ 
& \biggl[ (\mat{M+dM}_n)^T \mat{y}_n - \alpha \mathbf{c}_n + \muA \mat{Q}^T \left( \mat{s} - \mat{R}\splitA - \lagA \right) \biggr]
\end{split}
\end{equation}{}%
and
%
\begin{equation}
{\splitA}^* = \max \left( \mat{a}_n + \boldsymbol{\lambda}_{n,1:\nendm}^{\mathbf{(A)}}, \mathbf{0}_\nendm \right)
\end{equation}
where $\boldsymbol{\lambda}_{n,1:\nendm}^{\mathbf{(A)}}$ is the vector composed of the $\nendm$ first elements of $\lagA$ and the $\max$ must be understood as a term-wise operator. In the absence of any penalization, the solution is obtained by making $\alpha = 0$ in the previous equations.		

\subsection{Resolution \wrt{} M}

	\subsubsection{Distance between the endmembers and reference spectral signatures}
Using \eqref{eq:penalization_vca}, the scaled augmented Lagrangian \eqref{eq:lagrangian_M} is
\begin{equation*}
\begin{split}
\mathcal{L}_{\muM} & \left(\row{m}_\ell ,\splitM  ,\lagM  \right) = %
 \frac{1}{2} \norm_2{\row{y}_\ell  - \row{m}_\ell \mathbf{A} - \row{\boldsymbol{\delta}}_\ell } \\%
& + \frac{\muM}{2} \Norm_fro{\mathbf{e}\row{m}_\ell  - \splitM  + \mat{F}_\ell + \lagM } \\ %
& + \frac{\beta}{2} \norm_2{\row{m}_\ell - \row{m}_{\ell,0}}
  + \mathcal{I}_{\mathcal{S}^+}\left(\splitM   \right).
\end{split}
\end{equation*}%
Thus
\begin{equation}
\begin{split}
\row{m}_\ell ^* =&  \biggl[ \left(\row{y}_\ell  - \row{\boldsymbol{\delta}}_\ell   \right) \mat{A}^T + \beta \row{m}_{\ell,0} + \biggr. \\
& \biggl.\muM \mat{e}^T \left( \splitM   - \mat{F}_\ell - \lagM \right) \biggr] \\%
& \left[ \mat{AA}^T + \muM \left(\mat{e}^T\mat{e} + \beta \right) \mat{I}_\nendm \right]^{-1}
\end{split}
\end{equation}%
and for $k = 1,\dotsc ,\nendm $
%
\begin{equation}
\label{eq:wl}
\mat{w}_{\ell,k}^{(\mathbf{M})*} = \max \left( [ \mat{e} \row{m}_\ell   + \mat{F}_\ell + \lagM ]_k, \mathbf{0}_{\nbpix +1} \right).
\end{equation}%
In the absence of any endmember penalization, the solution is obtained by making $\beta = 0$ in the previous equation.
	\subsubsection{Mutual distance between the endmembers}

Using \eqref{eq:penalization_dendm}, the scaled augmented Lagrangian \eqref{eq:lagrangian_M} is
\begin{equation*}
\begin{split}
\mathcal{L}_{\muM} & \left(\row{m}_\ell ,\splitM  ,\lagM  \right) = %
 \frac{1}{2} \norm_2{\row{y}_\ell  - \row{m}_\ell \mathbf{A} - \row{\boldsymbol{\delta}}_\ell } \\%
& + \frac{\muM}{2} \Norm_fro{\mathbf{e}\row{m}_\ell  - \splitM  + \mat{F}_\ell + \lagM } \\ %
& + \frac{\beta}{2} \sum_{k=1}^\nendm \norm_2{\row{m}_\ell  \mat{G}_k}
  + \mathcal{I}_{\mathcal{S}^+}\left(\splitM   \right).
\end{split}
\end{equation*}%
Thus
\begin{equation}
\begin{split}
\row{m}_\ell ^* =&  \biggl[ \left(\row{y}_\ell  - \row{\boldsymbol{\delta}}_\ell   \right) \mat{A}^T +%
\muM \mat{e}^T \left( \splitM   - \mat{F}_\ell - \lagM \right) \biggr] \\%
& \left[ \mat{AA}^T + \beta \sum_{k=1}^\nendm \mat{G}_k \mat{G}_k^T  + \muM \left(\mat{e}^T\mat{e} \right) \mat{I}_\nendm \right]^{-1}
\end{split}
\end{equation}%
with $\splitM$ given by \eqref{eq:wl}.

	\subsubsection{Volume penalization}
Since the penalty is expressed \wrt{} the variable $\mathbf{T}$, the optimization sub-problems related to the endmembers have to be re-written accordingly. Using the notations
\begin{align}
\mathbf{\Delta}_T &= \left[ \begin{array}{c|c|c} %
 \mathbf{dT}_1 \mathbf{a}_1 & \dotsc & \mathbf{dT}_\nbpix  \mathbf{a}_\nbpix  %
 \end{array} \right] \\
\mathbf{\bar{Y}}_1 &= [\bar{\mathbf{y}} | \cdots| \bar{\mathbf{y}}] \in \R[\nband]{\nbpix } \\
\mathbf{\bar{Y}}_2 &= [\bar{\mathbf{y}} | \cdots| \bar{\mathbf{y}}] \in \R[\nband]{\nendm}
\end{align}
we obtain
\begin{align*}
\Norm_fro{\mat{Y} - \mat{MA-\Delta}} &= %
\Norm_fro{\mat{U}\left(\mat{Y}_{\text{proj.}} - \mat{TA-\Delta}_T \right) + \mat{\bar{Y}}_1 -  2\mat{\bar{Y}}_2\mat{A}}\\
&= \Norm_fro{\mat{U}\left(\mat{Y}_{\text{proj.}} - \mat{TA-\Delta}_T \right)} \\
&+  2 \left\langle \mat{U}\left(\mat{Y}_{\text{proj.}} - \mat{TA-\Delta}_T \right) \biggl| \mat{\bar{Y}}_1 -  2\mat{\bar{Y}}_2\mat{A} \right\rangle \biggr. \\ 
&   + \Norm_fro{\mat{\bar{Y}}_1 - 2 \mat{\bar{Y}}_2\mat{A}}.
\end{align*}{}%
The only terms depending on $\mat{T}$ are 
\begin{equation*}
 \Norm_fro{\mat{Y}_{\text{proj.}} - \mat{TA}} + 2 \biggl\langle \underbrace{ 
\mathbf{\Delta}_T -\mat{U}^T(\mat{\bar{Y}}_1 -  2\mat{\bar{Y}}_2\mat{A})}_{\mathbf{S}} \biggl| \mat{TA} \biggr\rangle \biggr.
\end{equation*}
with
\begin{equation*}
\left\langle \mathbf{S} \middle| \mat{TA} \right\rangle = \Tr{\mat{S}^T\mat{TA}} = %
\sum_{n=1}^\nbpix  \biggl( \sum_{j=1}^{\nendm-1} s_{jn} \row{t}_j\mathbf{a}_n \biggr).
\end{equation*}%
For $k = 1,\dotsc ,\nendm -1$, the resulting sub-problems are
\begin{equation*}
\row{t}_k ^* = %
\argmin{\row{t}_k }\left\{ \begin{array}{c}
\frac{1}{2} \norm_2{\row{y}_k ^{\text{proj.}} - \row{t}_k  \mathbf{A}} %
+ \sum_{n=1}^\nbpix  ( s_{kn} \row{t}_k \mathbf{a}_n ) \medskip \\
+ \frac{\beta}{2(\nendm-1)!^2} ( \row{t}_k \mathbf{f}_k  )^2 \medskip\\ %
\text{s.t.} \quad g_k (\row{t}_k ) \succeq \Zero[2(\nbpix +1)]{\nendm}
\end{array}  \right\}.
\end{equation*}
Introduce the splitting variables $\splitT$ such that
\begin{equation}
g_k (\row{t}_k ) = \splitT \quad \forall k = 1,\dotsc ,\nendm-1.
\end{equation}
According to \eqref{eq:penalization_volume}, the scaled augmented Lagrangian is
\begin{equation}
\begin{split}
\mathcal{L}_{\muT}&(\row{t}_k ,\splitT  ,\lagT) = \\ %
&\frac{1}{2} \norm_2{\row{y}_k ^{\text{proj.}} - \row{t}_k  \mathbf{A}} + %
\sum_{n=1}^\nbpix  \Bigl( s_{nk} \row{t}_k \mathbf{a}_n \Bigr)  \\ %
& + \frac{\beta}{2(\nendm-1)!^2} \Bigl( \row{t}_k \mathbf{f}_k  \Bigr)^2
+ \mathcal{I}_{\mathcal{S}^+}\left(\splitT  \right) \\ %
& + \frac{\muT}{2} \Norm_fro{g_k (\row{t}_k ) - \splitT   + \lagT}.
\end{split}
\end{equation}
Finally,
\begin{equation}
\begin{split}
\row{t}_k ^* = & \biggl[ \left(\row{y}_k ^{\text{proj.}} - \row{s}_k  \right)\mat{A}^T - 2\muT \left( \nbpix \row{z}_k  + \sum_{n=1}^\nbpix  \row{dt}_{n,k}  \right) \biggr. \\
& \left. + \muT \left( \row{t}_k ^- + \row{t}_k ^+ + \sum_{n=1}^\nbpix  (\row{t}_{n,k}^- + \row{t}_{n,k}^+)\right) \right. \\
& \biggl. +  \muT \begin{bmatrix}
1 & -1 & \mat{1}_\nbpix ^T & -\mat{1}_\nbpix ^T
\end{bmatrix} \Bigl( \splitT   - \lagT \Bigr)\biggr] \\ %
& \left[ \mathbf{AA}^T + \frac{\beta}{(\nendm-1)!^2}\mathbf{f}_k  \mathbf{f}_k^T + 2(\nbpix +1) \muT \mathbf{I}_\nendm \right]^{-1}
\end{split}
\end{equation}%
where $\mathbf{Z} = \mathbf{V}\bar{\mathbf{Y}}_2$ and for $p = 1,\dotsc ,\nendm$
%
\begin{equation}
\mat{w}_{k,p}^{(\mathbf{T})*} = \max \left( \left[ g_k (\row{t}_k ) + \lagT \right]_p, \mathbf{0}_{2(\nbpix +1)} \right).
\end{equation}

\subsection{Resolution with respect to dM}
Using \eqref{eq:penalization_dM}, the scaled augmented Lagrangian \eqref{eq:lagrangian_dM} is
\begin{equation*}
\begin{split}
\mathcal{L}_{\mudM}& \Bigl( \mathbf{dM}_n,\splitdM, \mathbf{\Lambda_{dM_n}} \Bigr) = \\ %
& \frac{1}{2}\norm_2{\mathbf{y}_n -  (\mathbf{M}+\mathbf{dM}_n)\mathbf{a_n}} \\
& + \frac{\mudM}{2} \Norm_fro{\mathbf{dM}_n + \mathbf{M} - \splitdM + \lagdM} \\
& + \frac{\gamma}{2} \Norm_fro{\mathbf{dM}_n} + \mathcal{I}_{\mathcal{S}^+} \Bigl( \splitdM \Bigr).
\end{split}
\end{equation*}
Hence
\begin{equation}
\begin{split}
\mathbf{dM}_n ^* = %
&\biggl[ (\mathbf{y_n} - \mathbf{M}\mathbf{a}_n)\mathbf{a}_n^T + %
\mudM \Bigl( \splitdM - \mathbf{M} - \lagdM \Bigr) \biggr] \\%
&\Bigl[ \mathbf{a}_n\mathbf{a}_n^T + (\mudM + \gamma)\mathbf{I}_\nendm  \Bigr]^{-1}
\end{split}
\end{equation}
and for $p = 1,\dotsc ,\nendm$
%
\begin{equation*}
\mathbf{w}_{n,p}^{(\mathbf{dM})*} = \max \left( [\mathbf{dM}_n + \mathbf{M} + \lagdM ]_p, \mathbf{0}_\nendm \right). 
\end{equation*}
\end{appendices}
\begin{figure}[h!]
\centering
\subfloat[1][Sphene]{
\includegraphics[keepaspectratio,height=0.3\textheight , width=0.13\textwidth]{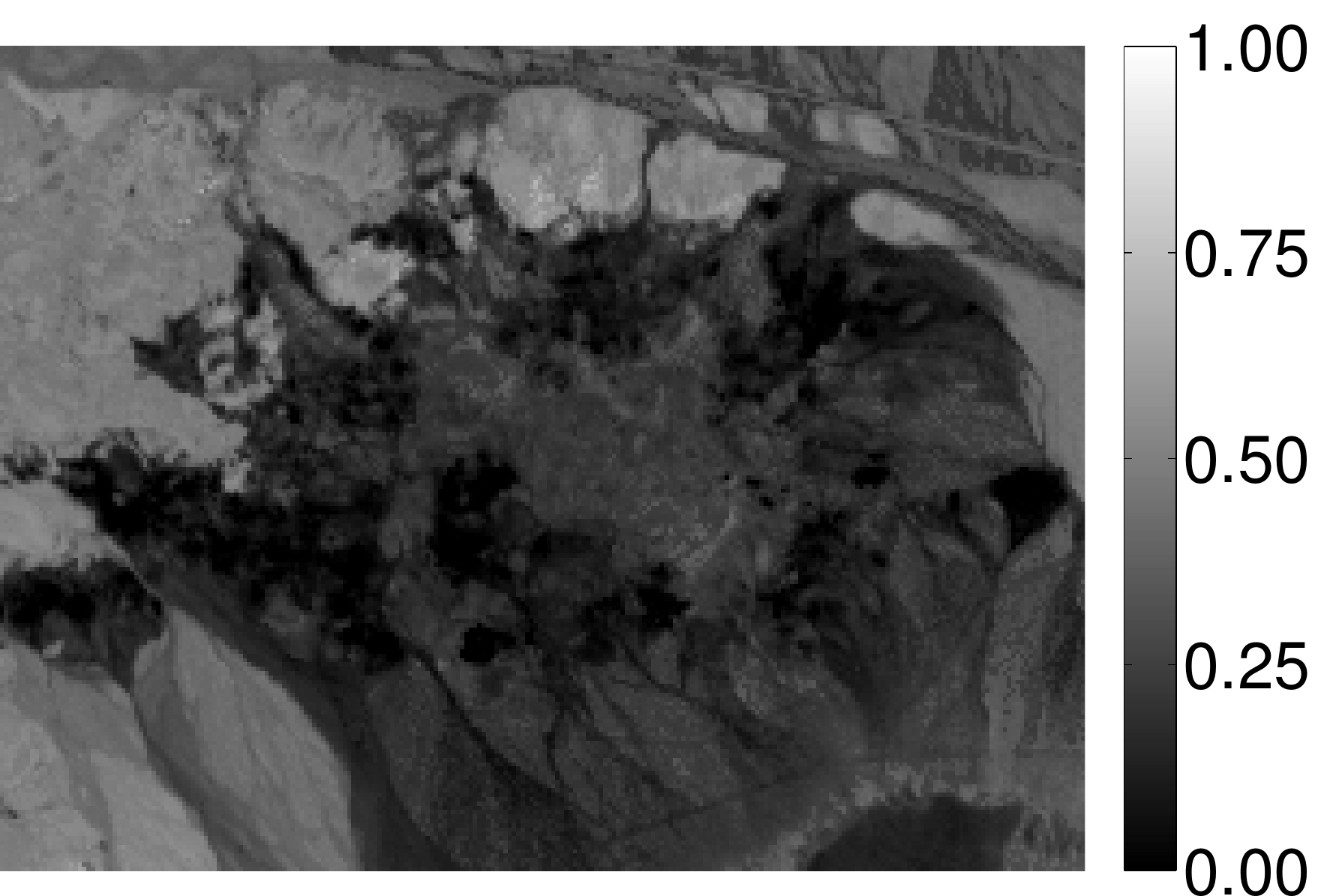}
\label{fig:cuprite_abundance1}}
\subfloat[2][Alunite]{
\includegraphics[keepaspectratio,height=0.3\textheight , width=0.13\textwidth]{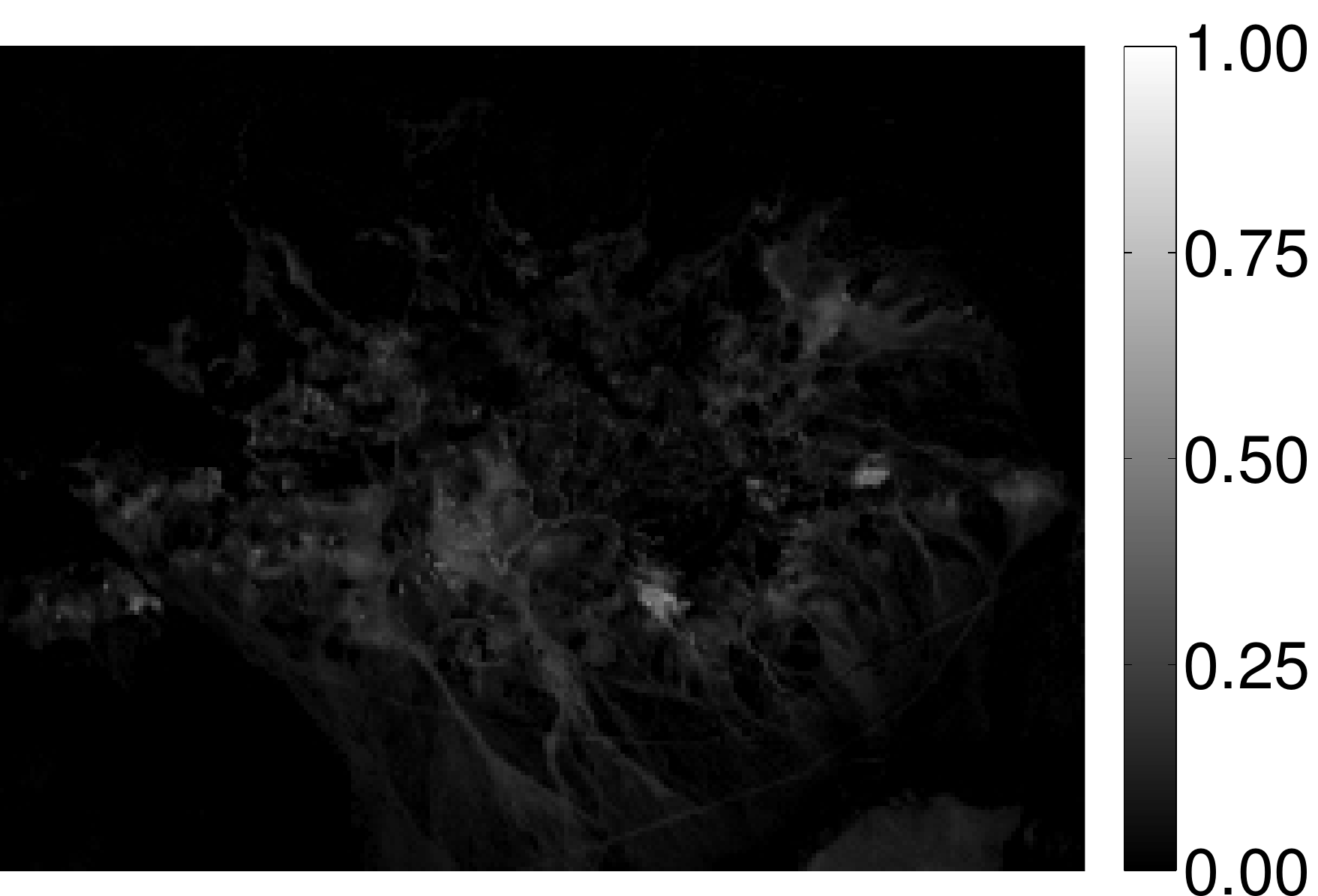}
\label{fig:cuprite_abundance2}}
\subfloat[3][Dumortierite]{
\includegraphics[keepaspectratio,height=0.3\textheight , width=0.13\textwidth]{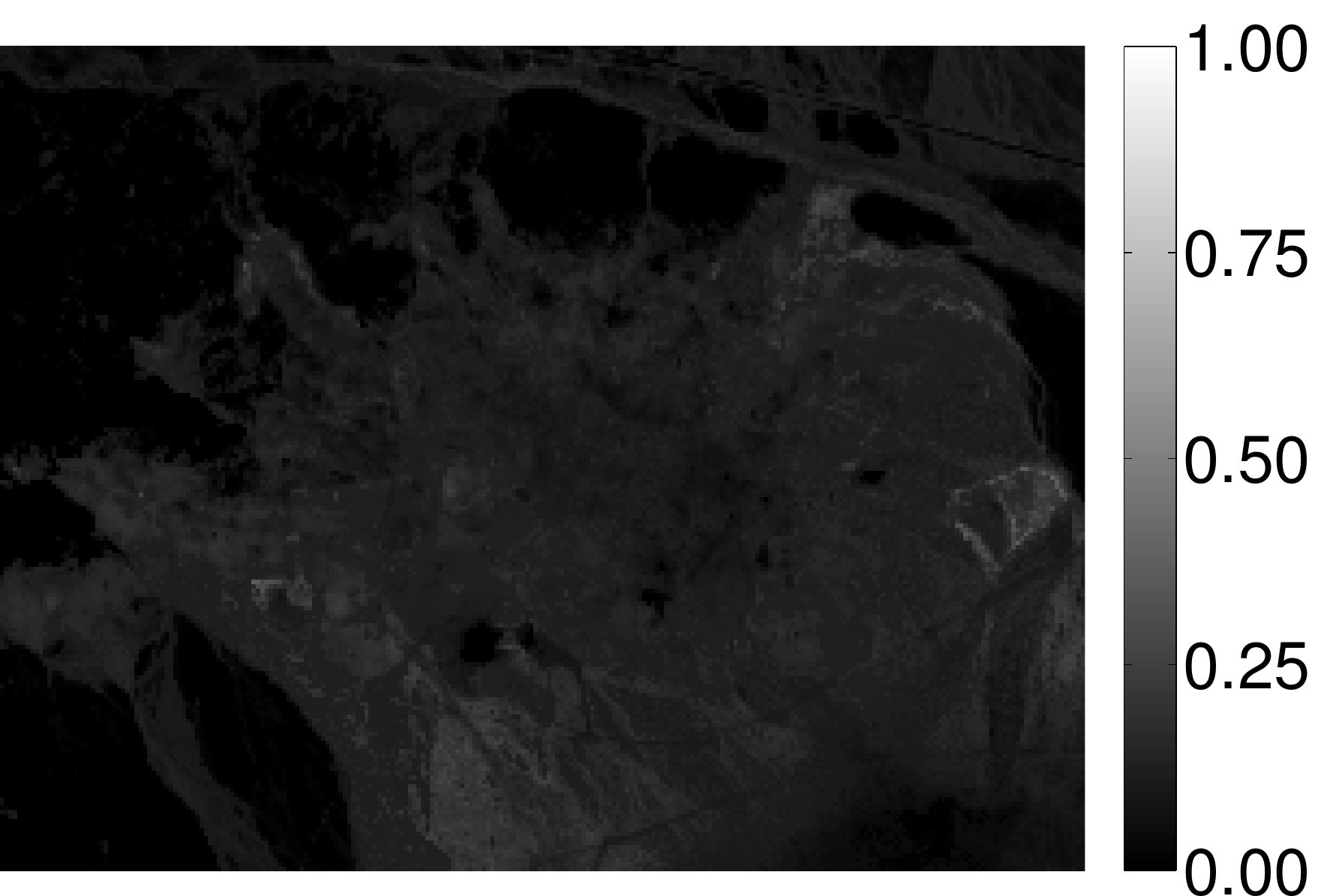}
\label{fig:cuprite_abundance3}}
\\
\subfloat[4][Montmorillonite]{
\includegraphics[keepaspectratio,height=0.3\textheight , width=0.13\textwidth]{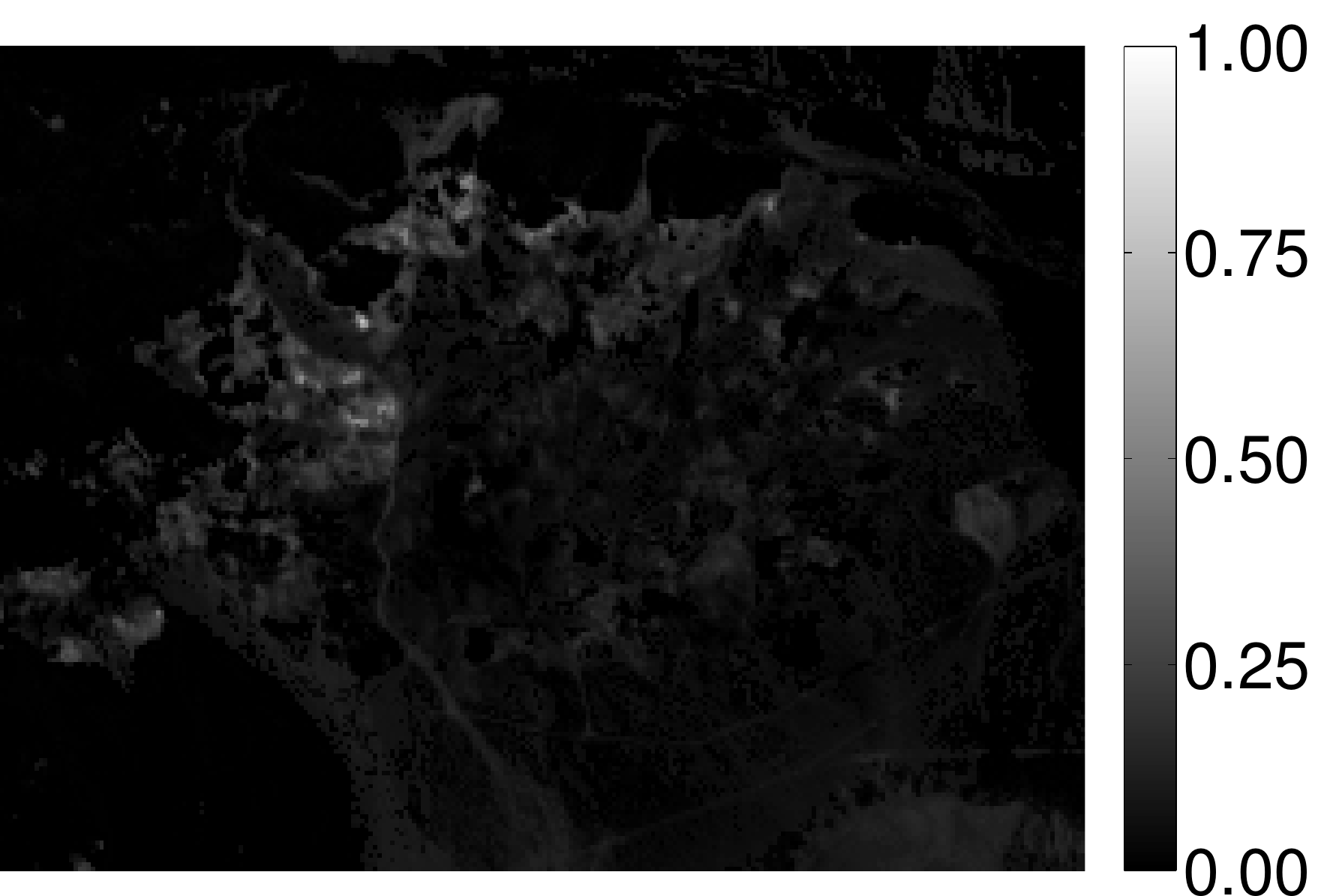}
\label{fig:cuprite_abundance4}}
\subfloat[5][Andradite]{
\includegraphics[keepaspectratio,height=0.3\textheight , width=0.13\textwidth]{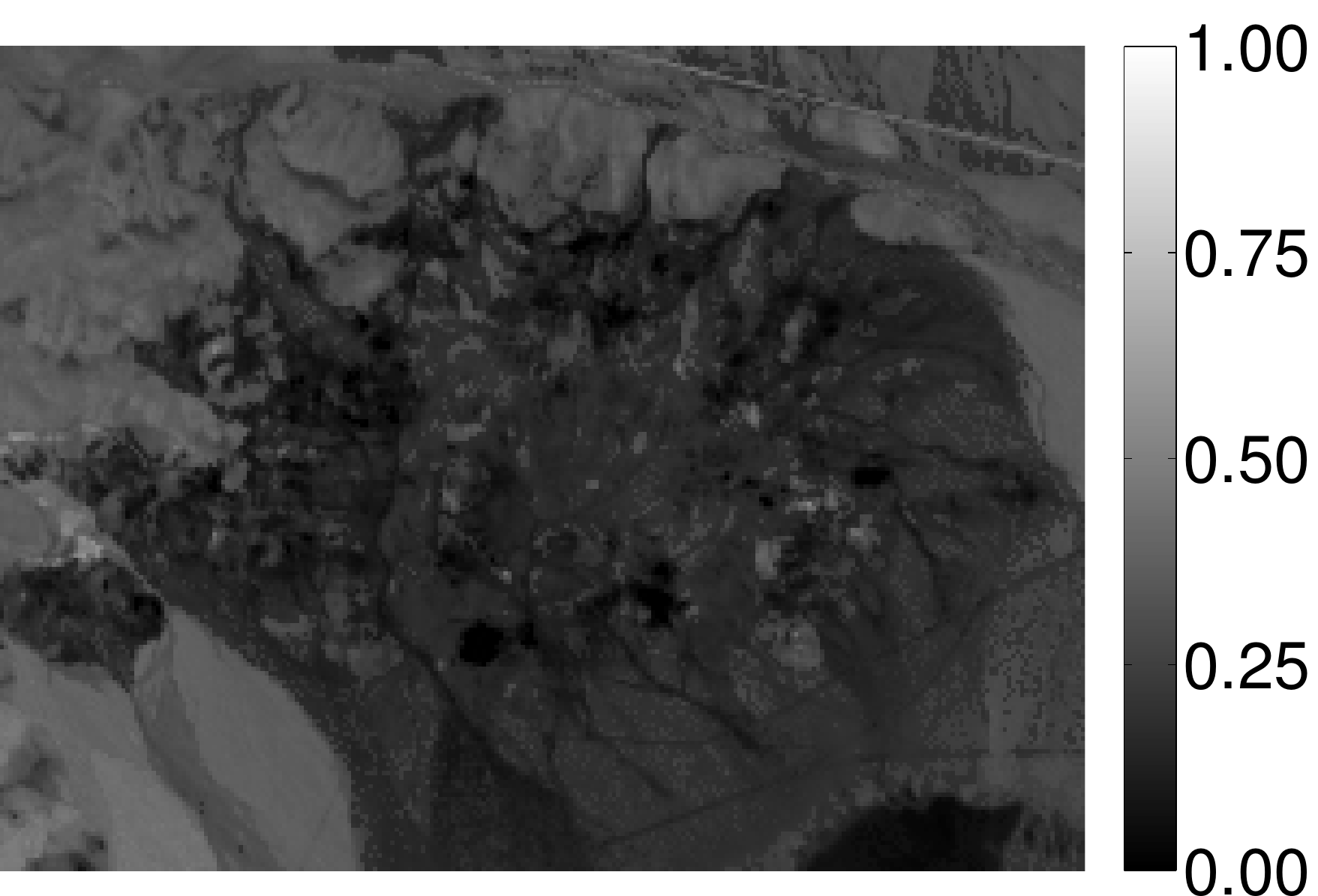}
\label{fig:cuprite_abundance5}}
\subfloat[6][Pyrope]{
\includegraphics[keepaspectratio,height=0.3\textheight , width=0.13\textwidth]{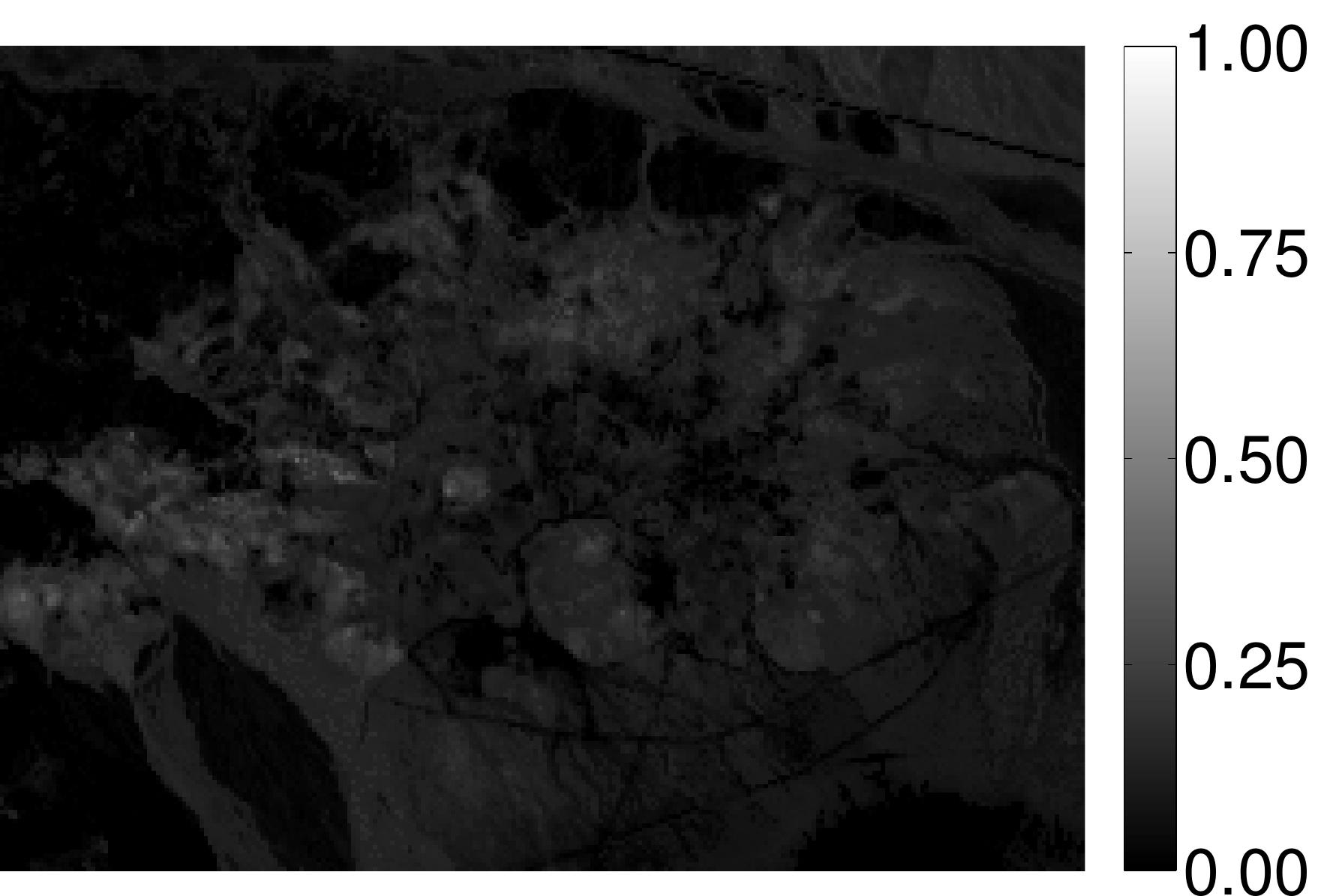}
\label{fig:cuprite_abundance6}}
\\
\subfloat[7][Buddingtonite]{
\includegraphics[keepaspectratio,height=0.3\textheight , width=0.13\textwidth]{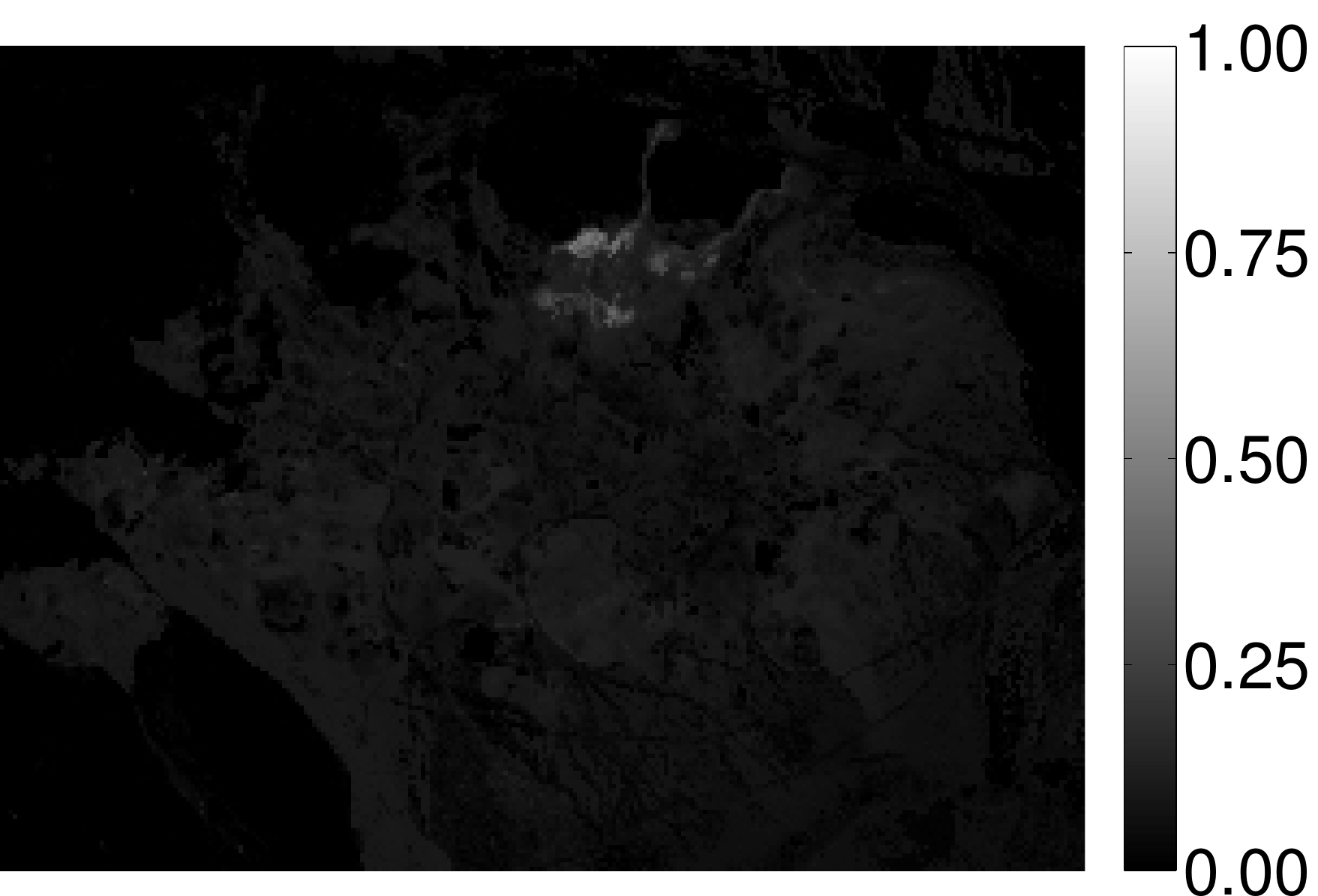}
\label{fig:cuprite_abundance7}}
\subfloat[8][Muscovite]{
\includegraphics[keepaspectratio,height=0.3\textheight , width=0.13\textwidth]{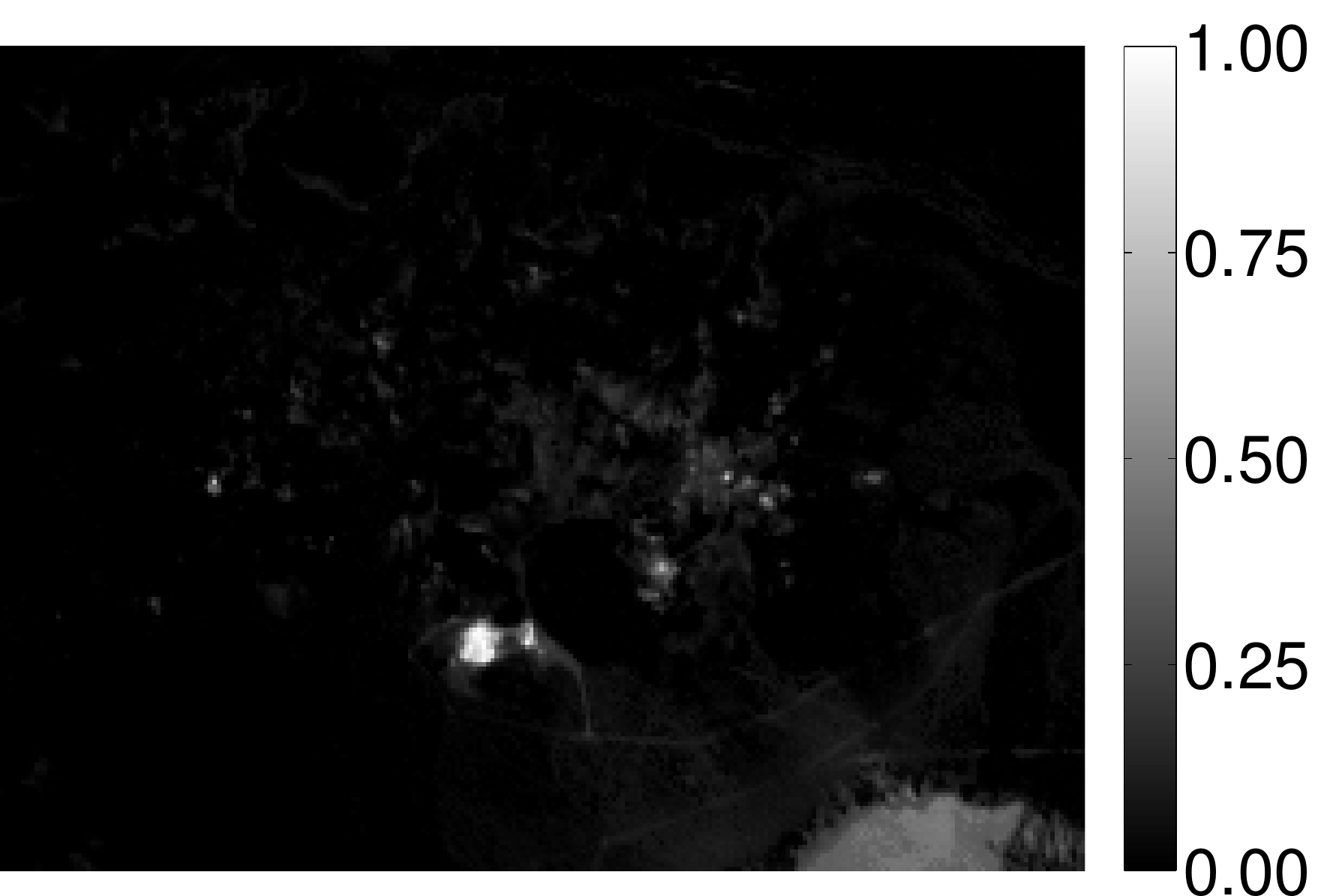}
\label{fig:cuprite_abundance8}}
\subfloat[9][Nontronite]{
\includegraphics[keepaspectratio,height=0.3\textheight , width=0.13\textwidth]{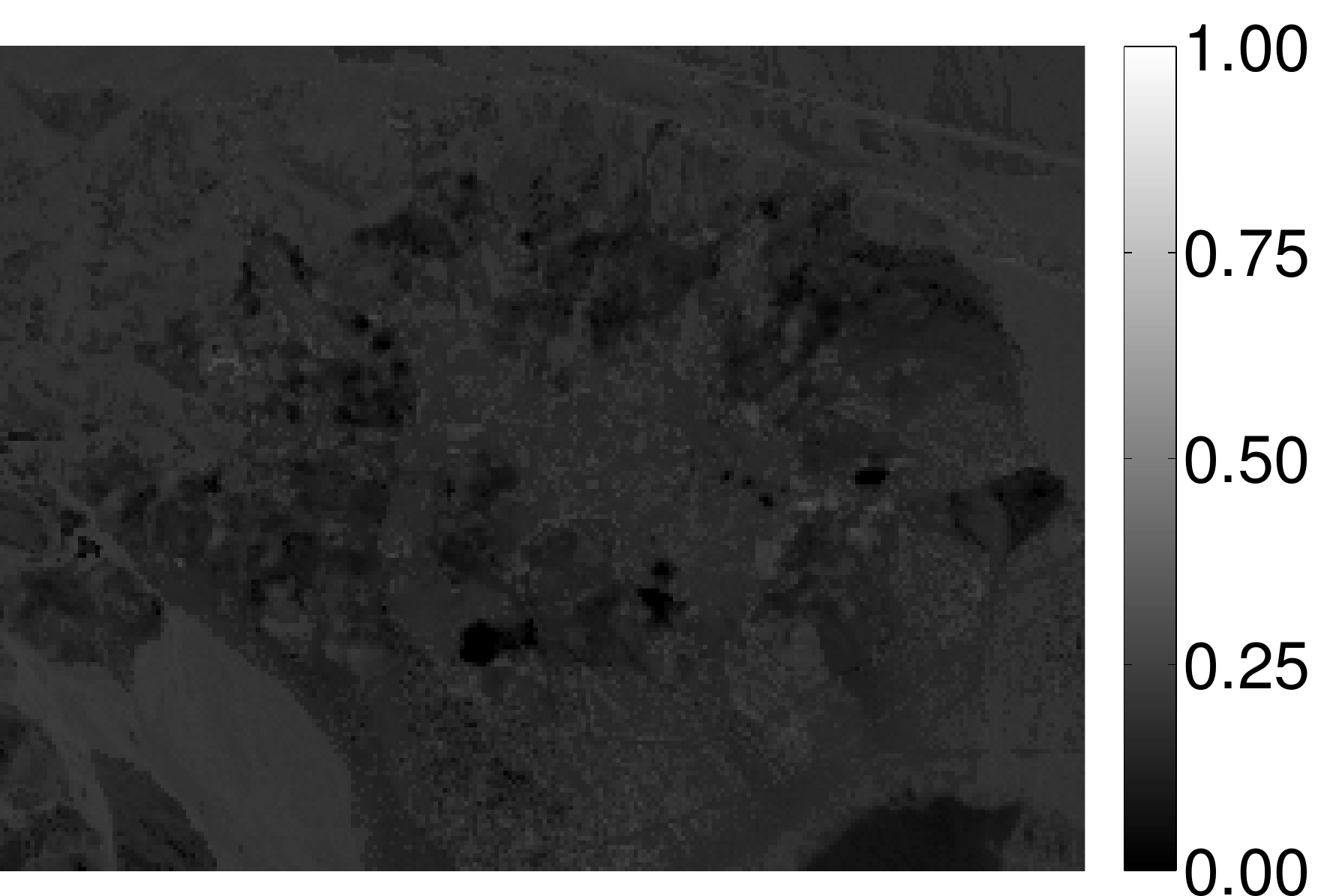}
\label{fig:cuprite_abundance9}}
\\
\subfloat[10][Kaolinite]{
\includegraphics[keepaspectratio,height=0.3\textheight , width=0.13\textwidth]{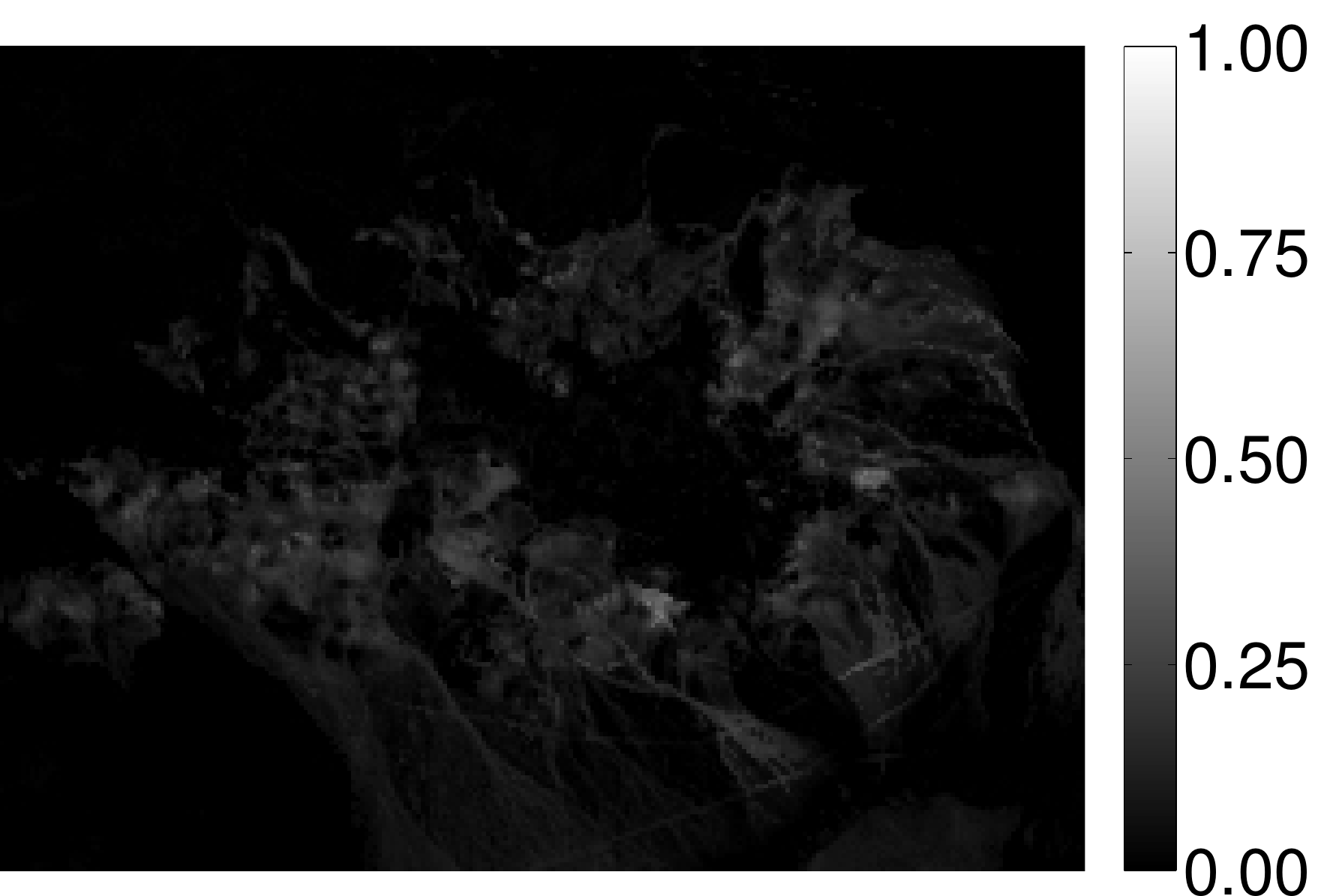}
\label{fig:cuprite_abundance10}}
\caption{ssvcaBCD/ADMM abundance results (Cuprite scene). The given identification is based on a visual comparison with the results obtained in \cite{Nascimento2005}.}
\label{fig:cuprite}
\end{figure}

\begin{figure}[h!]
\centering
\subfloat[1][Sphene]{
\includegraphics[keepaspectratio,height=0.3\textheight , width=0.13\textwidth]{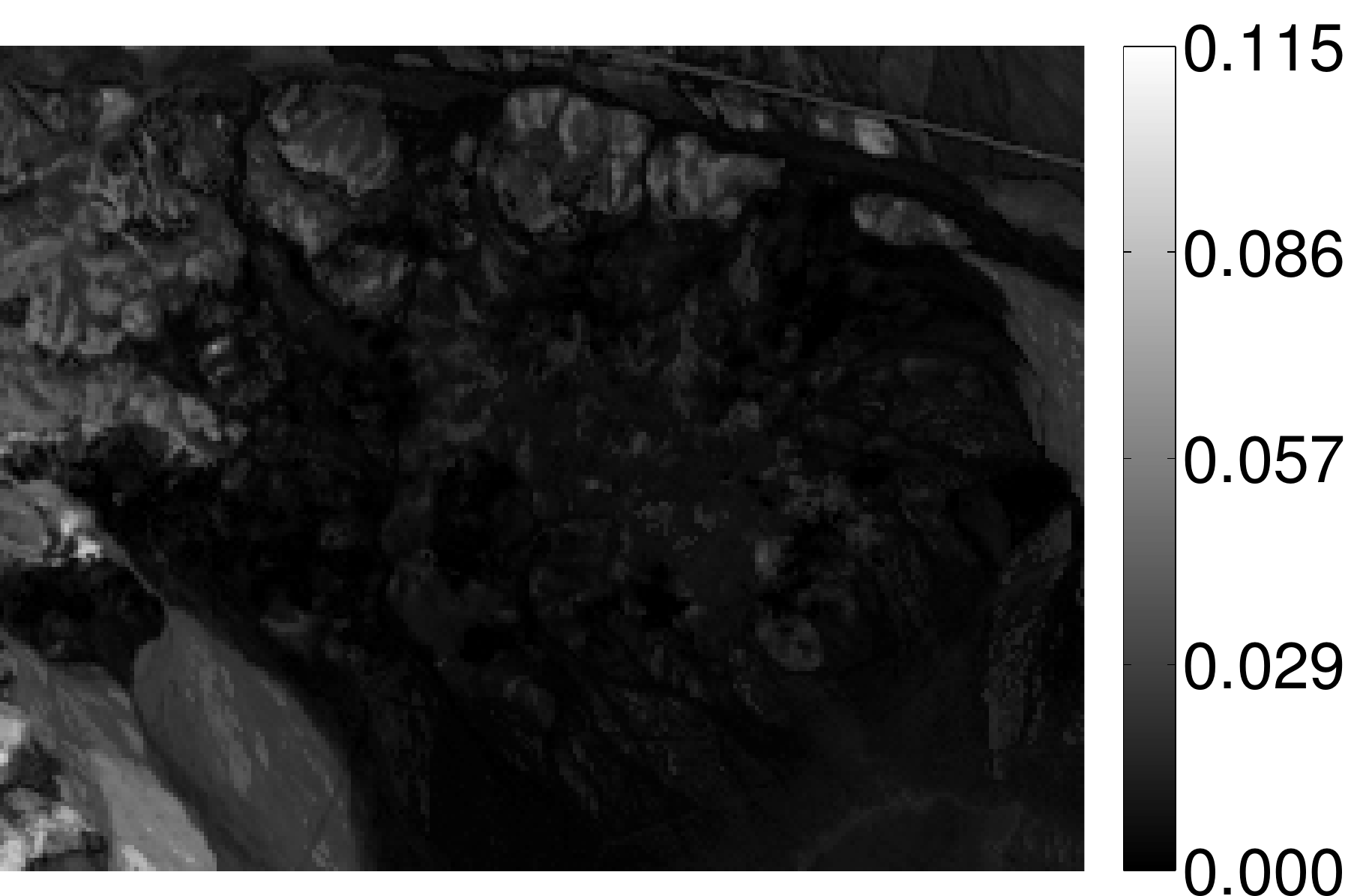}
\label{fig:cuprite_var1}}
\subfloat[3][Andradite]{
\includegraphics[keepaspectratio,height=0.3\textheight , width=0.13\textwidth]{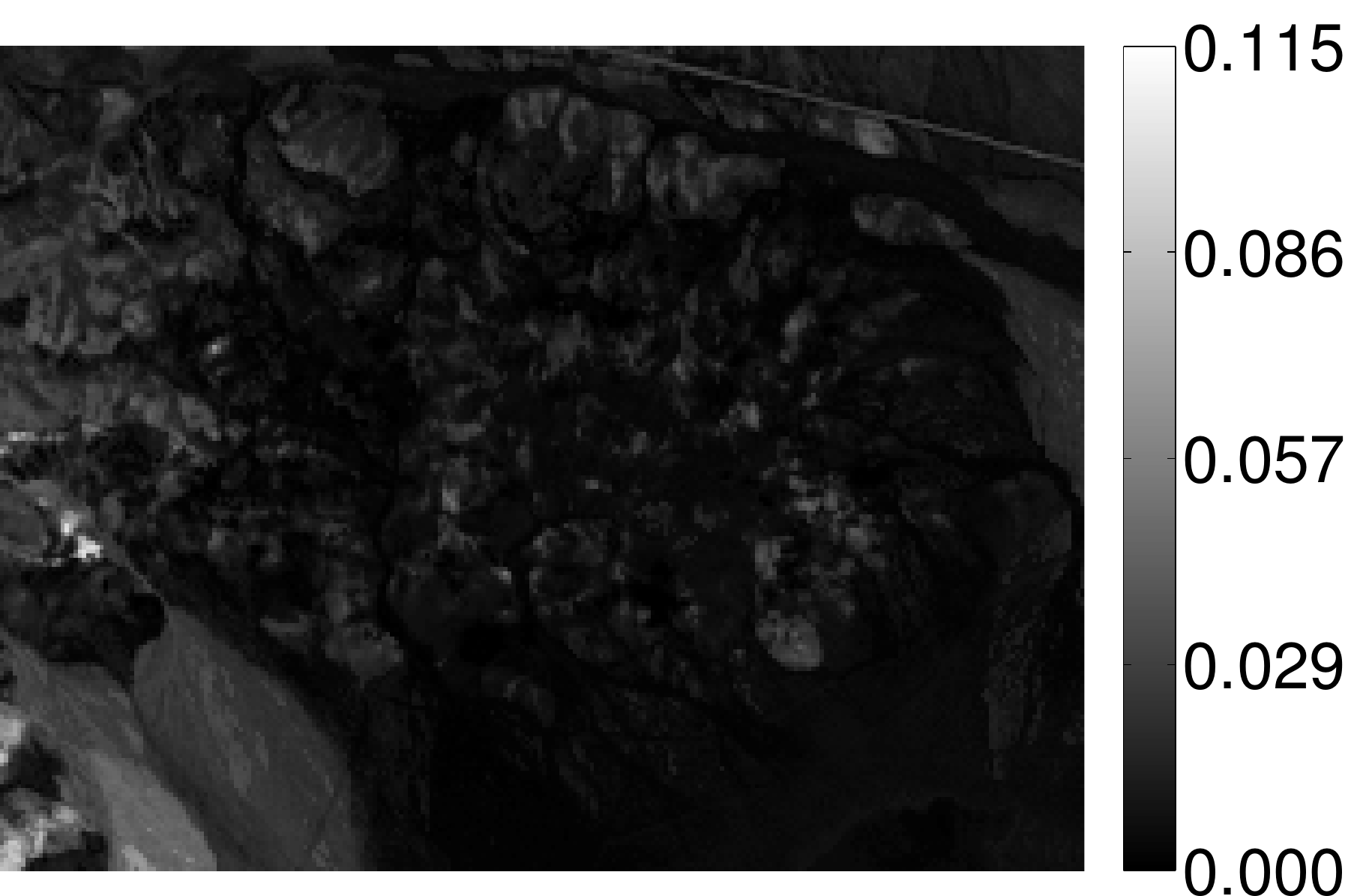}
\label{fig:cuprite_var5}}
\subfloat[6][Muscovite]{
\includegraphics[keepaspectratio,height=0.3\textheight , width=0.13\textwidth]{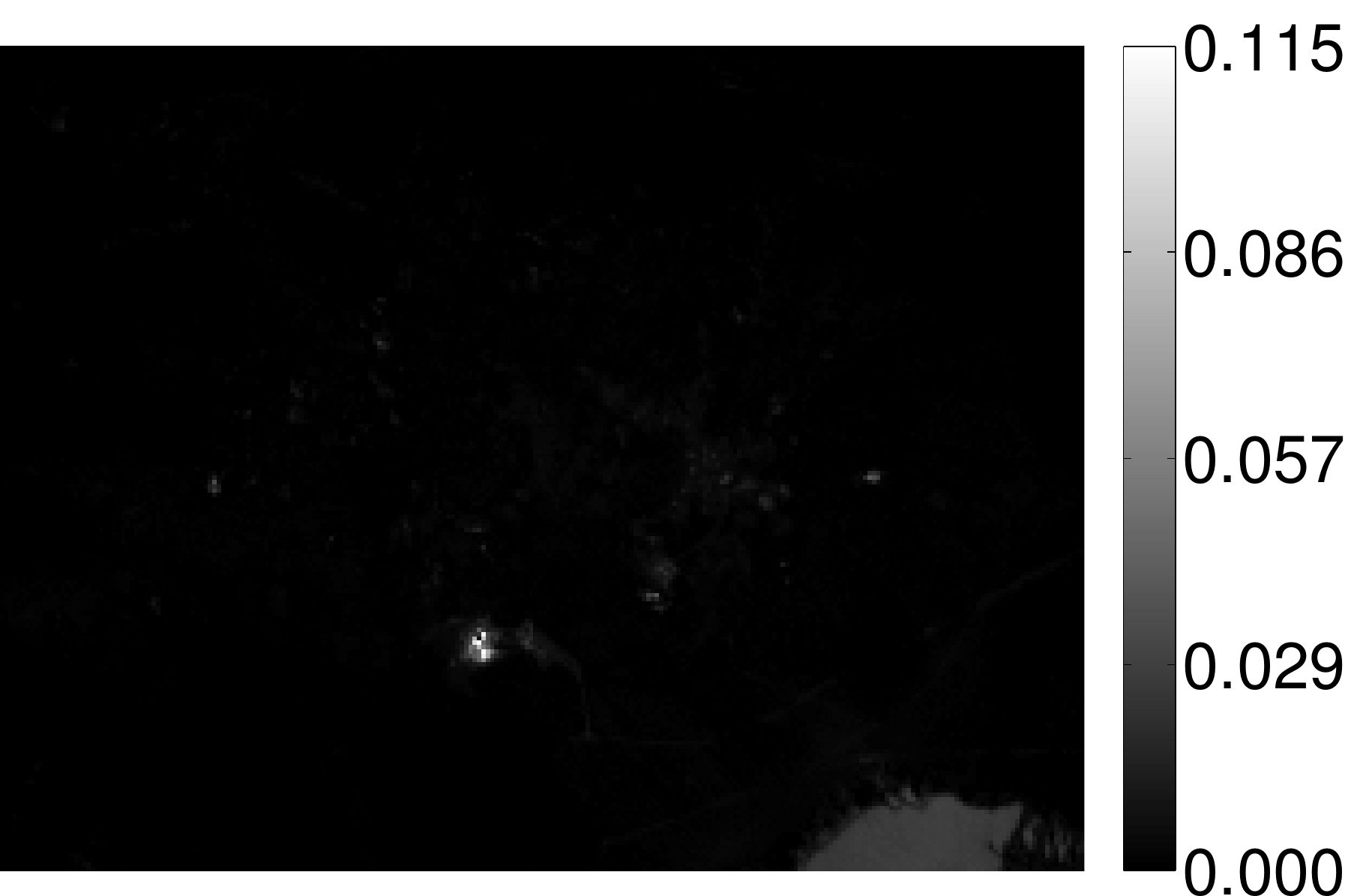}
\label{fig:cuprite_var8}}
\caption{Spatial distribution of the variability \wrt{} the endmembers presenting the most significant level of variability. The variability is presented in terms of energy ($\frac{1}{\sqrt{\nband}} \lVert \mathbf{dm}_{n,k} \rVert_{2}$ for the $k$th endmember in the $n$th pixel) for visualization purpose.}
\label{fig:cuprite_var}
\end{figure}

\begin{figure}[h!]
\centering
\subfloat[1][Sphene]{
\includegraphics[keepaspectratio,height=0.2\textheight , width=0.23\textwidth]{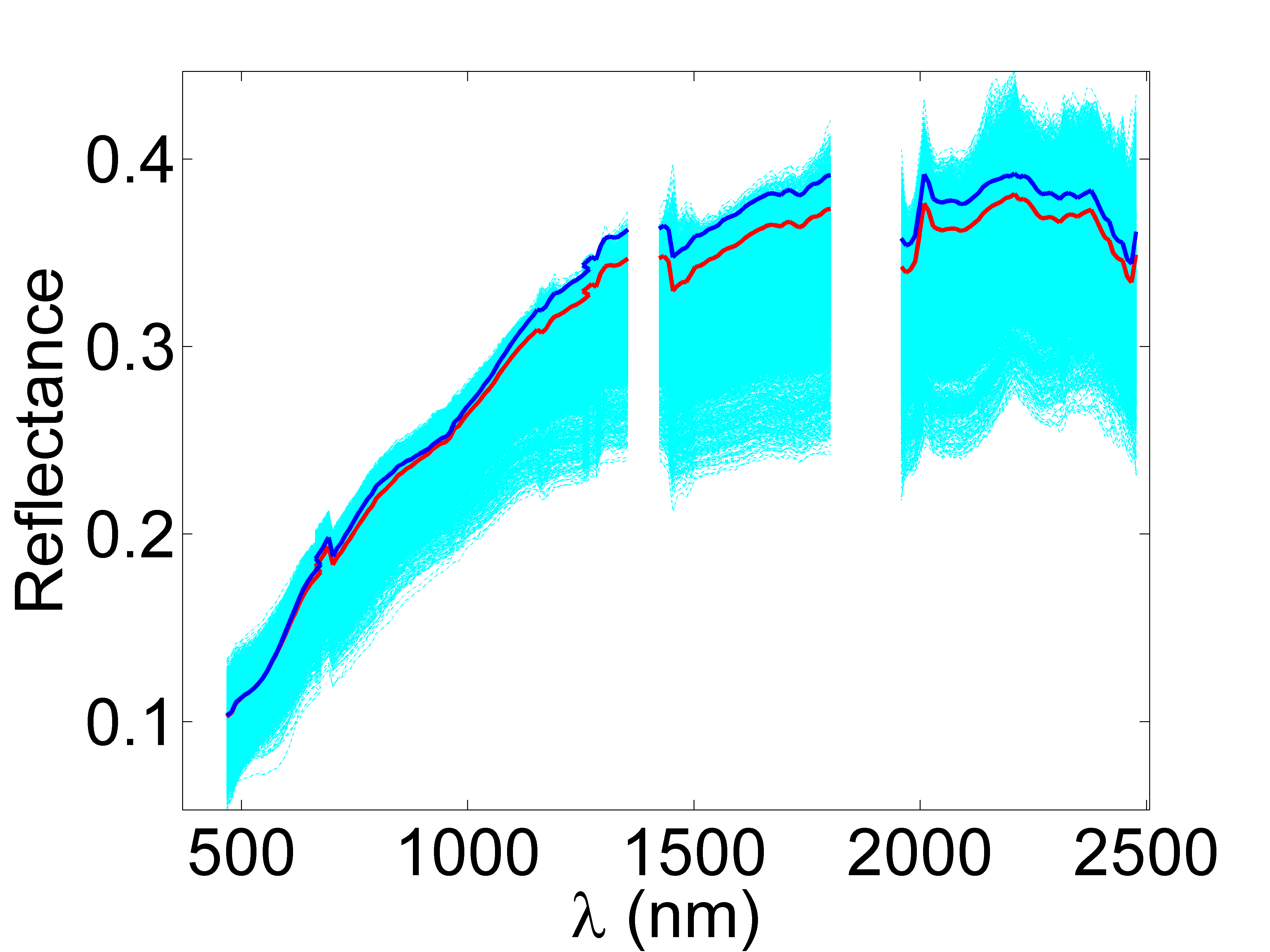}
\label{fig:cuprite_endm1}}
\subfloat[2][Alunite]{
\includegraphics[keepaspectratio,height=0.2\textheight , width=0.23\textwidth]{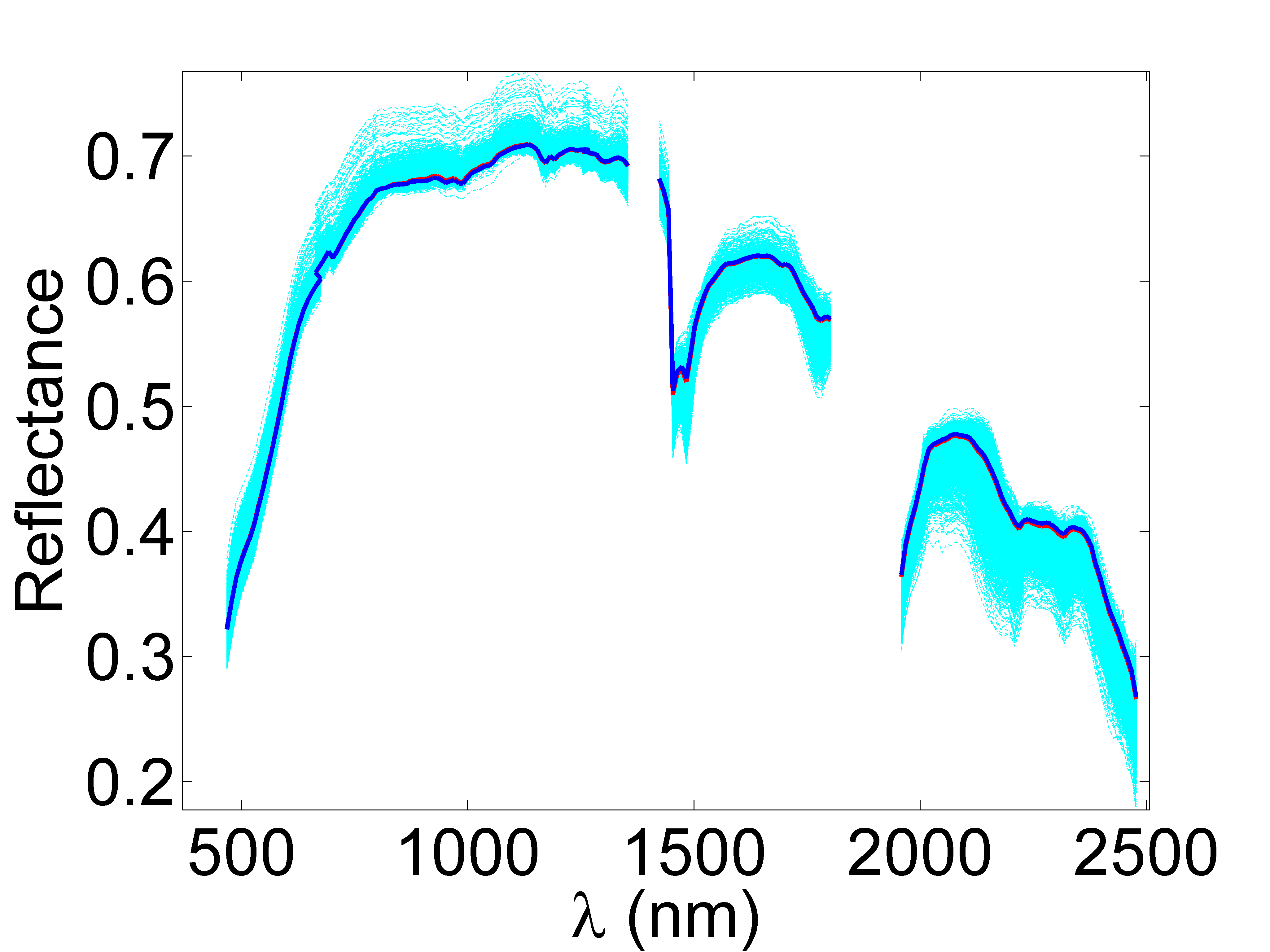}
\label{fig:cuprite_endm2}}
\\
\subfloat[3][Dumortierite]{
\includegraphics[keepaspectratio,height=0.2\textheight , width=0.23\textwidth]{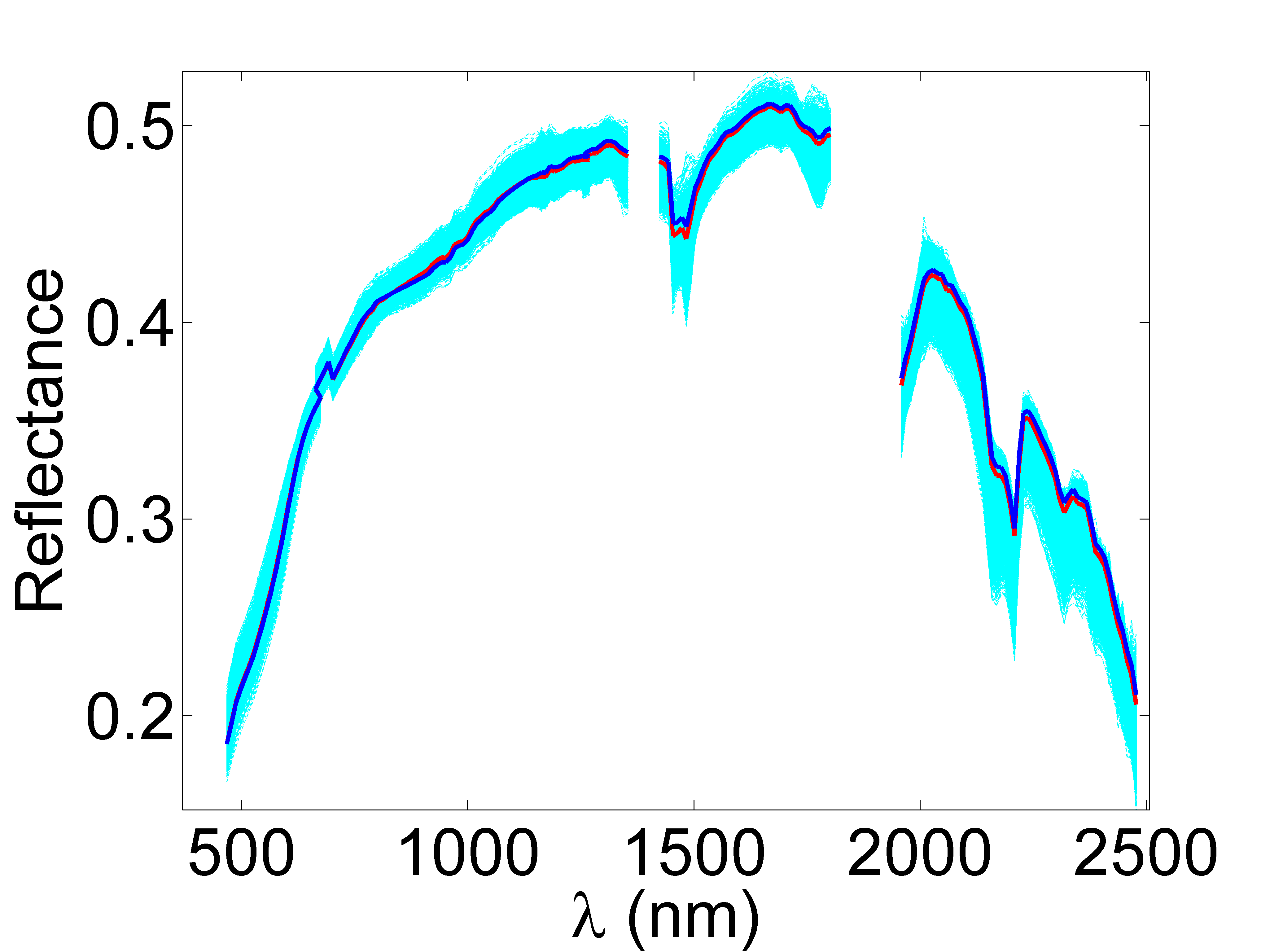}
\label{fig:cuprite_endm3}}
\subfloat[4][Montmorillonite]{
\includegraphics[keepaspectratio,height=0.2\textheight , width=0.23\textwidth]{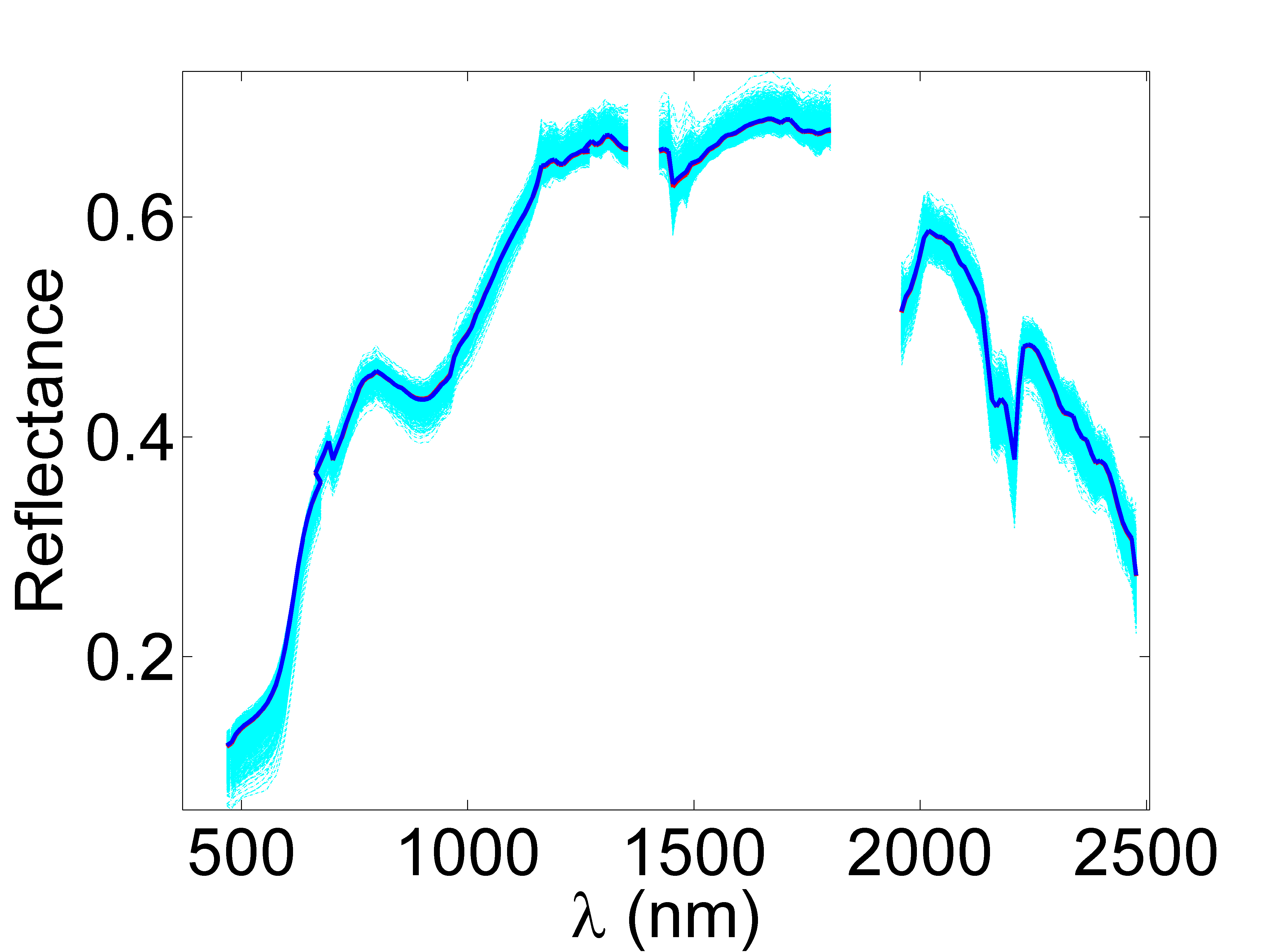}
\label{fig:cuprite_endm4}}
\\
\subfloat[5][Andradite]{
\includegraphics[keepaspectratio,height=0.2\textheight , width=0.23\textwidth]{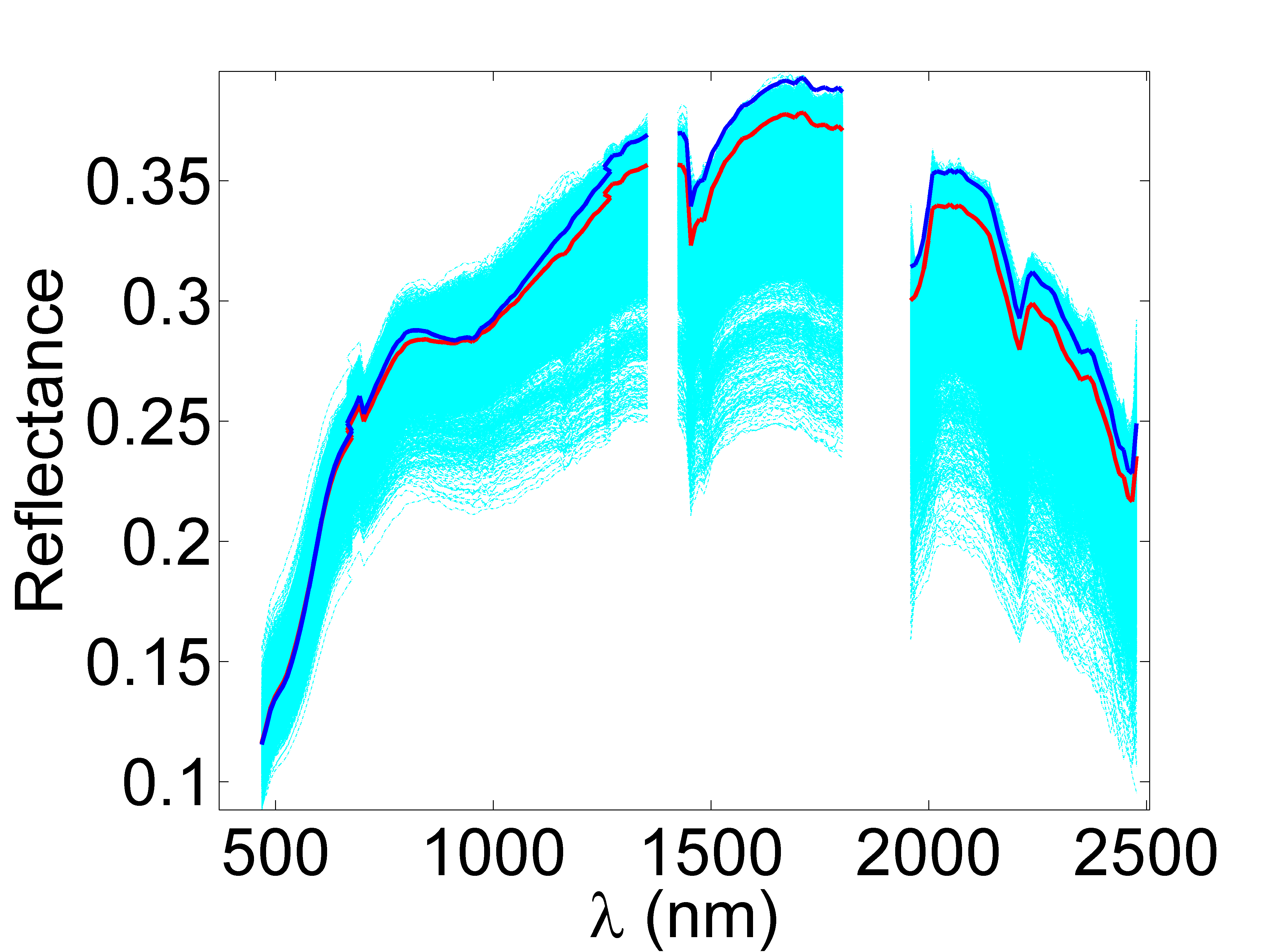}
\label{fig:cuprite_endm5}}
\subfloat[6][Pyrope]{
\includegraphics[keepaspectratio,height=0.2\textheight , width=0.23\textwidth]{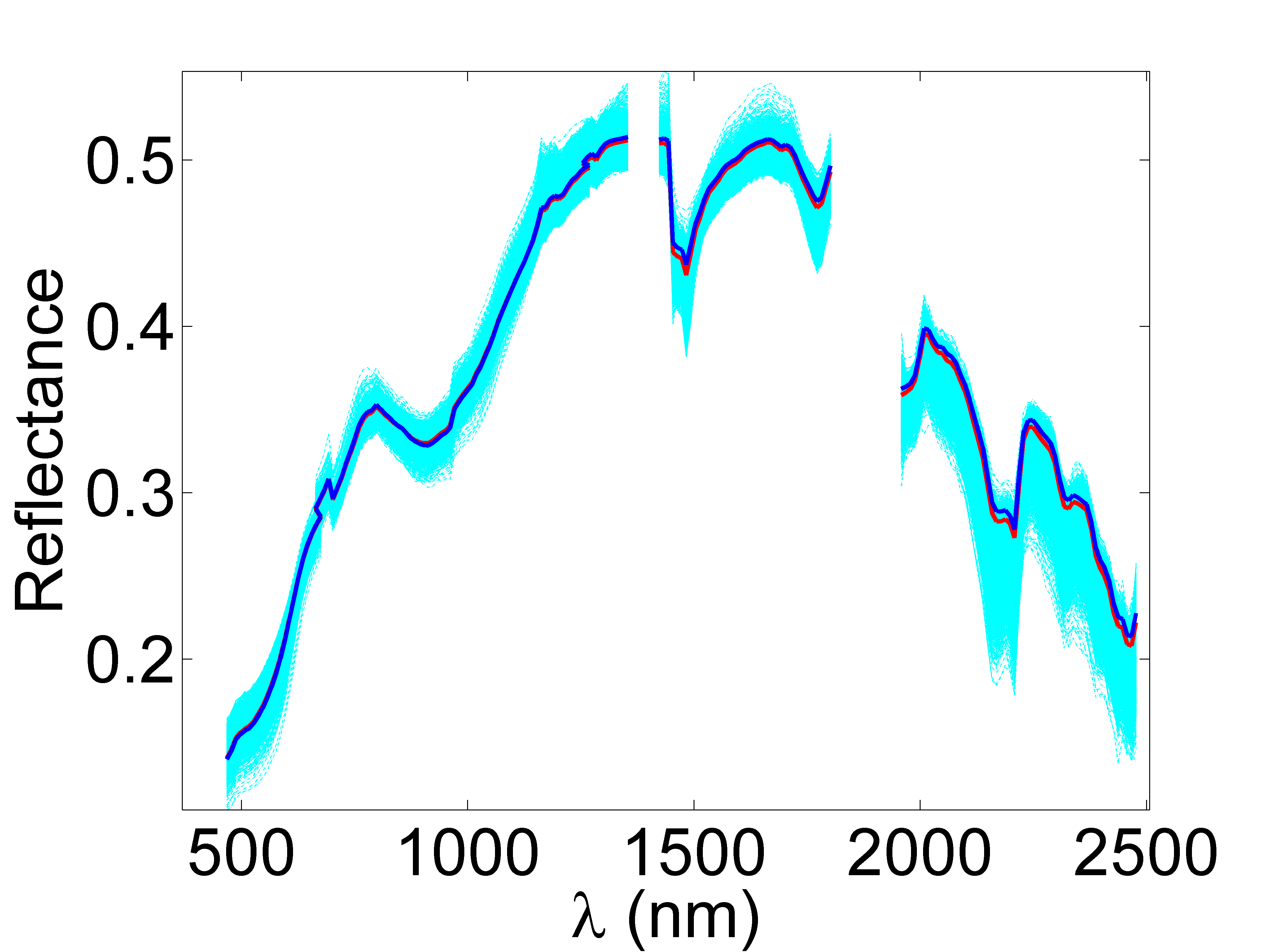}
\label{fig:cuprite_endm6}}
\\
\subfloat[7][Buddingtonite]{
\includegraphics[keepaspectratio,height=0.2\textheight , width=0.23\textwidth]{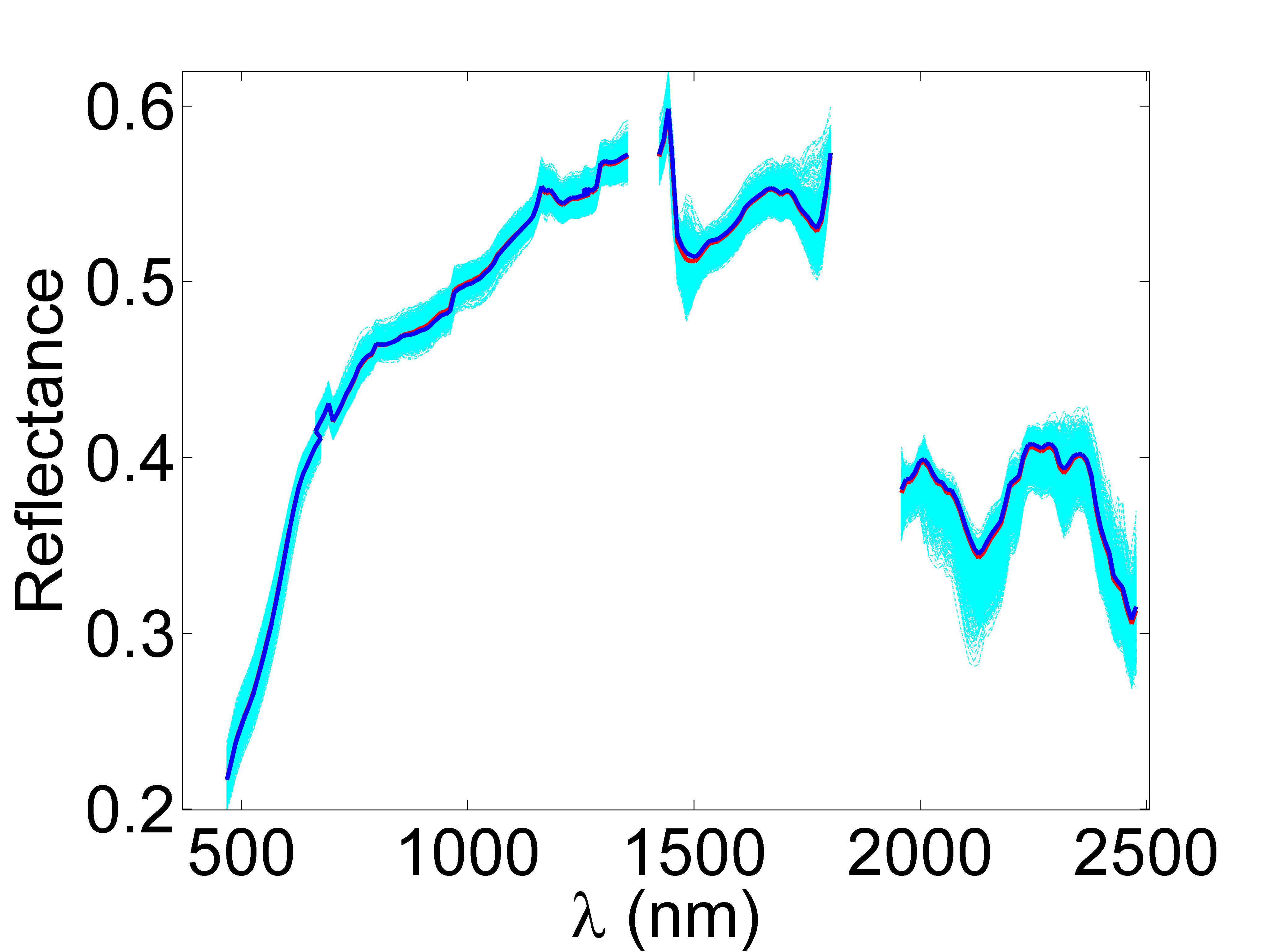}
\label{fig:cuprite_endm7}}
\subfloat[8][Muscovite]{
\includegraphics[keepaspectratio,height=0.2\textheight , width=0.23\textwidth]{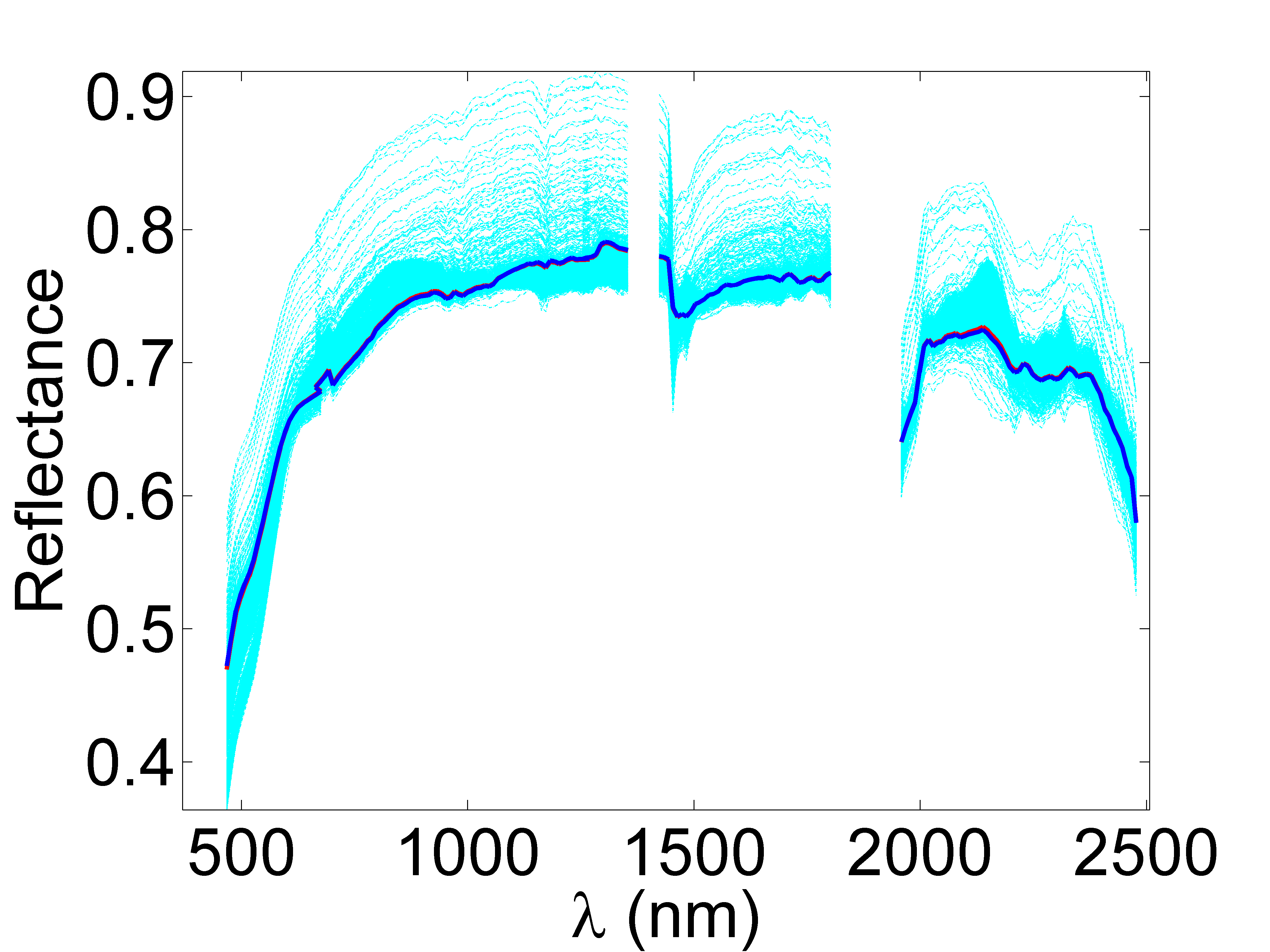}
\label{fig:cuprite_endm8}}
\\
\subfloat[9][Nontronite]{
\includegraphics[keepaspectratio,height=0.3\textheight , width=0.23\textwidth]{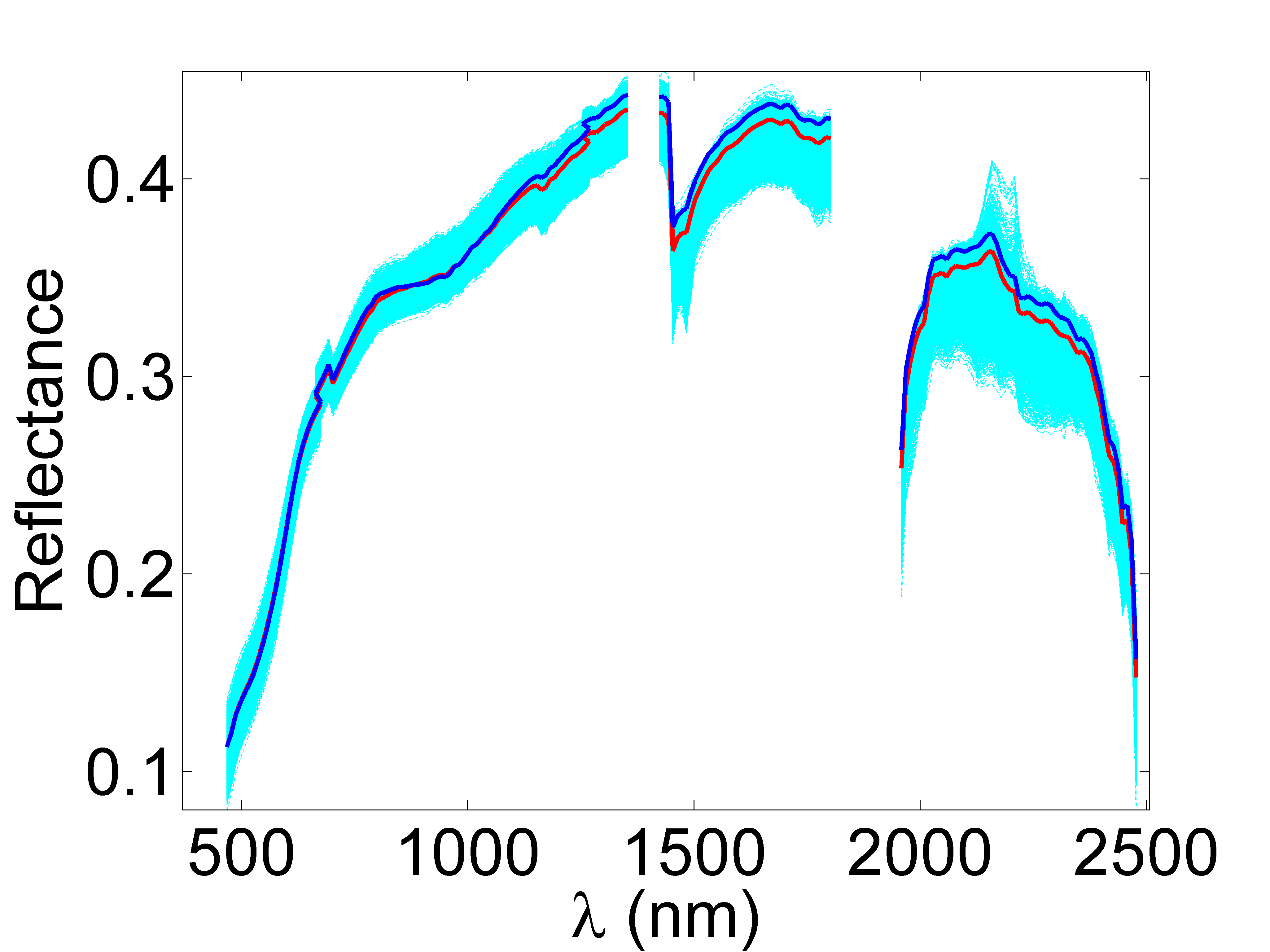}
\label{fig:cuprite_endm9}}
\subfloat[10][Kaolinite]{
\includegraphics[keepaspectratio,height=0.3\textheight , width=0.23\textwidth]{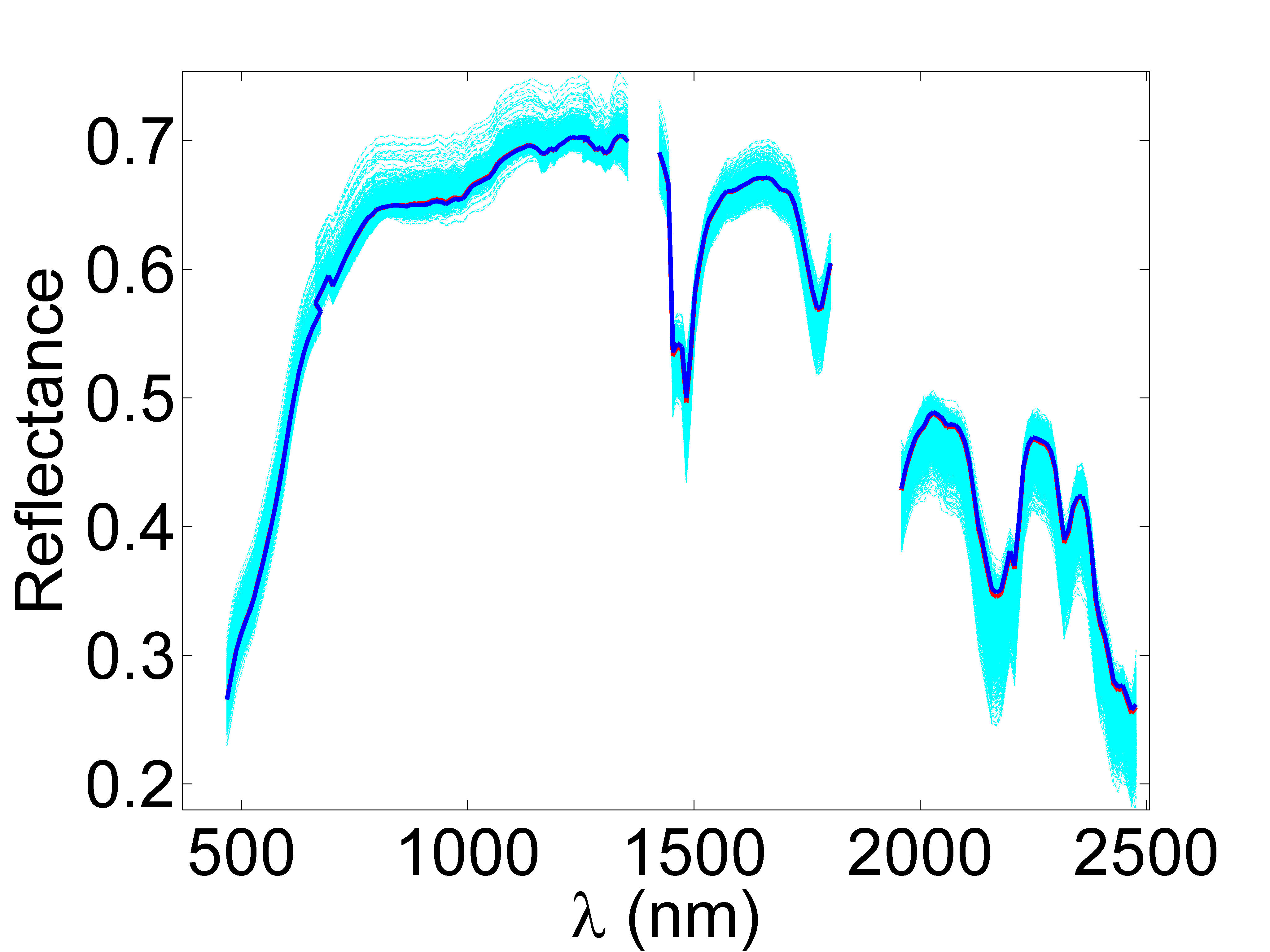}
\label{fig:cuprite_endm10}}
\caption{ssvcaBCD/ADMM endmembers (Cuprite scene). The BCD/ADMM-estimated endmembers (red lines) are given with typical examples of the estimated variability (cyan dotted lines), and the given identification is based on a visual comparison with the results obtained in \cite{Nascimento2005}. The endmembers estimated by the VCA algorithm are given in blue lines for comparison.}
\label{fig:cuprite_endm}
\end{figure}

%
%

\bibliographystyle{IEEEtran}
\nocite{Alfonso2010,Alfonso2011,Almeida2013,Almeida_arXiv2013,Boyd2010,Johnson2013,Matakos2013,Combettes2010,
Ammanouil2014,Ammanouil_12_2014,Heinz2000,Heinz2001,Chan2011tgrs,Nascimento2005,Dobigeon2009,Somers2012jstars2%
,Roberts1998,Goenaga2013,Jing2010,Boyd_book,Zare2014IEEESPMAG,Chen2014,Bioucas2009,Miao2007,Arngren2011,Berman2004,Altmann2011whispers,Fevotte2015,Halimi2011,Dobigeon2014,Vengazones2014WIHSPERS}

\bibliography{strings_all_ref2,all_ref2}

\begin{IEEEbiography}[{\includegraphics[width=1in,height=1.25in,clip,keepaspectratio]{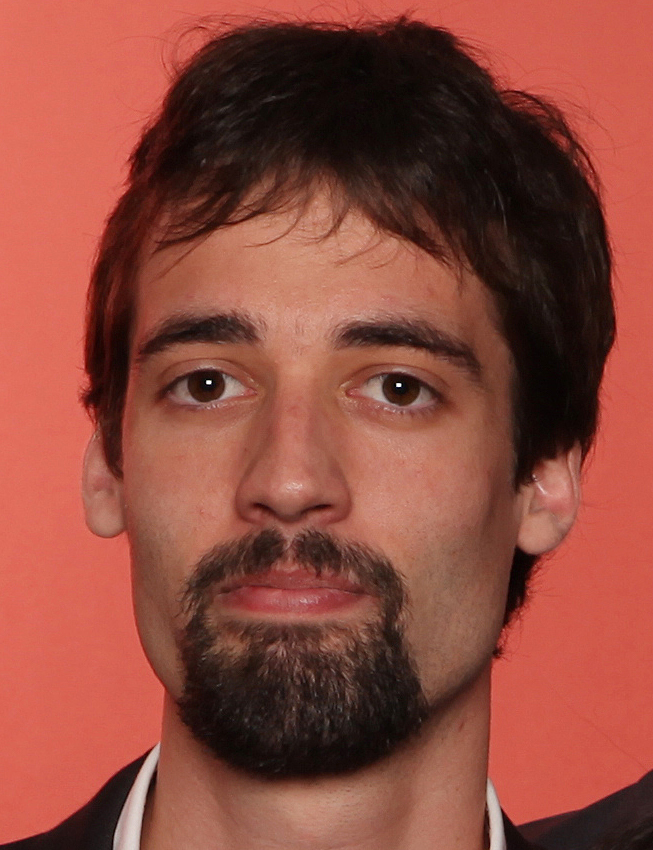}}]{Pierre-Antoine Thouvenin} received the state engineering degree in electrical engineering from ENSEEIHT, Toulouse, France, and the M.Sc. degree in signal processing from the National Polytechnic Institute of Toulouse (INP Toulouse), both in 2014. He is currently working toward the Ph.D. degree within the Signal and Communications Group of the IRIT Laboratory, Toulouse, France. His current research interests include hyperspectral unmixing and optimization techniques.
\end{IEEEbiography}

\begin{IEEEbiography}[{\includegraphics[width=1in,height=1.25in,clip,keepaspectratio]{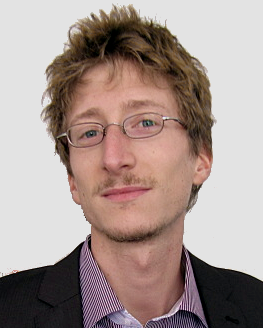}}]{Nicolas Dobigeon} (S’05--SM’08--SM’13) received the state engineering degree in electrical engineering from ENSEEIHT, Toulouse, France, and the M.Sc. degree in signal
processing from the National Polytechnic Institute of Toulouse (INP Toulouse), both in June 2004, as well as the Ph.D. degree and Habilitation \`a Diriger des Recherches in Signal Processing from the INP Toulouse in 2007 and 2012, respectively.
He was a Post-Doctoral Research Associate with the Department of Electrical Engineering and Computer Science, University of Michigan, Ann Arbor, MI, USA, from 2007 to 2008. 

Since 2008, he has been with the National Polytechnic Institute of Toulouse (INP-ENSEEIHT, University of Toulouse) where he is currently an Associate Professor. He conducts his research within the Signal and Communications Group of the IRIT Laboratory and he is also an affiliated faculty member of the Telecommunications for Space and Aeronautics (TeSA) cooperative laboratory.
His current research interests include statistical signal and image processing, with a particular interest in Bayesian inverse problems with applications to remote sensing, biomedical imaging and genomics.
\end{IEEEbiography}

\begin{IEEEbiography}[{\includegraphics[width=1in,height=3in,clip,keepaspectratio]{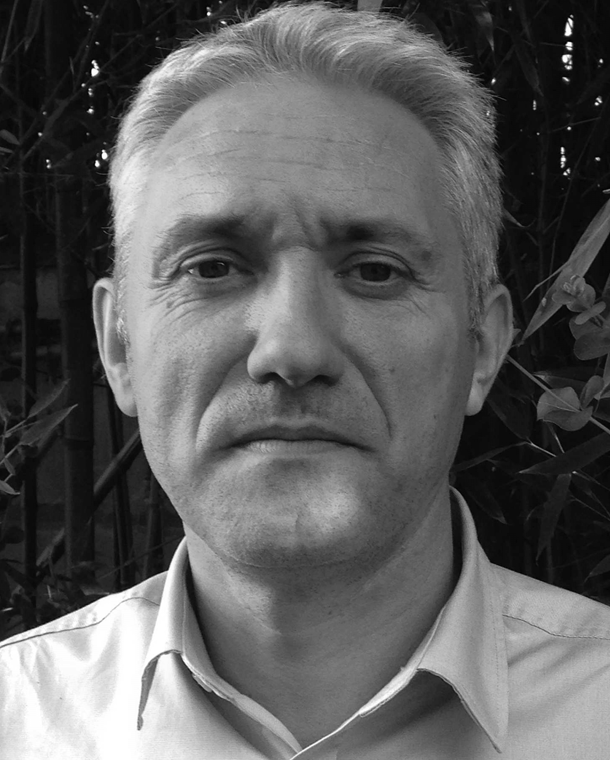}}]{Jean-Yves Tourneret} (SM’08) received the ing\'enieur degree in electrical engineering from ENSEEIHT, Toulouse in 1989 and the Ph.D. degree from the INP Toulouse in 1992. He is currently a professor in the University of Toulouse (ENSEEIHT) and a member of the IRIT laboratory (UMR 5505 of the CNRS). His research activities are centered around statistical signal and image processing with a particular interest to Bayesian and Markov chain Monte Carlo (MCMC) methods. 

He has been involved in the organization of several conferences including the European conference on signal processing EUSIPCO’02 (program chair), the international conference ICASSP’6 (plenaries), the statistical signal processing workshop SSP’12 (international liaisons), the International Workshop on Computational Advances in Multi-Sensor Adaptive Processing CAMSAP 2013 (local arrangements), the statistical signal processing workshop SSP’2014 (special sessions), the workshop on machine learning for signal processing MLSP2014 (special sessions). He has been the general chair of the CIMI workshop on optimization and statistics in image processing hold in Toulouse in 2013 (with F. Malgouyres and D. Kouam\'e) and of the International Workshop on Computational Advances in Multi-Sensor Adaptive Processing CAMSAP 2015 (with P. Djuric). He has been a member of different technical committees including the Signal Processing Theory and Methods (SPTM) committee of the IEEE Signal Processing Society (2001-2007, 2010-present). He has been serving as an associate editor for the IEEE Transactions on Signal Processing (2008-2011) and for the EURASIP journal on Signal Processing (since July 2013).
\end{IEEEbiography}

\vfill

\end{document}